\documentclass[reprint,aip,rsi,amsmath,amssymb,nofootinbib]{revtex4-1}

\usepackage[normalem]{ulem}
\usepackage{graphicx,natbib}
\usepackage{color,soul}
\usepackage{subfigure}
\usepackage{dcolumn}
\usepackage{bm}
\usepackage{hyperref}
\usepackage{url}
\usepackage{multirow}
\hypersetup{
    colorlinks=true,
    linkcolor=blue,
    filecolor=blue,      
    urlcolor=blue,
    citecolor=blue,
}
\usepackage{float}
\newcommand{\etal}{\textit{et al. }}

\begin{document}

\title{Theoretical Evaluation of Electronic  Density-of-states and Transport Effects on Field Emission from $n$-type Ultrananocrystalline Diamond Films}

\author{Oksana Chubenko}
\email[E-mail address: ]{chubenko@gwu.edu}
\affiliation{Department of Physics, The George Washington
University, 725 21st St. NW, Washington, DC 20052, USA}

\author{Stanislav S. Baturin}
\affiliation{PSD Enrico Fermi Institute, The University of Chicago, 5640 S. Ellis Ave, Chicago, IL 60637, USA}

\author{Sergey V. Baryshev}
\affiliation{Department of Electrical and Computer Engineering, Michigan State University, 428 S. Shaw Lane, East Lansing, MI 48824, USA}

\begin{abstract}
In the nitrogen-incorporated ultrananocrystalline diamond ((N)UNCD) films, representing an $n$-type highly conductive two-phase material comprised of $sp^3$ diamond grains and $sp^2$-rich graphitic grain boundaries, the current is carried by a high concentration of mobile electrons within the large-volume grain boundary networks. Fabricated in a simple thin-film planar form, (N)UNCD was found to be an efficient field emitter capable of emitting a significant amount of charge starting at the applied electric field as low as a few V/$\mu$m which makes it a promising material for designing electron sources. Despite the semimetallic conduction, field emission (FE) characteristics of this material demonstrate a strong deviation from the Fowler-Nordheim law in a high-current-density regime when (N)UNCD field emitters switch from a diode-like to resistor-like behavior. Such phenomenon resembles the current-density saturation effect in conventional semiconductors. In the present paper, we adapt the formalism developed for conventional semiconductors to study current-density saturation in (N)UNCD field emitters. We provide a comprehensive theoretical investigation of ($i$) the influence of partial penetration of the electric field into the material, ($ii$) transport effects (such as electric-field-dependent mobility), and ($iii$) features of a complex density-of-states structure (position and shape of $\pi-\pi^*$ bands, controlling the concentration of charge carriers) on the FE characteristics of (N)UNCD. We show that the formation of the current-density saturation plateau can be explained by the limited supply of electrons within the impurity $\pi-\pi^*$ bands and decreasing electron mobility in high electric field. Theoretical calculations are consistent with experiment.
\end{abstract}
\maketitle

\section{\label{sec:introduction}Introduction}

Field emission (FE) of electrons from the nitrogen-incorporated ultrananocrystalline diamond ((N)UNCD), simply fabricated in a form of thin films, has become a subject of numerous experimental studies. \cite{ Wu_2000, Corrigan_2002, Ikeda_2009, Lin_2011, Baryshev_2014} Due to a set of key features defined by the unique diamond/graphite phase nanostructure, these materials can be operated in conditions desirable for efficient electron sources i.e. at low applied electric fields and in a moderate vacuum environment. (N)UNCD field emitters demonstrate a metal-like behavior in the low-current regime. At some critical value of the emission current (and/or critical applied electric field), the behavior of the FE characteristic changes and any further increase of current, due to an applied electric field, occurs at a slower rate which eventually results in a saturation plateau. Such an effect is also observed in other carbon-derived polycrystalline materials with mixed $sp^2$/$sp^3$ hybridization\cite{Liao_1998, Xu_2000, Ducati_2002, Varshney_2011, Cahay_2014} and has been speculated to be due to different phenomena, including: electron tunneling through multiple barriers, \cite{Liao_1998} an intermediate state between the field emission and thermionic emission regimes, \cite{Xu_2000, Ducati_2002} or due to the space-charge effect. \cite{Xu_2000, Cahay_2014} Moreover, the observed saturation effect is a long-known characteristic feature of conventional semiconductors \cite{Arthur_1965, Baskin_1971, Serbun_2013} which has been explored theoretically. \cite{Stratton_1955, Stratton_1962, Baskin_1971}

The original theory describing electron FE from flat metal surfaces by electron tunnelling through a simple triangular potential barrier was developed by Fowler and Nordheim\cite{Fowler_1928} in 1928 and shortly after reconsidered by Nordheim\cite{Nordheim_1928} for Schottky's
image-rounded triangular barrier, and was extended by Murphy and Good\cite{Murphy_1956} in 1956.  Stratton\cite{Stratton_1955, Stratton_1962} later adapted the theory to explain FE from semiconductors assuming a spatially constant Fermi level (zero-current approximation). Baskin \etal\cite{Baskin_1971} reconsidered Stratton's equations and combined the surface tunneling theory with the bulk parameters of the material obtained from a self-consistent solution of Poisson's and Ohm's equations. They showed that both $p$-type and $n$-type semiconductors have a region on a current-voltage characteristic where they deviate from the Fowler-Nordheim (FN) law and linked this behavior to the partial penetration of the electric field into the semiconductor. This effect results in the change of carrier concentration in the near-surface region (creation of the inversion or accumulation layer underneath the surface in a $p$-type and $n$-type semiconductor, respectively) also called the space-charge region. Moreover, they emphasized the significant influence of the field-dependent carrier mobility on the FE characteristics. In Refs.~\onlinecite{Huang_1997, Liu_2006}, this approach was used to explain the behavior of FE characteristics from Si field emitters.

For conventional semiconductors, the surface space-charge layer, which arises in a material due to the change in electrostatic potential between the bulk and the surface, has been studied in depth.\cite{Kingston_1955, Seiwatz_1958, Stratton_1962, Baskin_1971, Huang_1997, Lin_1998} In the energy-band diagram, this effect is modeled as a variation of the position of the CB edge with respect to the Fermi level near the surface of the material. (N)UNCD films and other carbon-derived materials have a unique density-of-states structure represented by a combination of $\pi-\pi^*$ and $\sigma-\sigma^*$ bands associated with $sp^2$- and $sp^3$-hybridized carbon, respectively, present in the fundamental $\Sigma-\Sigma^*$ band gap of diamond.\cite{Robertson_1987, Zapol_2001} When exposed to strong electric fields, the downward band bending may cause the formation of a space-charge layer near the surface of $n$-type (N)UNCD. The carrier concentration and the carrier accumulation rate will be defined by the density-of-states structure and the shape of $\pi-\pi^*$ bands, in particular. Literature review suggests that the effect of the density of states on the FE characteristics of (N)UNCD has not yet been explored.

It should be mentioned that the theories behind FE relate the electric field applied to a material and the current density, but not the experimentally measured emission current. Therefore, current density is a crucial characteristic required to be known in order to compare experimental measurements to theoretical predictions and to enable proper comparison of FE properties of different cathodes. UNCD-based electron sources belong to a class of large-area field emitters.\cite{Forbes_2009} When assembled in a parallel-plate configuration, the current density is conventionally evaluated as an experimentally measured current normalized by the total area of the cathode or anode (by the smallest of the two areas) which remains constant during all measurements. It was observed \cite{Xu_1993, Xu_1994, Zhu_1998} that in polycrystalline diamond films the electron FE originates from the discrete spots randomly distributed over the surface of the cathode. Recently, we have quantitatively shown \cite{Chubenko_2017} that the surface density of emission sites in (N)UNCD films strongly depends on the magnitude of the applied electric field. It means that the emission area, which contributes to the current density, varies with the electric field as well. The methodology\cite{Chubenko_2017} developed for determining the FE area from large-area electron emitters makes it possible to study the current-density saturation effect present in $n$-type (N)UNCD films and to confirm or reject various hypotheses, which were proposed in the past, on the nature of this effect.

The purpose of the present work is to adapt the formalism developed for conventional semiconductors in order to explain the current-density saturation effect in $n$-type polycrystalline-diamond-based electron sources which have a characteristic density-of-states structure. In particular, we show how the FE characteristics of (N)UNCD depend on ($i$) the position of the Fermi level at the surface of the material, ($ii$) the shape and magnitude of the density of $\pi-\pi^*$ states, and ($iii$) the field-dependent mobility.

The paper is organized as follows. In Section~\ref{sec:conventional_scs}, we briefly review the  Stratton-Baskin-Lvov-Fursey formalism, describing the processes and effects in the bulk and at the surface of a semiconductor exposed to a high applied electric field. In Section~\ref{sec:saturation_UNCD}, we adapt the approach described in Section~\ref{sec:conventional_scs} to calculate the current-voltage characteristics of (N)UNCD films. In Section~\ref{sec:comparison} we compare theoretical results to experimental measurements. For convenience,  we provide a brief summary of our recent findings \cite{Chubenko_2017, Chubenko_2017_IVNC, Chubenko_2017_IVNC_2} on the field-dependent FE area of (N)UNCD films. We discuss the formation of experimentally estimated current-density saturation limits and the difficulties associated with evaluation of the FE area from planar polycrystalline diamond materials. Conclusions and outlooks are summarized in Section~\ref{sec:conclusions}.

\section{\label{sec:conventional_scs}Current-Density Saturation in Conventional Semiconductors}

\subsection{Semiconductor in Applied Electric Field}

\begin{figure}[b]
\centering
\includegraphics[width=3.4in]{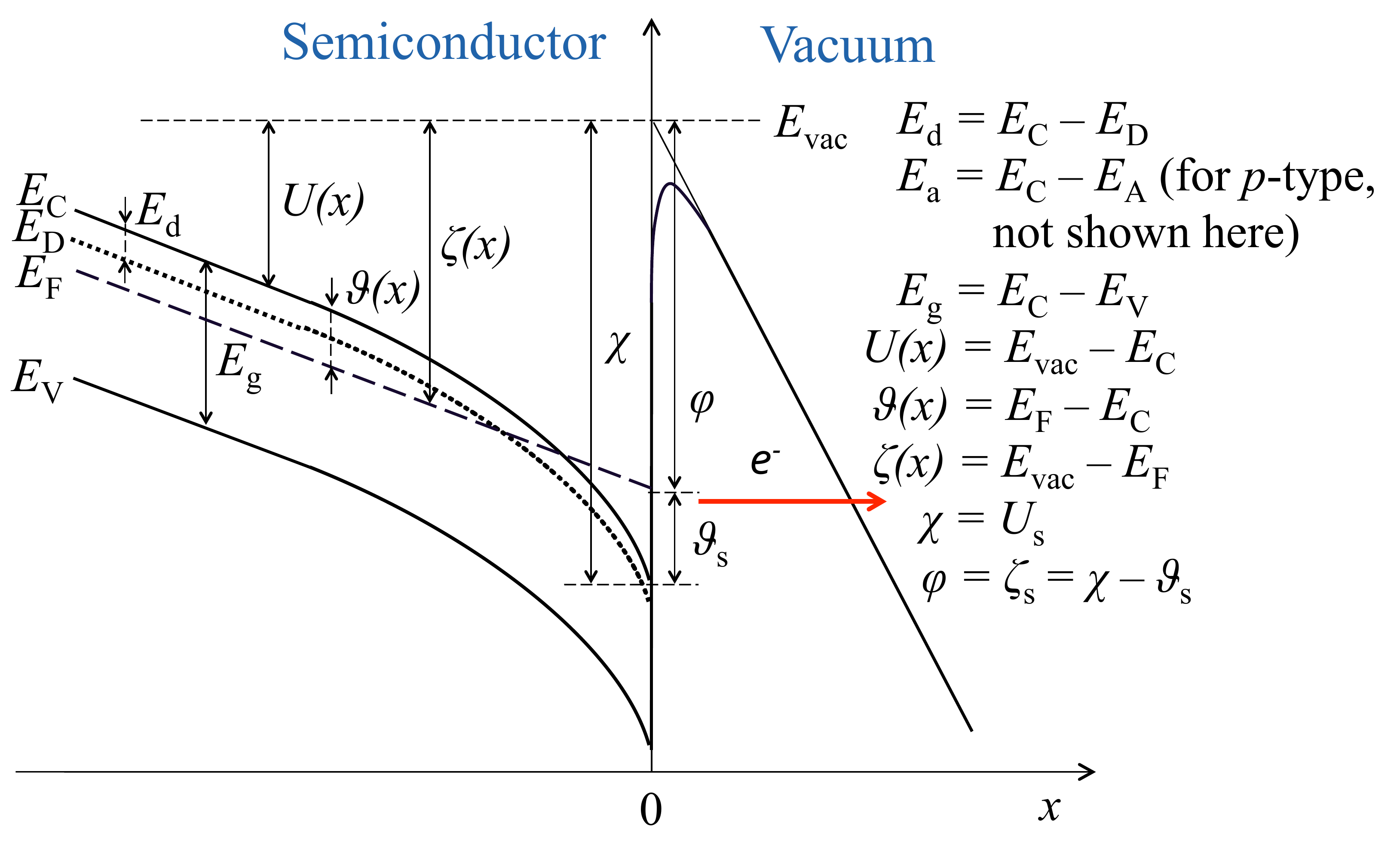}
\caption{Energy-band diagram of an $n$-doped semiconductor in a high electric field.}
\label{diagram_Si}
\end{figure}

We consider a one-dimensional model\cite{Baskin_1971} of a uniformly-doped semi-infinite semiconductor, i.e. the influence of a substrate on the FE process is neglected. When the voltage is applied between a semiconducting field emitter and an anode, the resulting electric field penetrates into the semiconductor and causes the bending of valence and conduction bands. For example, in $n$-type semiconductors a narrow accumulation layer appears at the surface of a field emitter (see Fig.~\ref{diagram_Si}), giving rise to the concentration of free electrons in the CB. Therefore, the FE current density in a low-field regime is limited by the tunnelling probability of CB electrons in the near-surface region through the field-deformed potential barrier and thus demonstrates a metal-like behavior.\cite{Baskin_1971,Liu_2006} 

As the applied electric field increases and penetrates deeper into the interior, the FE current density becomes limited by the supply of electrons. Moreover, scattering mechanisms that involve optical phonons become significant, limiting the hot-carrier mobility at high electric fields.\cite{Baskin_1971} This is the saturation regime.

Definition of all energies used in the calculations is given in Fig.~\ref{diagram_Si}, where we use an energy diagram of an $n$-type semiconductor as an example. $E_\text{C}$ and $E_\text{V}$ are the conduction band minimum (CBM) and the valence band maximum (VBM), respectively. $E_\text{F}$ and $E_\text{D}$ are the Fermi level and the donor level (for $p$-type semiconductor, the acceptor level $E_\text{A}$ placed close to the VBM is used instead), respectively. $E_\text{d}$ is the energy of a donor level with respect to the CBM and $E_\text{g}$ is the band-gap energy. $U(x)$ and $\zeta(x)$ are the functions characterizing the position of the CBM and the Fermi level within the material, respectively, with respect to the vacuum level $E_{\text{vac}}$. $\chi$ and $\varphi$ are the electron affinity and the work function, respectively. Function $\vartheta (x)$ defines the position of a CBM with respect to the Fermi level and thus determines the carrier concentration at any point in the material. The subscript $s$ will be used to distinguish between the values measured at the surface (i.e. at $x=0$) and those measured in the interior. 

\subsection{\label{sec:Poisson}Poisson's Equation}

To describe the space-charge, or the band-bending, region in a conventional semiconductor, an approach similar to the one described in Ref.~\onlinecite{Baskin_1971} can be used. In Poisson's equation,
\begin{equation}
\frac{d^2V}{dx^2}=-\frac{\rho}{\kappa\varepsilon_0},
\end{equation}
the potential difference $V$ is related to the potential energy $U$ as 
\begin{equation}
V=U/q,
\end{equation}
where $q$, the electron charge, allows for converting between V to eV. $\kappa$ is the dielectric constant of a semiconductor and $\varepsilon_0$ is the permittivity of free space. The charge density $\rho$ in the semiconductor bulk is given by (see, for example, Ref.~\onlinecite{Li_Semiconductor})
\begin{equation}
\rho=-q(n-p+N_\text{a}^--N_\text{d}^+).
\end{equation}
Here the volume densities of electrons $n$, holes $p$, negative acceptor ions $N_\text{a}^-$, and positive donor ions $N_\text{d}^+$ are defined as 
\begin{equation}
\begin{split}
n&= N_\text{C} F_{1/2}\big[(E_\text{F}-E_\text{C})/k_\text{B}T\big],\\
p&= N_\text{V} F_{1/2}\big[(E_\text{V}-E_\text{F})/k_\text{B}T\big],\\
N_\text{a}^- &= \frac{N_\text{a}}{1+2 \exp\big[(E_\text{A}-E_\text{F})/k_\text{B}T\big]},\\
N_\text{d}^+ &= \frac{N_\text{d}}{1+2 \exp\big[(E_\text{F}-E_\text{D})/k_\text{B}T\big]},
\end{split}
\label{concentrations}
\end{equation}
where $
N_\text{C} = 2 \big{(}2 \pi m_\text{e}^* k_\text{B}T/h^2\big{)}^{3/2}$ is the effective density of states in the CB and $
N_\text{V} = 2 \big{(}2 \pi m_\text{h}^* k_\text{B}T/h^2\big{)}^{3/2}$ is the effective density of states in the VB. Here, $m_\text{e}^*$ and $m_\text{h}^*$ are the density-of-states effective mass of electrons and holes, respectively, $k_\text{B}$ is Boltzmann's constant, $T$ is the lattice temperature, and $h$ is  Planck's constant. $F_{1/2}(\eta)=2\pi^{-1/2}\int_0^{\infty}x^{1/2}\big[1+\exp(x-\eta)\big]^{-1}dx$ is the Fermi-Dirac integral of the order $1/2$. $N_\text{a}$ and $N_\text{d}$ are the concentrations of acceptor and donor atoms, respectively.

It is convenient to introduce a dimensionless variable 
\begin{equation}
y(x)\equiv \frac{\vartheta(x)}{k_\text{B}T}=\frac{E_\text{F}-E_\text{C}}{k_\text{B}T}.
\label{y_def}
\end{equation}
In terms of $y$, Eqs.~\ref{concentrations} can be rewritten as
\begin{equation}
\begin{split}
n(y)&= N_\text{C} F_{1/2}(y),\\
p(y)&= N_\text{V} F_{1/2}(-E_\text{g}/k_\text{B}T-y),\\
N_\text{a}^-(y) &= \frac{N_\text{a}}{1+2 \exp(-E_\text{a}/k_\text{B}T-y)},\\
N_\text{d}^+(y) &= \frac{N_\text{d}}{1+2 \exp(E_\text{d}/k_\text{B}T+y)}.
\end{split}
\label{concentrations2}
\end{equation}
The resulting equation to solve is
\begin{equation}
\frac{d^2(U/q)}{dx^2}=-\frac{\rho\big[y(x)\big]}{\kappa\varepsilon_0}.
\label{Poisson}
\end{equation}

The relation between the spatial coordinate $x$ and the variable $y$ can be obtained by the Ohm's law,
\begin{equation}
j=q\big[n(y)\mu_\text{e}+p(y)\mu_\text{h}\big]\frac{d(E_\text{F}/q)}{dx},
\label{j}
\end{equation}
where $j$ is the current density, and $\mu_\text{e}$ and $\mu_\text{h}$ denote the electron and hole mobilities, respectively. The low-field carrier mobility in semiconductors depends on the temperature and on the doping density, $\mu=\mu(T,N)\equiv\mu_0$. The doping-dependence of both majority-\cite{Masetti_1983, Reggiani_2002} and minority-carrier mobility\cite{Caughey_1967, Reggiani_2002, Swirhun_1986_EDL, Swirhun_1986} at room temperature are described by the empirical expression
\begin{equation}
\mu_0(N)=\mu_{\text{min}}+\frac{\mu_{\text{max}}-\mu_{\text{min}}}{1+(N/C_\text{r})^{\alpha}}-\frac{\mu_1}{1+(C_\text{s}/N)^{\delta}},
\label{mu(N)}
\end{equation}
where $\mu_{\text{min}}$, $\mu_{\text{max}}$, $\mu_1$, $C_\text{r}$, $C_\text{s}$, $\alpha$, and $\delta$ are the parameters obtained by fitting experimental data (e.g., see data for Si in Ref.~\onlinecite{Masetti_1983} ). 

Moreover, it was revealed experimentally\cite{Ryder_1953, Gunn_1956, Prior_1959, Schweitzer_1965} that the mobility of electrons and holes in semiconductors become field-dependent when the internal electric field $\mathcal{F}$ reaches its critical value $\mathcal{F}_0$. Thus, the linear relation between the current density $j$ and the electric field $\mathcal{F}$, $j=qn\mu_0\mathcal{F}$ (or between the drift velocity $v_\text{d}$ and the electric field $\mathcal{F}$, $v_\text{d}=\mu_0\mathcal{F}$), fails at $\mathcal{F}>\mathcal{F}_0$. The theoretical interpretation of this phenomenon\cite{Shockley_1951} is that the carrier mobility in semiconductors is limited by the scattering by acoustic phonons at weak fields ($\mathcal{F}\sim 10^2$ V cm$^{-1}$) and by the scattering by optical phonons at larger fields ($\mathcal{F}\sim 10^3-10^4$ V cm$^{-1}$). In the first case, the mobility remains constant with the electric field, and in the second case, it decreases according to the parametric expression\cite{Caughey_1967} 
\begin{equation}
\mu(\mathcal{F})=\frac{\mu_0}{\big[1+(\mathcal{F}/\mathcal{F}_0)^{\gamma}\big]^{1/\gamma}},
\label{mu(F)}
\end{equation}
where $\gamma=1$ for holes and $\gamma=2$ for electrons. Note that in this section we use $\mathcal{F}$ to denote the internal electric field. $F_\text{s}$ and $E$ will be used below to denote the surface electric field in vacuum and the applied electric field, respectively.

Using the relation between the internal electric field $\mathcal{F}$ and the electrostatic potential $V$,
\begin{equation}
\mathcal{F}\equiv-\frac{dV}{dx}=-\frac{d(U/q)}{dx}=\frac{d(E_\text{C}/q)}{dx},
\label{FU}
\end{equation}
and Eq. \ref{y_def}, Eq. \ref{j} can be rewritten as
\begin{equation}
j=q\big[n(y)\mu_\text{e}(\mathcal{F})+p(y)\mu_\text{h}(\mathcal{F})\big](k_{\text{B}}T/q\frac{dy}{dx}+\mathcal{F}).
\end{equation}
Finally, we obtain
\begin{equation}
dx=\frac{k_\text{B}T/q}{j/\Big\{q\big[n(y)\mu_\text{e}(\mathcal{F})+p(y)\mu_\text{h}(\mathcal{F})\big]\Big\}-\mathcal{F}}dy.
\label{xy}
\end{equation}

With Eq.~\ref{FU} and Eq.~\ref{xy}, Poisson's equation~\ref{Poisson} becomes\cite{Baskin_1971}
\begin{equation}
\frac{d\mathcal{F}}{dy}=-\frac{k_\text{B}T/q}{\kappa\varepsilon_0}\frac{\rho(y)}{\mathcal{F}-j/\Big\{q\big[n(y)\mu_\text{e}(\mathcal{F})+p(y)\mu_\text{h}(\mathcal{F})\big]\Big\}}.
\label{Poisson_final}
\end{equation}
This differential equation can be solved numerically for a fixed $j$ using the boundary condition
\begin{equation}
\mathcal{F}(y)|_{y=y_\text{b}}=\mathcal{F}_\text{b},
\end{equation}
where $\mathcal{F}_\text{b}$ can be found from Ohm's law in the bulk,
\begin{equation}
j=q\big[n(y_\text{b})\mu_\text{e}(\mathcal{F}_\text{b})+p(y_\text{b})\mu_\text{h}(\mathcal{F}_\text{b})\big]\mathcal{F}_\text{b}.
\end{equation}
Here $y_\text{b}$ corresponds to the position of the Fermi level in the bulk material and can be obtained from the charge neutrality condition in thermal equilibrium which is mathematically expressed as $\rho(y)|_{y=y_\text{b}}=0$. 

\subsection{Stratton's Equations}

In the simplified form, Stratton's equations\cite{Stratton_1962} for the electron current densities $j_{\text{em}}^-$ and $j_{\text{em}}^+$ emitted from the semiconductor surfaces with negative ($y_\text{s}<0$) and positive ($y_\text{s}>0$) Fermi energies, respectively, under the electric field at the surface $F_\text{s}$ can be written as
\begin{equation}
\begin{split}
j_{\text{em}}^-(F_\text{s},y_\text{s})&=A_1\exp\big{[}-B_1\frac{\chi^{3/2}}{F_\text{s}}v(Y_1)\big{]}\exp(y_\text{s})\\ &\mbox{}\quad\times\Big[1-C_1\frac{\chi^{1/2}}{F_\text{s}}t(Y_1)k_\text{B}T\Big]^{-1},
\end{split}
\label{jneg_eq}
\end{equation}
where $A_1=4\pi m_0 q (k_\text{B}T)^2/h^3$, $B_1=8\pi\sqrt{2m_0}/(3qh)$, and $C_1=4\pi\sqrt{2 m_0}/(qh)$. Numerically, $A_1=1.08\times 10^7$ A cm$^{-2}$ for $T=300$ K,  $B_1=6.83\times 10^7$ V eV$^{-3/2}$cm$^{-1}$, and $C_1=1.02\times 10^8$ eV$^{-1/2}$cm$^{-1}$. $\chi$ is measured in eV and $F_\text{s}$ in V cm$^{-1}$. $v$ and $t$ are special elliptic functions (known as Nordheim functions) of the argument $Y_1=C\sqrt{F_\text{s}}/\chi$, where $C=\sqrt{q^3/(4\pi\epsilon_0)}\sqrt{(\kappa-1)/(\kappa+1)}=3.79\times 10^{-4}\sqrt{(\kappa-1)/(\kappa+1)}$ eV cm$^{1/2}$ V$^{-1/2}$. 
\begin{equation}
\begin{split}
j_{\text{em}}^+(F_\text{s},y_\text{s}) &= \frac{A_2}{t^2(Y_2)}\frac{F_\text{s}^2}{\varphi(y_\text{s})}\exp\big{[}-B_1\frac{\varphi^{3/2}(y_\text{s})}{F_\text{s}}v(Y_2)\big{]}\\
&\mbox{}\quad\times\Big\{1
-\exp\big[-B_2\frac{\varphi^{1/2}(y_\text{s})}{F_\text{s}}y_\text{s} t(Y_2)\big]\\
&\mbox{}\quad\times\Big[1+B_2\frac{\varphi^{1/2}(y_\text{s})}{F_\text{s}}y_\text{s} t(Y_2)\Big]\Big\},
\end{split}
\label{jpos_eq}
\end{equation}
where $A_2=q^3/(8\pi h)=1.54\times 10^{-6}$ A eV V$^{-2}$ and $B_2=4\pi\sqrt{2m_0}k_{\text{B}}T/(qh)=2.65\times 10^6$ V eV$^{-1/2}$ cm$^{-1}$. $Y_2=C\sqrt{F_\text{s}}/\varphi(y_\text{s})$. We use an approximate expression for the Nordheim function, $v(Y)$, given by \cite{Forbes_2006}
\begin{equation}
\begin{split}
v(Y)\approx 1-Y^2+\frac{1}{3}Y^2\ln(Y),
\end{split}
\end{equation}
and assume that $t(Y)\approx1$.
 
In Eqs.~\ref{jneg_eq} and \ref{jpos_eq}, $j_{\text{em}}$ can be used as a fixed parameter\cite{Baskin_1971} and $F_\text{s}$ can be plotted as a function of $y_\text{s}$ for different values of $j_{\text{em}}$. Solution examples of both Poisson's and Stratton's equations will be shown in insets of Fig. \ref{jposneg} in Section~\ref{sec:application_of_eqs}. 

\subsection{Combined Poisson's and Stratton's Equations}

Self-consistent graphical solutions of Poisson's and Stratton's equations give the value of $\mathcal{F}_\text{s}$ for each $j$ specified.\cite{Baskin_1971} Poisson's equation provides the solution for the internal electric field $\mathcal{F}(y_\text{s})$, i.e. at the interior boundary of the material/vacuum interface, whereas the Stratton's equations give the solution for the local electric field $F_\text{s}(y_\text{s})$ realized at the emitting surface outside in vacuum, such that $\mathcal{F}(y_\text{s})=F_\text{s}(y_\text{s})/\kappa$. All $j$ vs. $F_\text{s}$ curves shown in this work were obtained by this method. The local electric field at the surface is related to the magnitude of the applied electric field $E$ through the enhancement factor $\beta$ as
\begin{equation}
F_\text{s} = \beta E.
\label{beta}
\end{equation} 
Eventually, $j$ vs. $E$ curves can be calculated and compared to the experimental results using $\beta$ as an adjustable parameter.

\section{\label{sec:saturation_UNCD}Electronic Density-of-States and Transport Effects on Field Emission from (N)UNCD Films}

\subsection{Subtleties of Electron Conduction in (N)UNCD Films}

It has been shown both theoretically by the tight-binding simulations \cite{Zapol_2001} and experimentally by the scanning tunneling-field emission microscopy \cite{Karabutov_1999, Harniman_2015} 
that the electron field emission in (N)UNCD films originates from the grain boundaries rather than from sharp grains. It was also shown\cite{Zapol_2001, Corrigan_2002} that the electronic properties of (N)UNCD films are controlled by the $sp^2$-bonded carbon atoms present in the grain boundaries of the material. A graphitic $sp^2$ phase places impurity bands ($\pi-\pi^*$ bands) inside the fundamental diamond band gap ($\Sigma-\Sigma^*$ bands) that effectively reduces the optical band gap of polycrystalline diamonds. The addition of nitrogen during the growing process increases the density of states in $\pi-\pi^*$ bands,\cite{Zapol_2001, Chen_2001} causes delocalization of electron states,\cite{Beloborodov_2006, Achatz_2006} and leads to an increase in both the grain size and grain-boundary width,\cite{Bhattacharyya_2001, Birrell_2002} giving rise to a number of unique properties. For instance, compared to the nearly insulating UNCD, (N)UNCD becomes a highly conductive material with the current being carried by a high concentration of highly mobile electrons within the large-volume grain boundary networks. Experimental measurements \cite{Bhattacharyya_2001, Williams_2004, Achatz_2006, Ikeda_2008} show that the conductivity is of $n$-type. With increasing nitrogen content, the conductivity in (N)UNCD becomes insensitive to temperature,\cite{Bhattacharyya_2001, Achatz_2006} showing a quasimetallic behavior.\cite{Achatz_2006} Conduction was explained to be due to the existence of delocalized interband-gap defect bands. This is different from low-conductive (N)UNCD films, in which electron transport was found to occur via a variable range hopping mechanism across localized states near the Fermi level.\cite{Achatz_2006, Beloborodov_2006}

(N)UNCD films are characterized by a high value of the field enhancement factor $\beta$, typically several hundred. The mechanism of the field enhancement in (N)UNCD films and other materials with mixed $sp^2/sp^3$ bonding is commonly attributed\cite{Ilie_2000, Carey_2001} to the dielectric inhomogeneities within the film originating from the difference between conductive spatially-localized $sp^2$-rich clusters surrounded by a more insulating $sp^3$ matrix. The field lines focus onto the localized conductive grain boundaries, providing large local electric fields and locally decreasing material work function.\cite{Ilie_2000} Therefore, the enhancement is determined by the geometry of grain boundaries and/or the presence of a space-charge region. \cite{Ilie_2000}
\begin{figure}[t]
\centering
\includegraphics[width=3.4in]{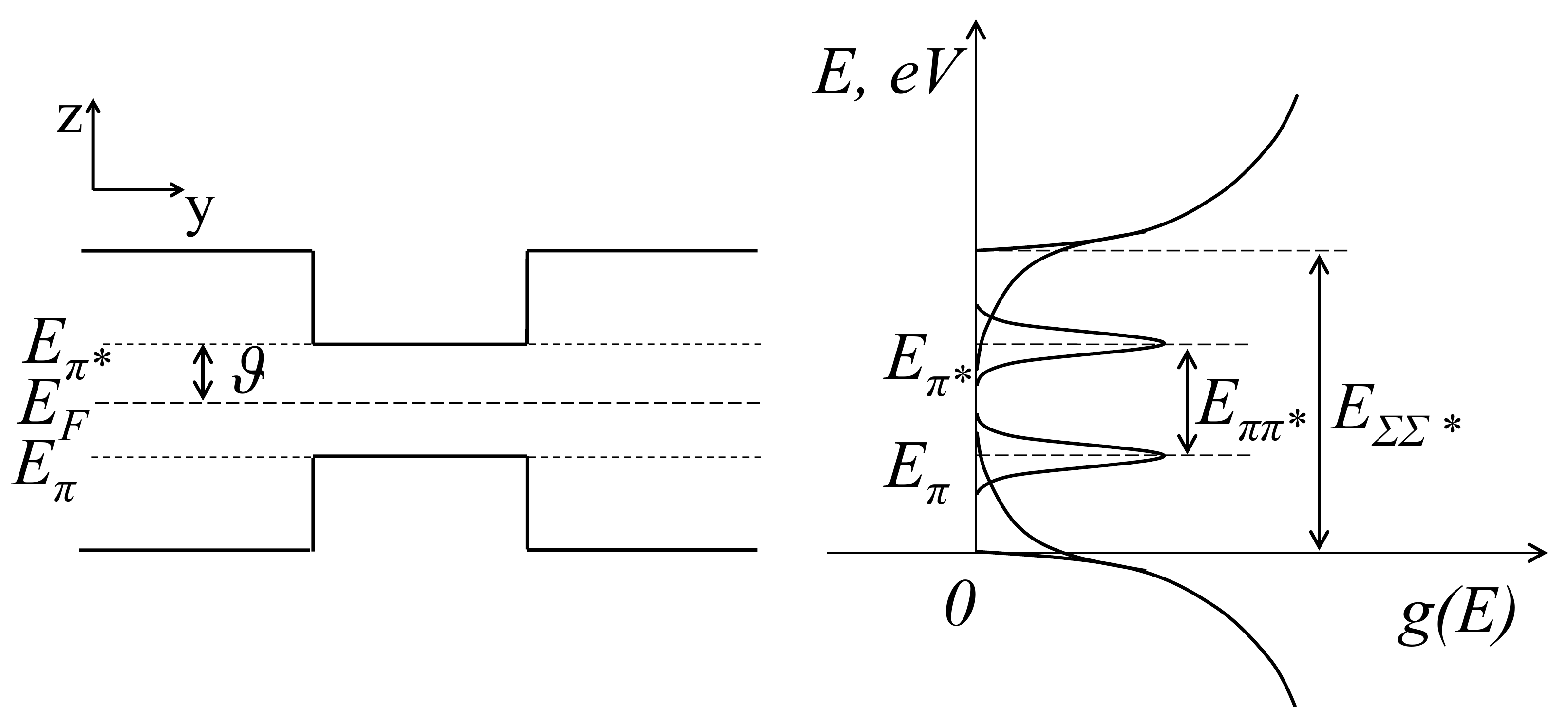}
\caption{Band-structure model and the density-of-states model of $sp^2$-rich polycrystalline diamond films. The narrow $pseudo$ band gap corresponds to the separation between $\pi$ and $\pi^*$ bands introduced by $sp^2$ carbon atoms, and the wide $fundamental$ band gap corresponds to the separation between $\Sigma$ and $\Sigma^*$ bands.}
\label{densities_model}
\end{figure}

\subsection{Density-of-States Model}

The energy-band structure of (N)UNCD films and other $sp^2$-rich polycrystalline materials is conventionally modeled \cite{Robertson_1987} as a combination of $\pi$ and $\sigma$ bands associated with $sp^2$ and $sp^3$ carbon, respectively, present in the band gap of diamond. It was shown \cite{Dasgupta_1991, Nesladek_1996, Zammit_1998} that the density of states in the $\pi-\pi^*$ bands can be approximated by Gaussian functions centered at $E_{\pi}$ and $E_{\pi^*}$ and described by the variance $w^2$ (see Fig.~\ref{densities_model}). The extended $\Sigma$ bands and $\sigma$ band tails can be represented by the square-root functions and the exponential functions decaying into the band-gap region, respectively. Assuming that the $\pi$ and $\pi^*$, $\sigma$ and $\sigma^*$, as well as $\Sigma$ and $\Sigma^*$ bands are symmetric about the diamond midgap energy, \cite{Robertson_1987} the expressions for densities of states can be written as
\begin{equation}
\begin{split}
g_{\pi}(E) &= N_{\pi_{\text{max}}}\exp\big[-\frac{(E-E_{\pi})^2}{2w^2}\big],\\
g_{\pi^*}(E) &= N_{\pi_{\text{max}}}\exp\big[-\frac{(E-E_{\pi^*})^2}{2w^2}\big],\\
g_{\sigma}(E) &= \begin{cases}
D\sqrt{E_\text{m}-E_{\Sigma\Sigma^*}}\exp\Big[\frac{-(E-E_{\Sigma\Sigma^*})-E_\text{m}}{E_0}\Big],\\
\qquad\qquad\qquad\qquad E>E_{\Sigma\Sigma^*}-E_\text{m},\\
D\sqrt{-E},\qquad E< E_{\Sigma\Sigma^*}-E_\text{m},
\end{cases}\\
g_{\sigma^*}(E) &= \begin{cases}
D\sqrt{E_\text{m}-E_{\Sigma\Sigma^*}}\exp\big[\frac{E-E_\text{m}}{E_0}\big],\quad E<E_\text{m},\\
D\sqrt{E-E_{\Sigma\Sigma^*}},\quad E>E_\text{m}.
\end{cases}
\end{split}
\label{densities1}
\end{equation}
Here all energies are measured with respect to the valence band maximum of diamond. $N_{\pi_{\text{max}}}$ is the magnitude of Gaussian functions at $E=E_{\pi,\pi^*}$. $E_{\pi,\pi^*}=E_{\Sigma\Sigma^*}/2\mp E_{\pi\pi^*}/2$, where $E_{\Sigma\Sigma^*}$ is the energy separation between the extended states (fundamental band gap), and $E_{\pi\pi^*}$ is the energy separation between the localized $\pi$ states (pseudo band gap) as shown in Fig.~\ref{densities_model}. Constant $D$ as well as $w$, $E_\text{m}$ (the energy at which $\sigma$ merges with $\Sigma$), $E_0$ (the energy that determines the slope of band tail decay), and $E_{\pi\pi^*}$ are the fitting parameters that can be extracted from experimental absorption spectra. 

Optical transitions under the photoexcitation with the energy $\hbar \omega$ are described by the absorption coefficient $\alpha$ that is given at $T=0$ K as \cite{Tauc,Yu}
\begin{equation}
\alpha(\hbar \omega)=\frac{K}{\hbar \omega}\int g_\text{i}(E)g_\text{f}(E+\hbar \omega) dE,
\end{equation}
where $g_{\text{i,f}}$ are the densities of the initial and final states. The constant $K$ is given by 
\begin{equation}
K = \frac{1}{4\pi\varepsilon_0}\frac{2\pi^2q^2\hbar P^2 L_{\text{loc}}^3}{m_0^2cn},
\label{K}
\end{equation}
where $\hbar$ is the reduced Planck's constant, $P$ is the optical matrix element, $L_{\text{loc}}$ is the localization length, $m_0$ is the free electron mass, $c$ in the speed of light, and $n$ is the refractive index of the material. We assume that $P$ has the same value as in a crystalline material, i.e. $P\approx h/a$,\cite{Tauc} where $a=2.448$ $\AA$ \cite{Alzahrani_2009} is the lattice constant of graphite. 
Using physically reasonable values for other parameters (the localization length $L_{\text{loc}}=10$ $\AA$ which is of the order of a grain-boundary width of (N)UNCD\cite{Beloborodov_2006} and a refractive index $n=2$ as that of graphite\cite{Pelton_1998}), $K$ can be estimated as $K\approx3.44\times10^{-37}$ eV$^2$cm$^5$. From this, the optical transitions between the $\pi-\pi^*$ and the $\pi-\sigma^*$ bands can be written in the following form
\begin{equation}
\begin{split}
\alpha_{\text{abs}}^{\pi\pi^*}(\hbar \omega) &\propto \frac{B_1}{\hbar \omega},\\
\alpha_{\text{abs}}^{\pi\sigma^*}(\hbar \omega) &\propto \frac{B_2}{\hbar \omega},
\end{split}
\label{alpha}
\end{equation}
where $B_1=KN_{\pi_{\text{max}}}^2$ and $B_2=KN_{\pi_{\text{max}}}D$ are often used\cite{Nesladek_1996, Zammit_1998, Achatz_2006_APL} as generalized fitting parameters available in the literature. This allows the density of states, defined in Eq.~\ref{densities1}, to be rewritten in terms of coefficient $K$ and fitting parameters $E_{\Sigma\Sigma^*}$, $E_{\pi\pi^*}$, $w$, $E_\text{m}$, $E_0$, $B_1$, and $B_2$. 

\begin{figure}[b]
\centering
\includegraphics[width=3.2in]{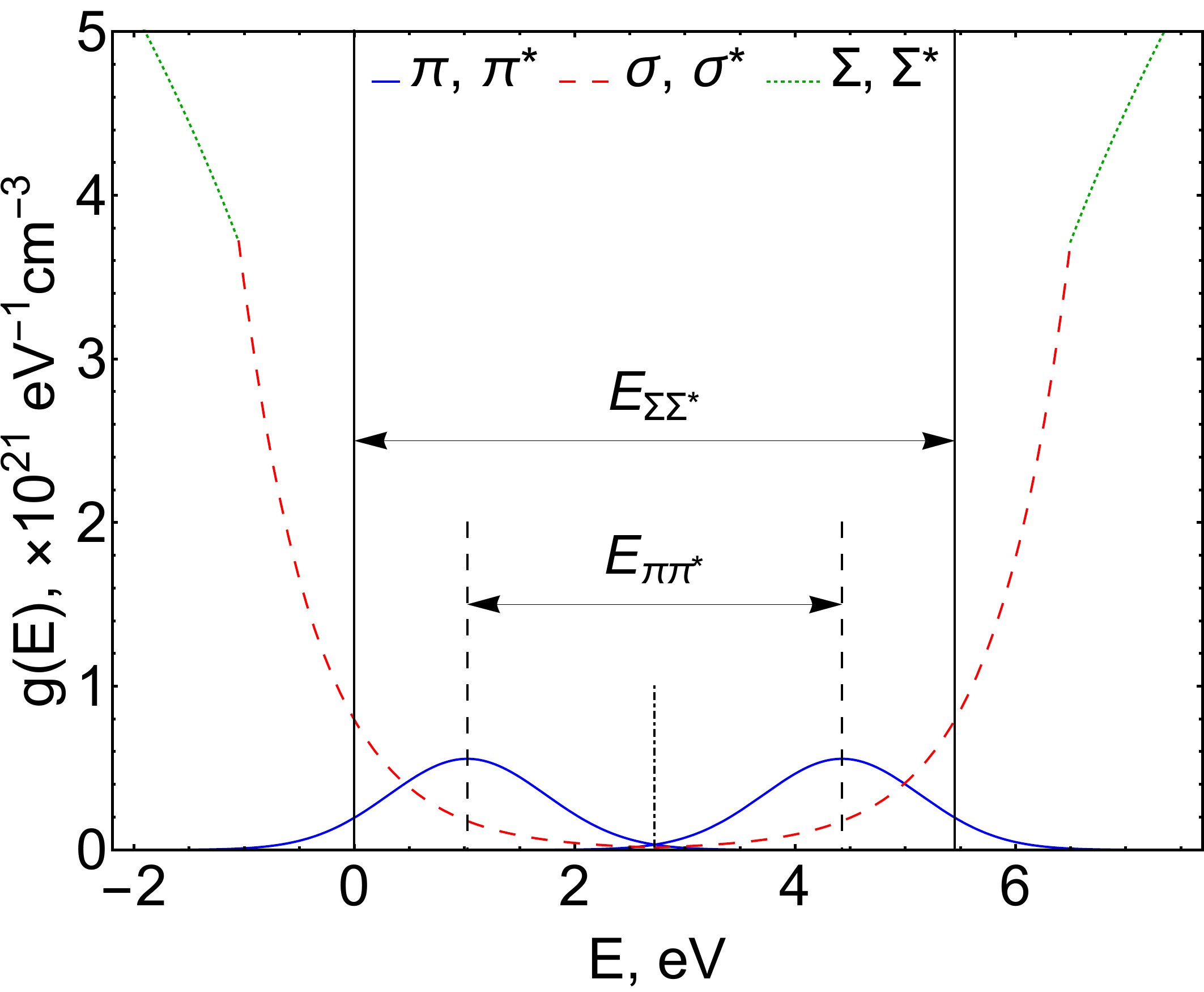}
\caption{Density of states in polycrystalline diamond films calculated with Eq.~\ref{densities1} using the density-of-states model parameters from Ref.~\onlinecite{Zammit_1998} for sample labeled 860-20: $E_{\Sigma\Sigma^*}=5.45$ eV, $E_{\pi\pi^*}=3.4$ eV, $w=0.71$ eV, $E_\text{m}=6.5$ eV, $E_0=0.68$ eV, $B_1=2.3\times 10^4$ cm$^{-1}$, and $B_2=1.5\times 10^5$ cm$^{-1}$ eV$^{-1/2}$.}
\label{distributions}
\end{figure}

In general, the density-of-states parameters depend on many aspects, including the concentration of nitrogen, conditions of material growth, and so on. In Table~\ref{CVD_parameters}, we summarize the parameters for density-of-states model for several carbon-derived materials (synthetic diamond films, amorphous carbon films etc. synthesized by chemical or physical vapor deposition (CVD or PVD) methods) available in the literature. The exemplary density-of-state curves for a polycrystalline diamond film calculated using the fitting parameters taken from Ref.~\onlinecite{Zammit_1998} for one particular sample labeled 860-20 are shown in Fig.~\ref{distributions}. For this purpose, we use the localization length $L_\text{loc}=3 \AA$ comparable with a grain-boundary width of undoped nanocrystalline diamond films.\cite{Csencsits_1996} As discussed in Ref.~\onlinecite{Bhattacharyya_2005}, the overlapping of interband-gap bands leads to the production of electron states at and above the Fermi level. As the amount of $sp^2$-phase in grain boundaries increases, the density of states at the midgap energy rises up, filling the energy gap. For strongly overlapping $\pi-\pi^*$ bands, the conductivity becomes semimetallic.

The effect of nitrogen on the position of the Fermi level in the fundamental band gap has been discussed in the literature.\cite{Zapol_2001, Robertson_1987} Since we rely on the density-of-states model to be symmetric about the fundamental-gap center,\cite{Robertson_1987, Zammit_1998, Achatz_2006_APL} the charge neutrality condition requires that the Fermi level  coincides with the midgap energy. Under this assumption, the midgap density of states and electron concentration are defined by the degree of overlapping between the $\pi$ and $\pi^*$ bands.

\begin{table*}
\caption{\label{CVD_parameters}Fitting parameters of the density-of-states model for some polycrystalline carbon materials.}
\begin{ruledtabular}
\begin{tabular}{lcccc} 
 & Polycrystalline diamond films\cite{Zammit_1998} & $\alpha$-C, $\alpha$-C:H\cite{Dasgupta_1991} & UNCD\cite{Achatz_2006_APL} & (N)UNCD\cite{Achatz_2006_APL}\\
\hline
$E_{\pi\pi^*}$, eV & $3.4\div 3.8$ & $3.2\div 3.9$ & 2.5  & 2.1\\
$w$, eV & $0.68\div 0.76$ & $0.48\div 0.73$& 0.4 & 0.53 \\
$E_\text{m}$, eV & $5.5\div 6.8$  & -& -& -\\
$E_0$, eV & $0.48\div 0.68$  & -& -& -\\
$B_1$, cm$^{-1}$ & $(6.1\div 23)\times 10^3$ & - & $10^5$ & $2.5\times 10^5$\\
$B_2$, cm$^{-1}$eV$^{-1/2}$ & $(57\div 150)\times 10^3$& - & - &- \\
\end{tabular}
\end{ruledtabular}
\end{table*}

\subsection{\label{sec:application_of_eqs}Application of Poisson's and Stratton's Equations to $sp^2$-rich Diamond Films}

Using the results of the previous section, the carrier charge density, which appears in Poisson's equation (Eq.~\ref{Poisson_final}), for the case of $sp^2$-rich polycrystalline materials can be found as
\begin{equation}
\rho=-q\big{[} (n_{\pi^*}+n_{\sigma^*})-(p_{\pi}+p_{\sigma}) \big{]},
\end{equation}
where
\begin{equation}
n_{\pi^*,\sigma^*} = 2\int\limits_{E_{\Sigma\Sigma^*}/2}^{\infty}g_{\pi^*,\sigma^*}(E)f(E)dE
\end{equation}
and
\begin{equation}
p_{\pi,\sigma} = 2\int\limits_{-\infty}^{E_{\Sigma\Sigma^*}/2}g_{\pi,\sigma}(E)\big{[}1-f(E)\big{]}dE.
\end{equation}
Introducing the dimensionless variable
\begin{equation}
y = \frac{\vartheta}{k_\text{B}T}= \frac{E_F-E_{\pi^*}}{k_\text{B}T},
\end{equation}
the Fermi-Dirac distribution $f(E)$ can be written as
\begin{equation}
f(E,y) = \frac{1}{1+\exp\big(\frac{E-E_{\pi^*}}{k_\text{B}T}-y\big)}.
\label{FDy}
\end{equation}
Assuming that the Fermi level in the bulk lies in between $\pi$ and $\pi^*$ peaks, we find $y_\text{b}=-E_{\pi\pi^*}/(2k_\text{B}T)$.

\begin{figure}[t]
\centering
\includegraphics[width=3.2in]{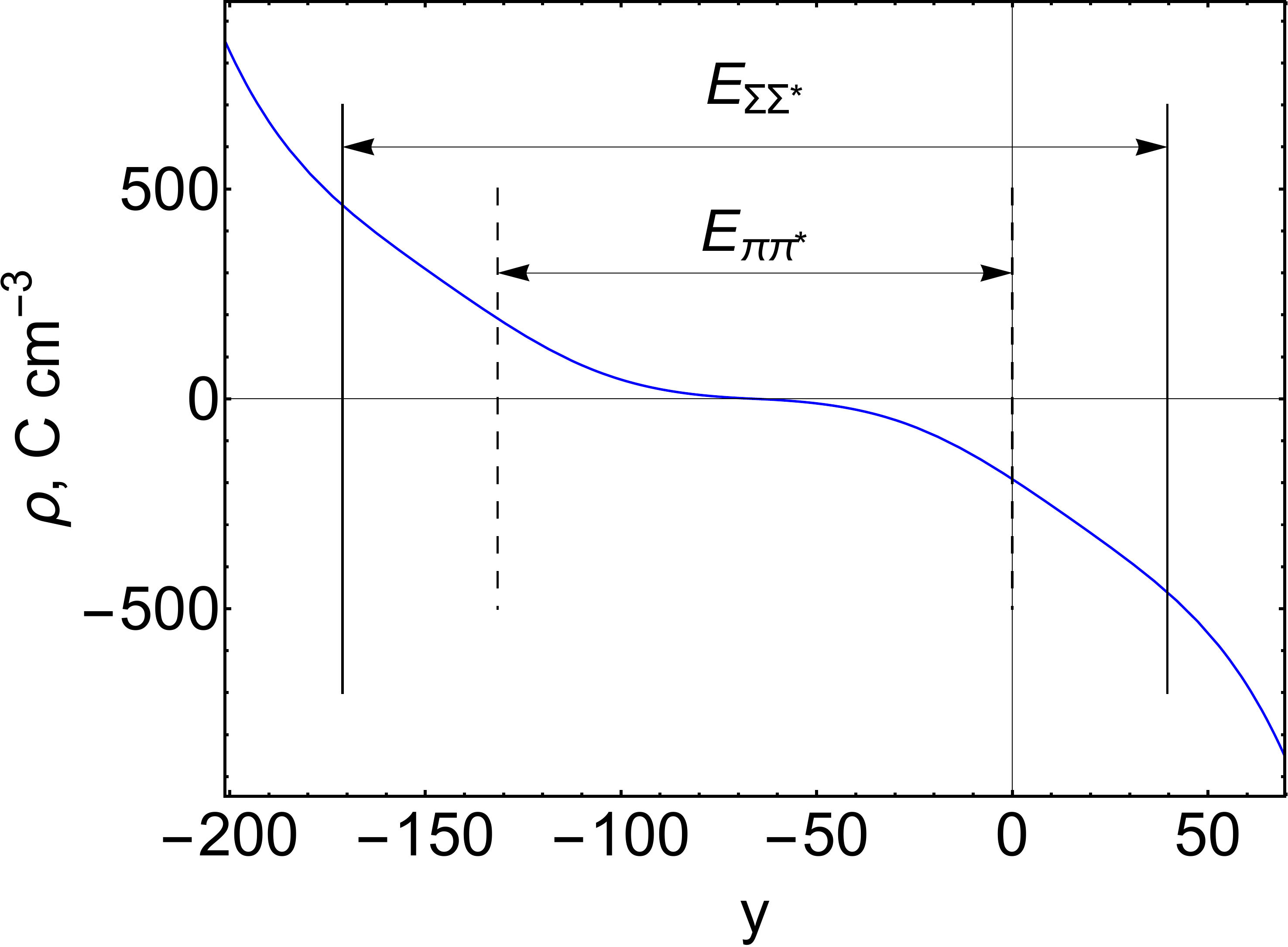}
\caption{The charge-density distribution as a function of position of the Fermi level in the pseudo band gap of polycrystalline diamond films calculated for the density of states shown in Fig.~\ref{distributions}. $y=0$ corresponds to the case when the Fermi level coincides with energy $E_{\pi^*}$. The value $y=-132$ corresponds to the intersection of Fermi level with $E_{\pi}$.}
\label{rho_vs_y}
\end{figure}

Substituting the density of states $g(E)$, defined by Eq.~\ref{densities1}, and the Fermi-Dirac distribution $f(E,y)$, defined by Eq.~\ref{FDy}, the expressions for the concentrations of electrons in $\pi^*$ and $\sigma^*$ bands become 
\begin{equation}
\begin{split}
n_{\pi^*}(y) &= 2\sqrt{\frac{B_1}{K}}\int\limits_{E_{\Sigma\Sigma^*}/2}^{\infty}\exp\big[-\frac{(E-E_{\pi^*})^2}{2w^2}\big] f(E,y)dE,\\
n_{\sigma^*}(y) &= 2\frac{B_2}{\sqrt{KB_1}}\Bigg{[}
\sqrt{E_\text{m}-E_{\Sigma\Sigma^*}}\\
&\mbox{}\quad\times\int\limits_{E_{\Sigma\Sigma^*}/2}^{E_m}\exp\big(\frac{E-E_\text{m}}{E_0}\big) f(E,y)dE\\
&\mbox{}\quad+\int\limits_{E_\text{m}}^{\infty}\sqrt{E-E_{\Sigma\Sigma^*}} f(E,y)dE\Bigg{]}.
\end{split}
\label{n_pi_sigma}
\end{equation}
The concentrations of holes in $\pi$ and $\sigma$ bands can be found as
\begin{equation}
\begin{split}
p_{\pi}(y) &= 2\sqrt{\frac{B_1}{K}}\int\limits_{-\infty}^{E_{\Sigma\Sigma^*}/2}\exp\Big[-\frac{(E-E_{\pi})^2}{2w^2}\Big]\\
&\mbox{}\quad\times\big{[}1-f(E,y)\big{]}dE,\\
p_{\sigma}(y) &= 2\frac{B_2}{\sqrt{KB_1}}\Bigg\{\sqrt{E_\text{m}-E_{\Sigma\Sigma^*}}\\
&\mbox{}\quad\times\int\limits_{E_{\Sigma\Sigma^*}-E_\text{m}}^{E_{\Sigma\Sigma^*}/2}\exp\Big[\frac{-(E-E_{\Sigma\Sigma^*})-E_\text{m}}{E_0}\Big]\\
&\mbox{}\quad\times\big[ 1-f(E,y)\big]dE \\
&\mbox{}\quad+\int\limits_{-\infty}^{E_{\Sigma\Sigma^*}-E_\text{m}}\sqrt{-E}\big[ 1-f(E,y)\big{]}dE\Bigg\}.
\end{split}
\label{p_pi_sigma}
\end{equation}
In Fig.~\ref{rho_vs_y}, we show the calculated $\rho$ vs. $y$ curve which assumes the density-of-states structure shown in Fig.~\ref{distributions}. The increase and broadening of $\pi$ and $\pi^*$ bands lead to the increasing charge density in the diamond band gap.

As mentioned above, it is rigorously established that the bulk electrical conduction properties of (N)UNCD films are determined by the $sp^2$ carbon represented by the $\pi$ and $\pi^*$ bands. Below, we show how the shape of these bands, their position in the fundamental diamond band gap, and the decrease of carrier mobility in strong electric fields affect field emission properties of $sp^2$-rich polycrystalline materials. For (N)UNCD films, only $\pi-\pi^*$ band density-of-states parameters are available in the literature.\cite{Achatz_2006_APL} To describe $\sigma-\sigma^*$ bands, we use the parameters $E_\text{m}$, $E_0$, and $B_2$ typical for polycrystalline diamond films per Ref.~\onlinecite{Zammit_1998}. The choice of these parameters though is not expected to affect the resulting FE characteristics significantly. For simplicity, we do not consider the contribution from the hopping mechanism which may take place due to the presence of localized states.\cite{Bhattacharyya_2005} Investigation of the temperature effects on FE properties of (N)UNCD is beyond the scope of this work.

\subsubsection{\label{sec:effect_of_mobility}Effect of Field-Dependent Mobility}

\begin{table}[b]
\caption {Parameters used to calculate FE characteristics of (N)UNCD films.}
\begin{ruledtabular}
\begin{tabular}{lcc} 
 Parameter & Value & Reference  \\
\hline
$E_{\Sigma\Sigma^*}$, eV & 5.45 & Ref.~\onlinecite{Zammit_1998}\\
$E_{\pi\pi^*}$, eV & 2.1 & Ref.~\onlinecite{Achatz_2006_APL} \\
$w$, eV & 0.53 & Ref.~\onlinecite{Achatz_2006_APL} \\
$E_\text{m}$, eV & 6.5 & Ref.~\onlinecite{Zammit_1998}\\
$E_0$, eV & 0.68 & Ref.~\onlinecite{Zammit_1998}\\
$B_1$, cm$^{-1}$ & $2.5\times 10^5$ & Ref.~\onlinecite{Achatz_2006_APL}\\
$B_2$, cm$^{-1}$eV$^{-1/2}$ & $1.5\times 10^5$ & Ref.~\onlinecite{Zammit_1998}\\
$\mu_0$, cm$^2$V$^{-1}$s$^{-1}$ & 1.5 & Ref.~\onlinecite{Williams_2004}\\
$\mathcal{F}_0$, V cm$^{-1}$ & 10$^4$\\
$L_\text{loc}$, $\AA$ & 10 & Ref.~\onlinecite{Beloborodov_2006}\\
$\kappa$ & 4.5 & Ref.~\onlinecite{Beloborodov_2006} \\
$\varphi$, eV & 3.6 & Ref.~\onlinecite{Quintero_2014}\\
\end{tabular}
\end{ruledtabular}
\label{UNCD_parameters}
\end{table}

It was noticed \cite{Conwell_High} that the dependence of electron mobility on the applied electric field has similar character for most conventional semiconductors, except that larger fields are required to produce deviations from Ohm's law (the region of the drift-velocity saturation) in materials with lower electron mobilities. The saturation of the drift velocity in semiconductors originates from the inelastic scatterings of highly energetic electrons on optical phonons. For the case of polycrystalline diamond films and (N)UNCD films, in particular, there is no available data for the dependence of the carrier mobility on the applied electric field. Therefore, we apply a simple expression commonly used for semiconductors \cite{Baskin_1971, Ryder_1953}
\begin{equation}
\mu(\mathcal{F})=\begin{cases}
\mu_0, \qquad \mathcal{F}<\mathcal{F}_0\\
\mu_0\sqrt{\frac{\mathcal{F}_0}{\mathcal{F}}}, \qquad \mathcal{F}>\mathcal{F}_0
\end{cases}
\end{equation}
to describe hypothetical field effects on carrier mobility after the electric field reaches a critical value $\mathcal{F}_0$ and to evaluate the importance of this effect (if exists) on the resulting current-density curve behavior. 

Electron mobilities in typical semiconductors significantly exceed the mobilities of electrons in $sp^2$-rich materials (compare: $\mu_{\text{e}0}^{\text{Ge}}\sim 3600$ cm$^2$V$^{-1}$s$^{-1}$ ($\mathcal{F}_0\sim 0.9\times 10^3$ V cm$^{-1}$)\cite{Ryder_1953}; $\mu_{\text{e}0}^{\text{Si}}\sim 1370$ cm$^2$V$^{-1}$s$^{-1}$ ($\mathcal{F}_0\sim 2.5\times 10^3$ V cm$^{-1}$)\cite{Ryder_1953}; $\mu_{\text{e}0}^{\text{(N)UNCD}}\sim 1.5$ cm$^2$V$^{-1}$s$^{-1}$ \cite{Williams_2004}). Therefore, it can be assumed that the deviation from Ohm's law in $sp^2$-rich materials may occur at relatively large electric fields $\mathcal{F}_0 \gtrsim 10^4$ V cm$^{-1}$. The density-of-states model parameters and material parameters, that are used to calculate FE characteristics of (N)UNCD, are summarized in Table~\ref{UNCD_parameters}.

\begin{figure}[t]
\centering
\subfigure[][]{\includegraphics[width=3.2in]{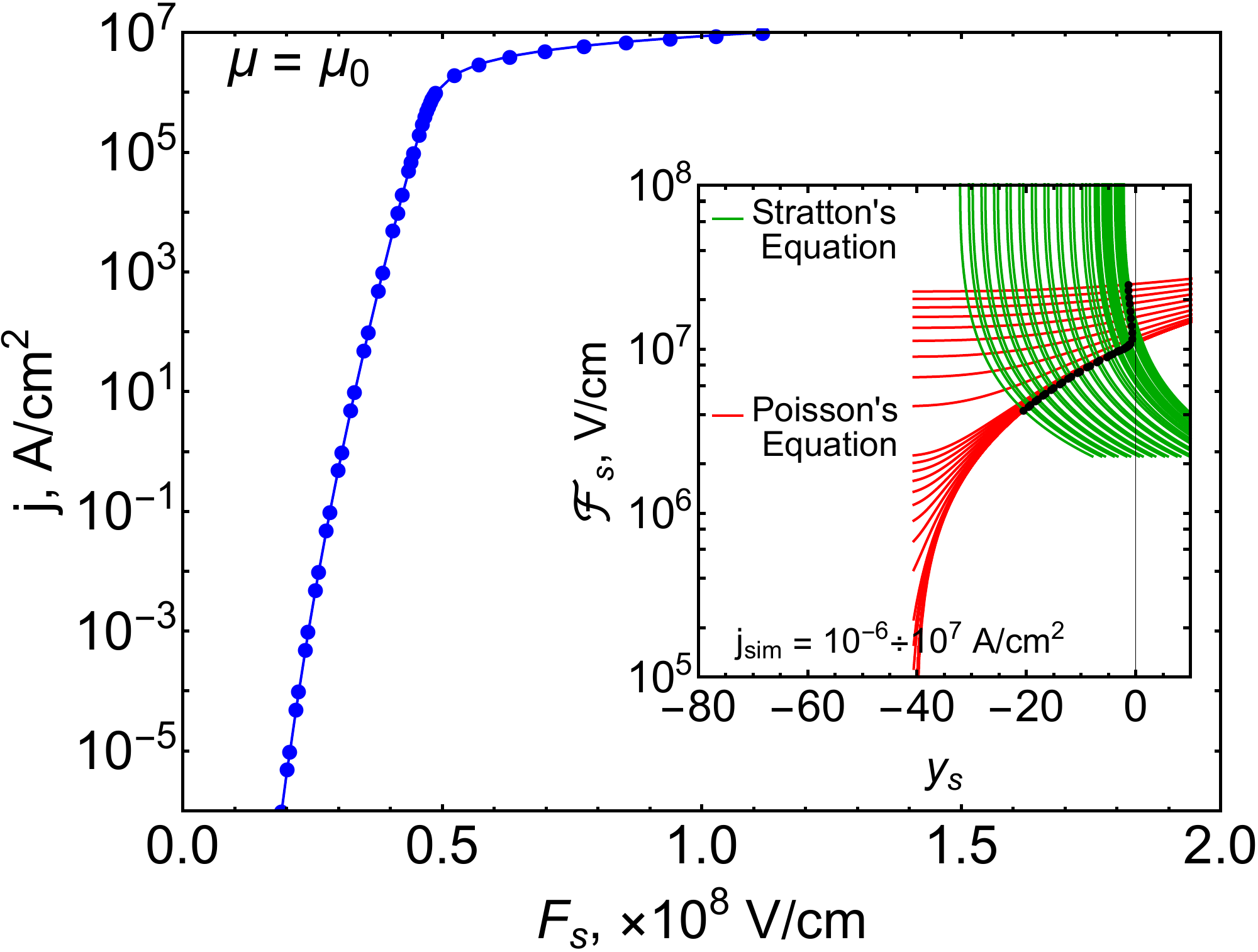}
\label{res_mu0}
}
\hfil
\subfigure[][]{\includegraphics[width=3.2in]{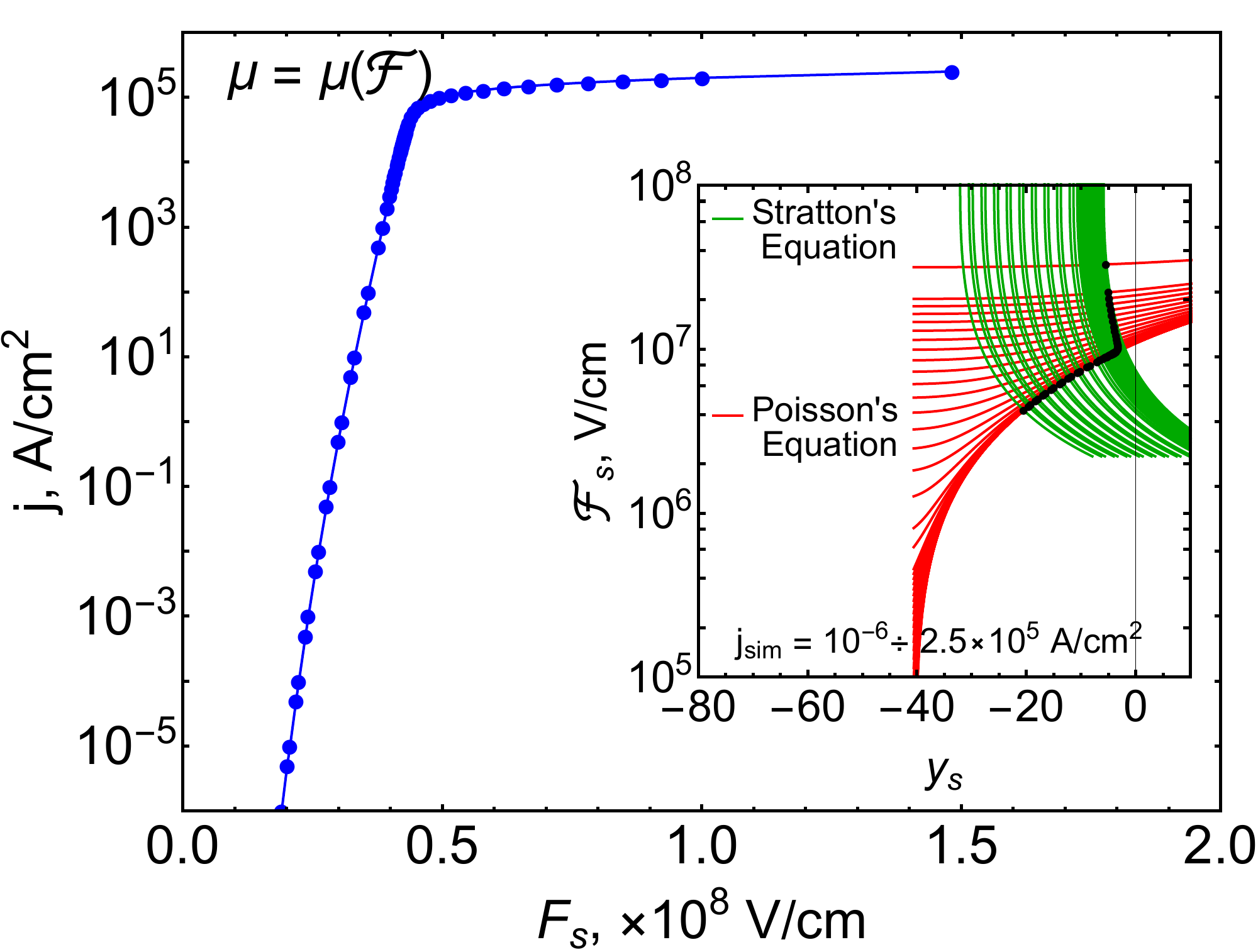}
\label{res_mu}
}
\caption{Formation of a current-density saturation plateau in (N)UNCD films assuming (a) field-independent and (b) field-dependent mobility.}
\label{jposneg}
\end{figure}

The comparison presented in Fig.~\ref{jposneg} reflects the significant effect which the field-dependent mobility may impose onto the electric characteristics of (N)UNCD. Obtained FE characteristics result from the self-consistent solution of Poisson's equation (Eq.~\ref{Poisson_final}, the red curves in the insets) and Stratton's equation (Eq.~\ref{jneg_eq}, the green curves in the insets). The intersections of the two sets of curves (red and green) provide the resulting current-density dependencies on the electric field. Plots in Fig.~\ref{res_mu0} and Fig.~\ref{res_mu} take field-independent and field-dependent mobility into account, respectively. For the set of density-of-states model parameters specified in Table~\ref{UNCD_parameters}, deviation of the current-density curve from the FN law and further saturation occur in both cases, but the remarkable difference exists. If the field-dependence of the mobility is taken into account, a saturation plateau lies at about $\sim$10$^5$ A cm$^{-2}$ which is two orders of magnitude lower as compared to the case considering the mobility independent of the electric field and yielding the saturation plateau as high as 10$^7$ A cm$^{-2}$. For $\mu=\mu_0$, the mobility is still finite. Therefore, the  presence of current-density saturation effect is due to the limited surface metallization rate (this rate is obviously higher than for $\mu=\mu_0\sqrt{\mathcal{F}_0/\mathcal{F}}$), in other words, it is due to the limited electron supply into the accumulation region.

This result emphasizes the importance of field effects on the transport properties in (N)UNCD and encourages further experimental study to help refine (confirm or rule out) existing models on field emission for semiconductors or semimetals, i.e. beyond FN-like formalism. We note that the choice of the mechanism affecting carrier mobility leads to saturation plateaus that may differ by several orders of magnitude. For instance, the current density saturating at 10$^7$ A cm$^{-2}$ is troublesome because it does not allow for distinguishing between the material-driven saturation and the saturation due to Child-Langmuir effect that becomes significant at a current density $\sim$10$^7$ A cm$^{-2}$.\cite{Dyke_1953, Barbour_1953}

In Figs.~\ref{sat_kappa} to \ref{sat_B1}, we present results that compare FE characteristics with field-independent and field-dependent mobility.

\subsubsection{Effect of Dielectric Constant}

\begin{figure}[b]
\centering
\subfigure[][]{\includegraphics[width=3.2in]{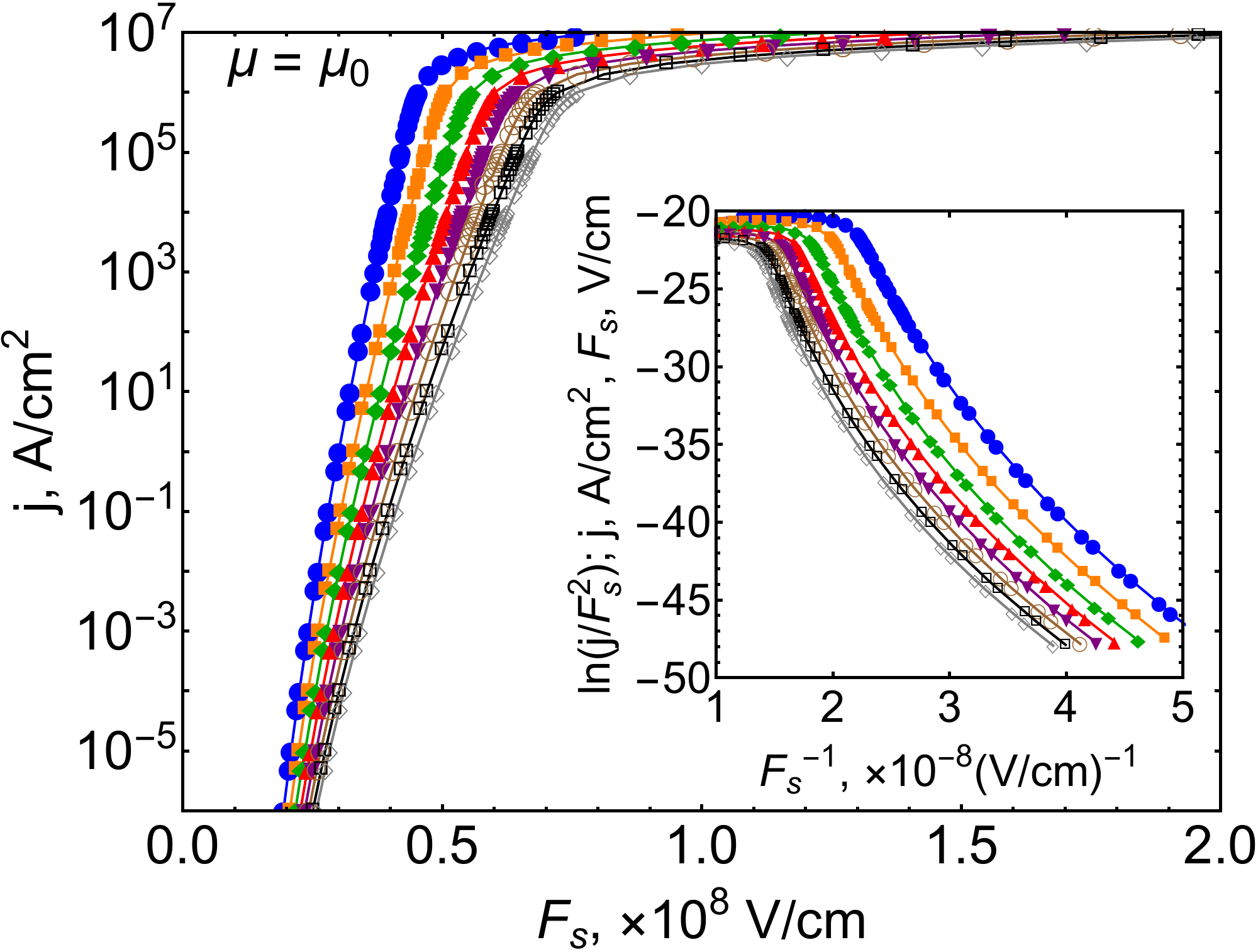}
\label{sat_kappa_mu0}
}

\subfigure[][]{\includegraphics[width=3.2in]{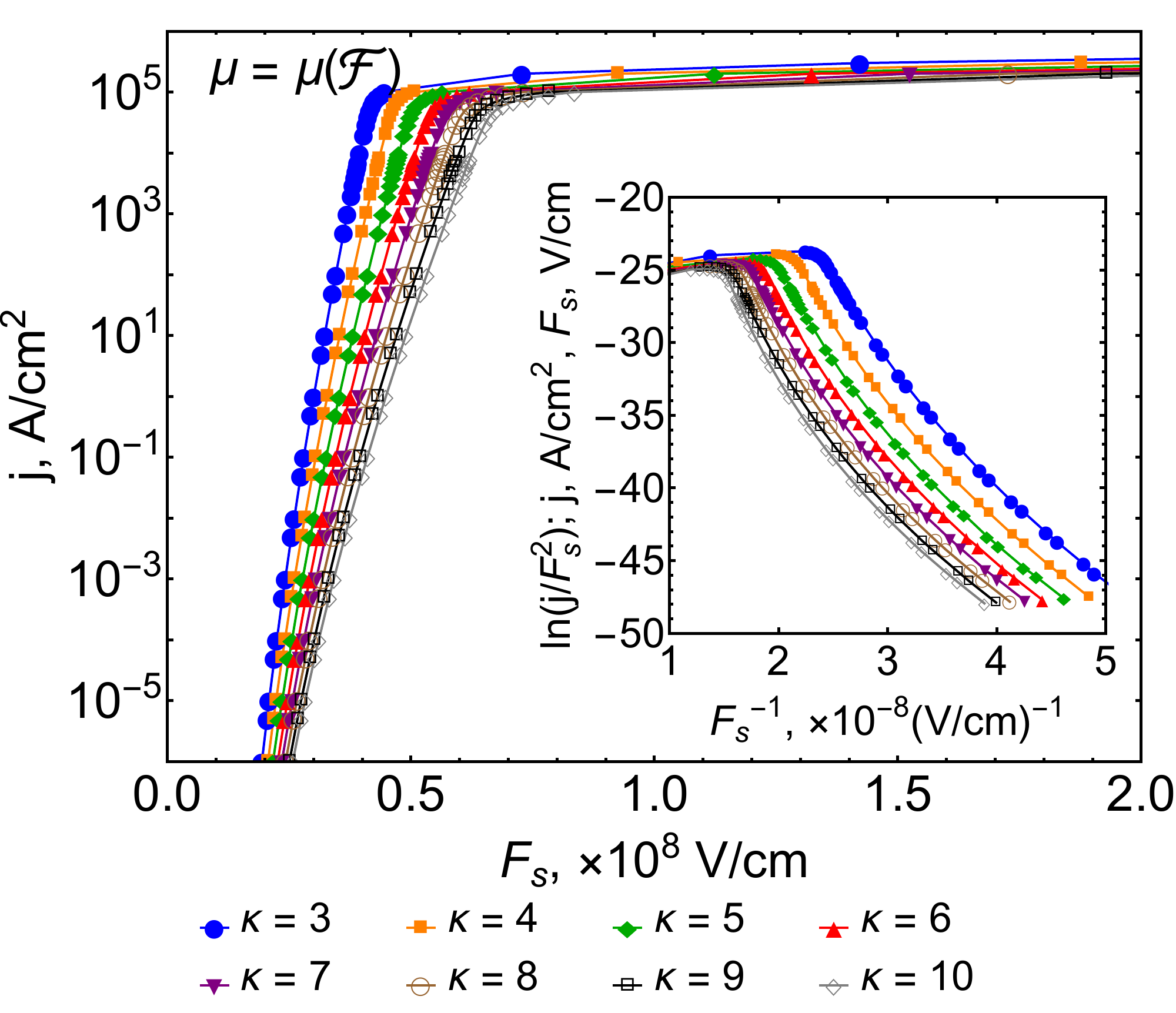}
\label{sat_kappa_mu}
}
\caption{Simulation results for various values of the dielectric constant $\kappa$ assuming (a) field-independent and (b) field-dependent mobility plotted in a semi-log scale and in the FN coordinates. Other parameters were set to the values specified in Table~\ref{UNCD_parameters}.}
\label{sat_kappa}
\end{figure}

The dielectric constant, $\kappa$, is defined as the ratio between the electric field in vacuum and the electric field in the material, i.e. $\kappa = F_\text{s}/\mathcal{F}$. Its effect on the FE current density is shown in Fig.~\ref{sat_kappa}. We plot the results in semi-log and FN, $\ln(j F_\text{s}^{-2})$ vs. $F_\text{s}^{-1}$, coordinates.

Fig.~\ref{sat_kappa} illustrates the effect of $\kappa$ on the $j$ vs. $F_\text{s}$ curve as $\kappa$ varies from 3 to 10. The lower the $\kappa$ the lower the surface field because field enhancement drops: this result means that as $\kappa$ decreases, the electric field line termination at the surface weakens (stronger field penetration into material) and local electric field decreases.
At the same time, as expected, $\kappa$ does not have any appreciable affect on the saturation plateau. Another interesting result seen in Figs.~\ref{sat_kappa_mu0} and \ref{sat_kappa_mu} is that the field-dependent mobility leads to faster switching from the FN to saturation regime due to a limited charge transport time in the bulk (drift velocity gets saturated). Dielectric constant $\kappa\approx 3\div6$ is commonly used to describe (N)UNCD and other polycrystalline carbon materials. \cite{Beloborodov_2006} We use the average value $\kappa=4.5$ in all calculations below.

\subsubsection{Effect of Density of States}

\begin{figure}[b]
\centering
\subfigure[][]{\includegraphics[width=3.2in]{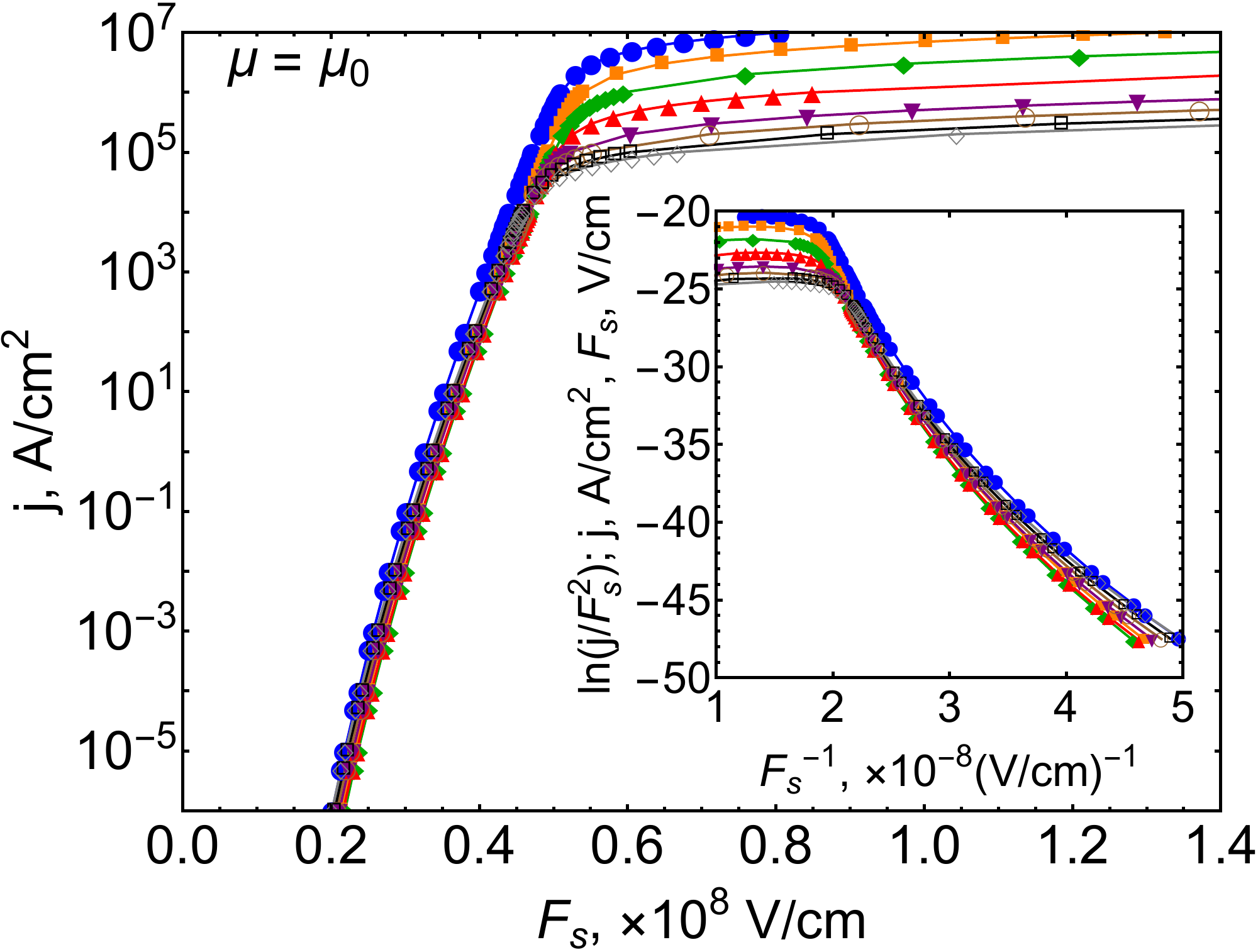}

}
\subfigure[][]{\includegraphics[width=3.2in]{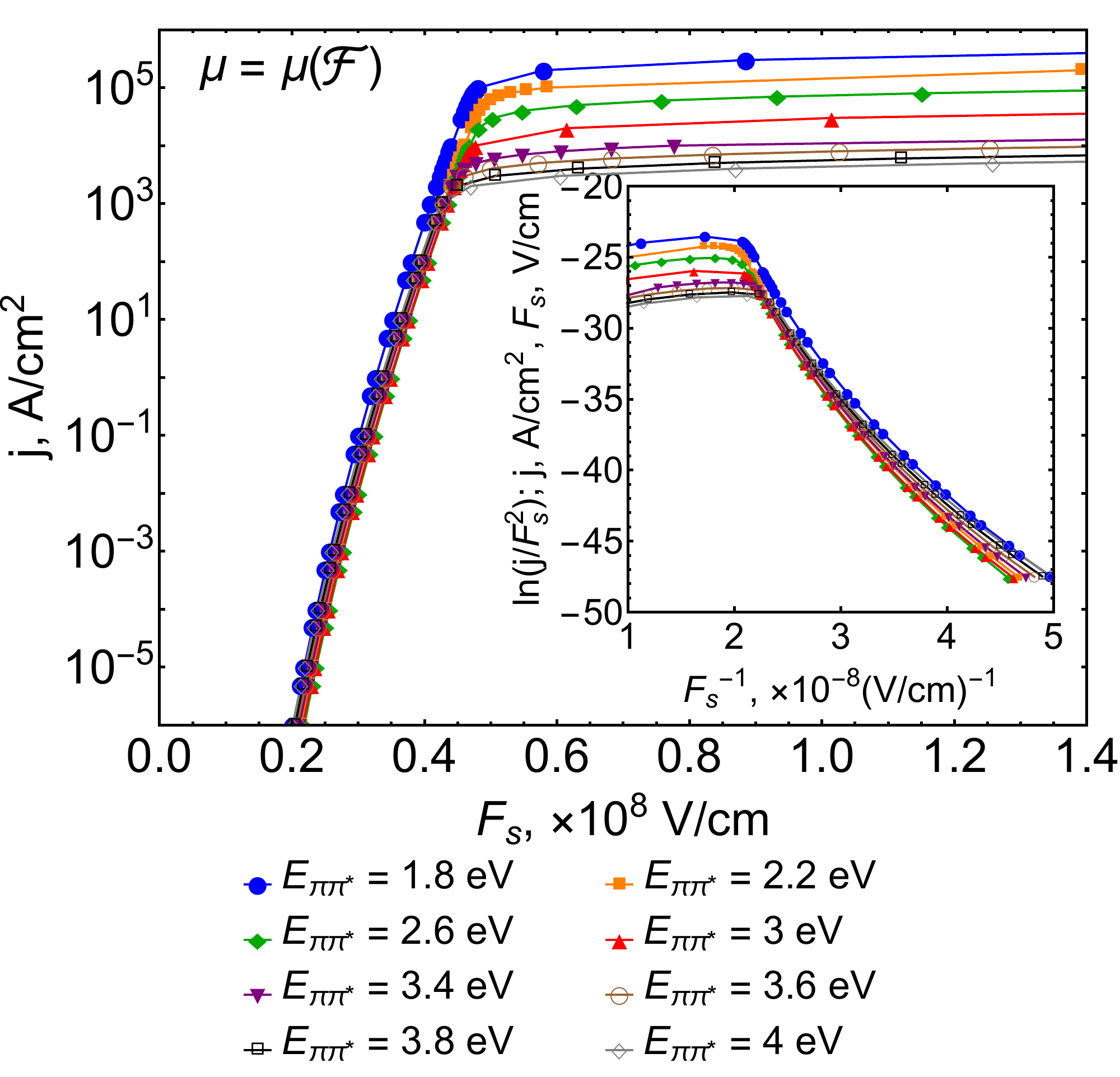}
}
\caption{Simulation results for various values of the pseudo-band-gap energy $E_{\pi\pi^*}$ assuming (a) field-independent and (b) field-dependent mobility plotted in a semi-log scale and in the FN coordinates. Other parameters were set to the values specified in Table~\ref{UNCD_parameters}.}
\label{sat_Epp}
\end{figure}

\begin{figure}[t]
\centering
\subfigure[][]{\includegraphics[width=3.2in]{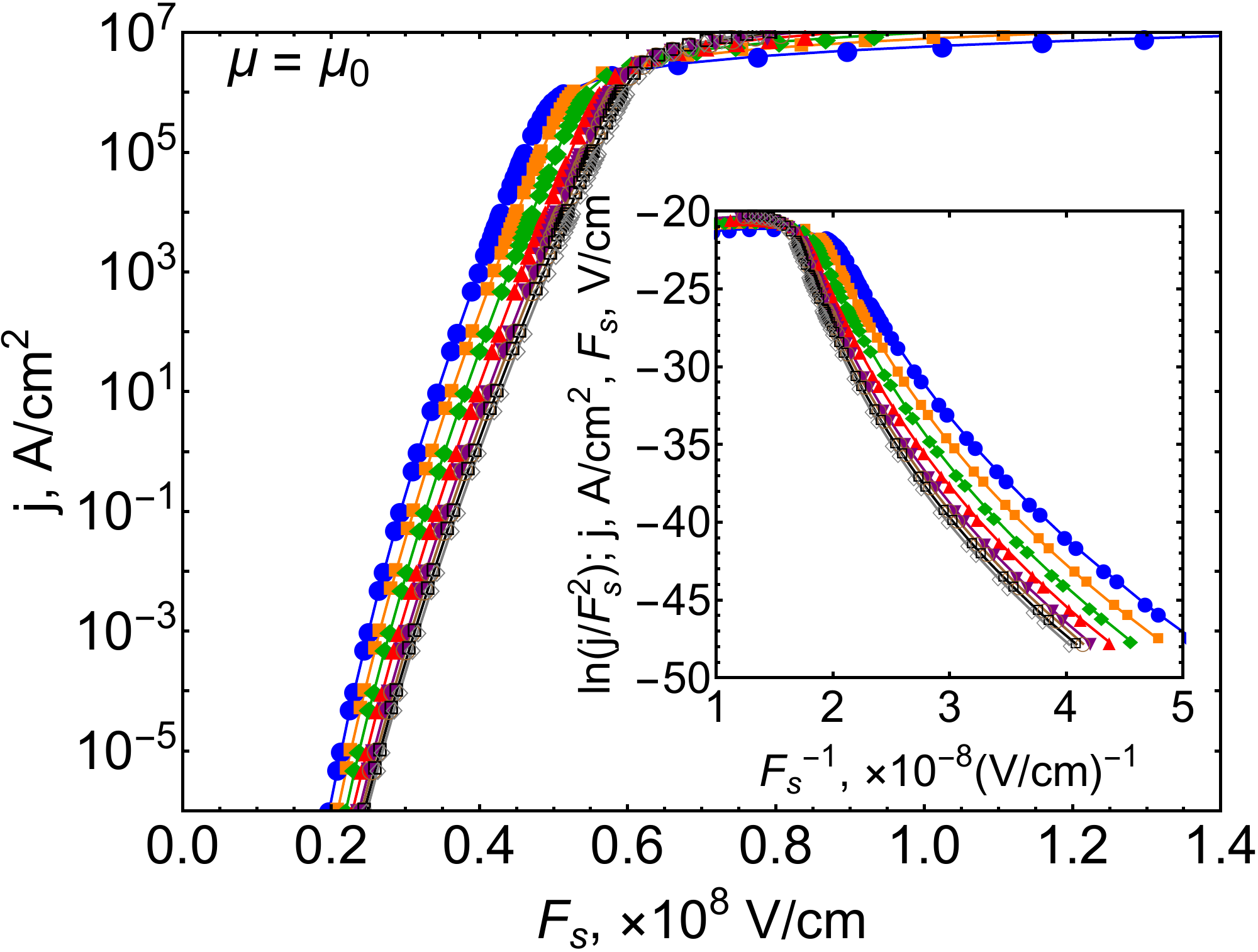}
}
\subfigure[][]{\includegraphics[width=3.2in]{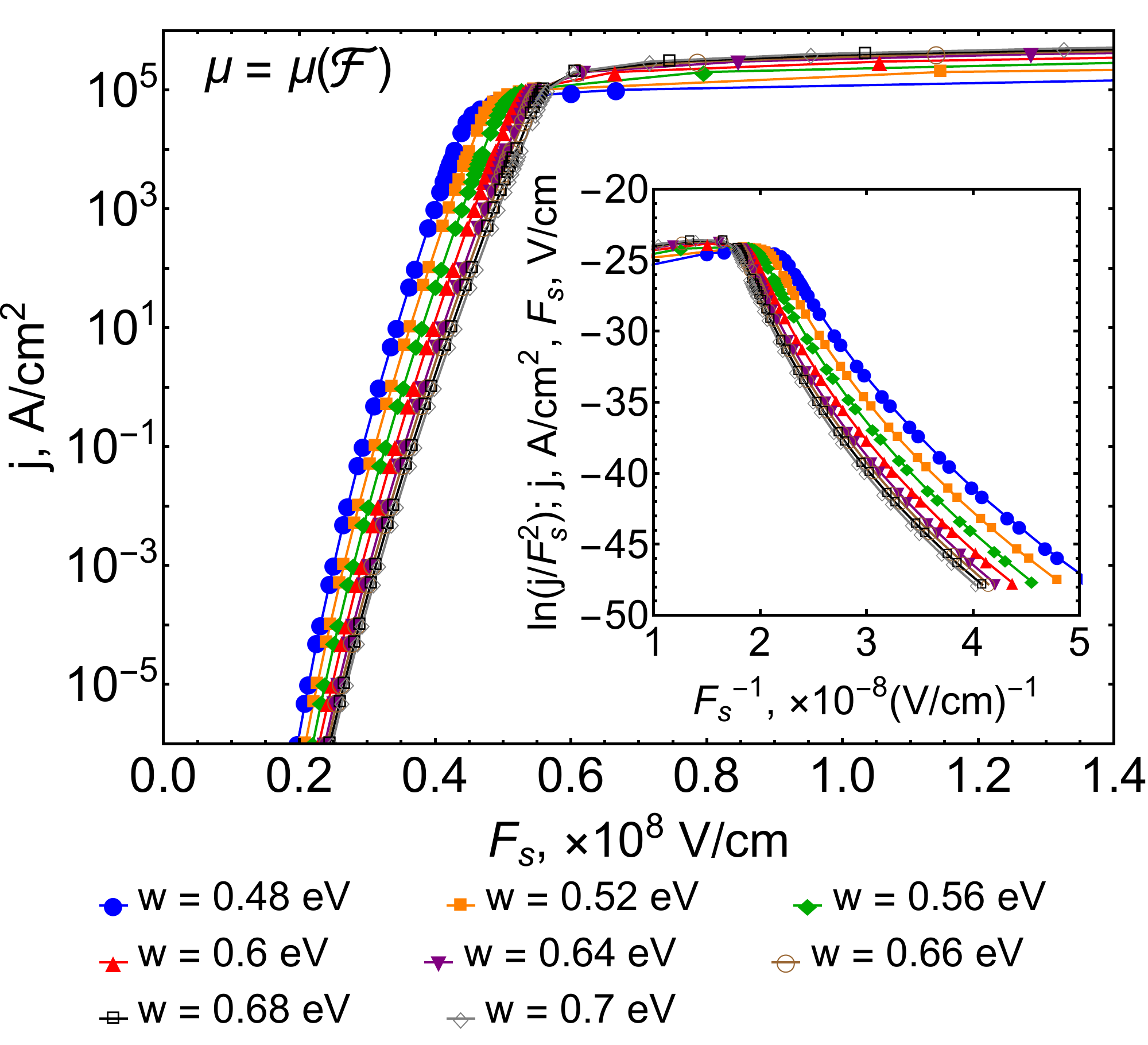}
}
\caption{Simulation results for various values of the width of $\pi$ and $\pi^*$ bands $w$  assuming (a) field-independent and (b) field-dependent mobility plotted in a semi-log scale and in the FN coordinates. Other parameters were set to the values specified in Table~\ref{UNCD_parameters}.}
\label{sat_w}
\end{figure}

\begin{figure}[!h]
\centering
\subfigure[][]{\includegraphics[width=3.2in]{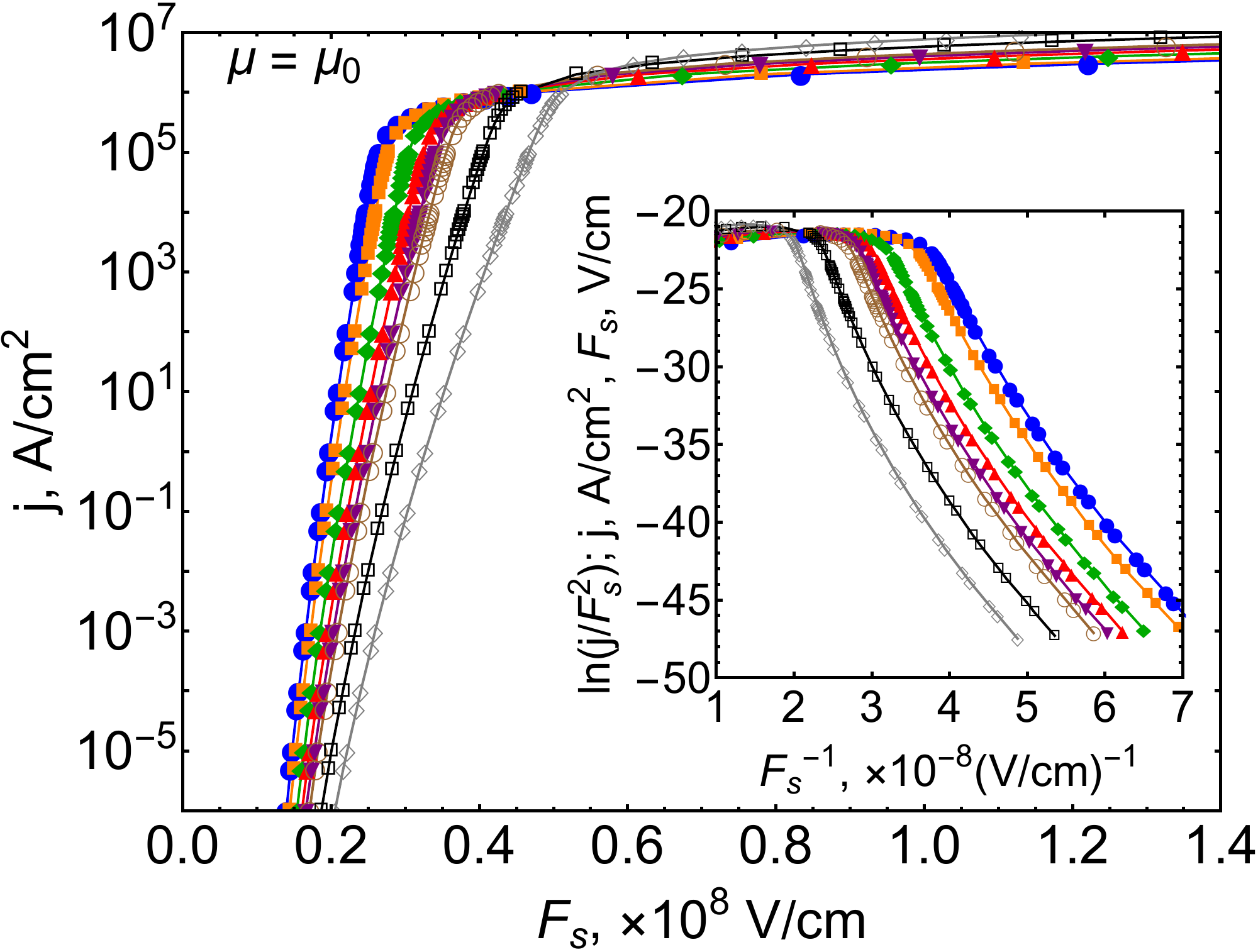}
}
\subfigure[][]{\includegraphics[width=3.2in]{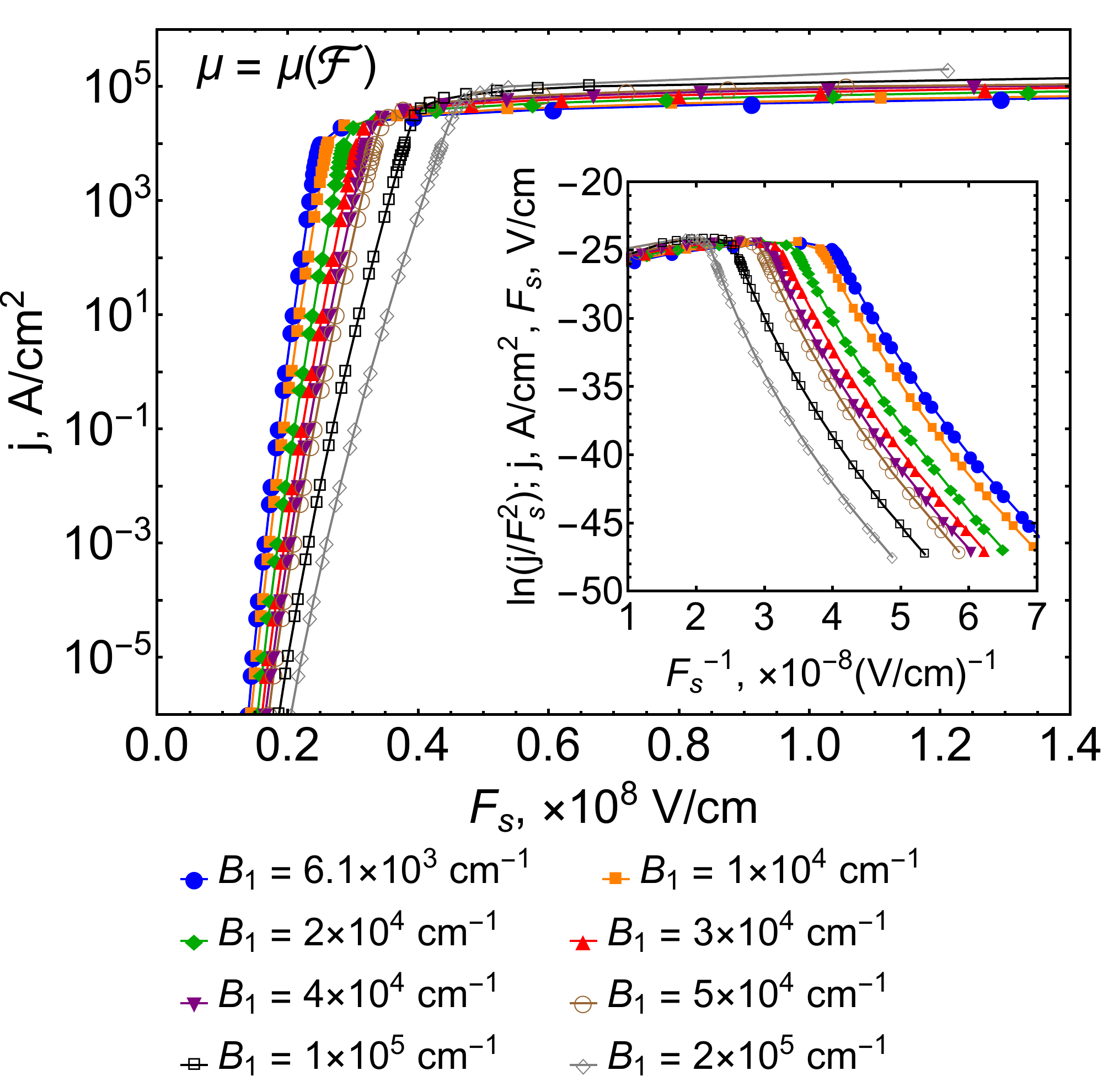}
}
\caption{Simulation results for various values of the preintegration factor $B_1$ assuming (a) field-independent and (b) field-dependent mobility plotted in a semi-log scale and in the FN coordinates. Other parameters were set to the values specified in Table~\ref{UNCD_parameters}.}
\label{sat_B1}
\end{figure}

It was experimentally shown \cite{Achatz_2006_APL} that the incorporation of nitrogen, the source of high electric conductivity, leads to reduced separation between the $\pi$ peaks ($E_{\pi\pi^*}$ parameter, pseudo band gap, decreases) and broadening of the $\pi$ peaks (increase of the $w$ parameter). Simultaneously, the normalization parameter $B_1$ increases. In Figs.~\ref{sat_Epp}--\ref{sat_B1}, we show how variation of these parameters modifies the resulting FE characteristics of (N)UNCD. In order to pinpoint the effect imposed by $E_{\pi\pi^*}$, $w$, and $B_1$, we calculate the $j$ vs. $F_\text{s}$ profiles by sweeping $E_{\pi\pi^*}$ or $w$ or $B_1$ while keeping all remaining parameters fixed at the values specified in Table~\ref{UNCD_parameters}.

It can be seen in Fig.~\ref{sat_Epp} that variation of parameter $E_{\pi\pi^*}$ does not have significant effect on the FN part of the current-density curve. At the same time, one can see that higher current densities can be obtained from the samples characterized by narrower pseudo band gaps thanks to larger electron supply. The same supply argument exists behind the observed saturation plateau scaling -- the smaller the pseudo band gap the larger the electron supply, the larger the saturation plateau.

The increase in $w$ (Fig.~\ref{sat_w}) and $B_1$ (Fig.~\ref{sat_B1}) implies that the electron concentration increases which is equivalent to the increase in conductivity. This leads to a significant shift of the FN-like part toward higher surface fields, which means that a higher electron concentration leads to stronger focusing of electric field lines on more conductive grain boundaries, providing higher field enhancement factor $\beta=F_\text{s}/E$. This result correctly predicts the trend, often observed in experiment, in that UNCD films with higher nitrogen content (i.e. higher conductivity) demonstrate lower turn-on electric fields. At the same time, the effect of $w$ and $B_1$ on the saturation plateau for the given pseudo band gap of 2.1 eV, while significant (about 10-fold effect), is not as strong as when $E_{\pi\pi^*}$ is varied (about 100-fold effect). From extended modelling (not shown here) we additionally conclude that the effects of $w$ and $B_1$ on the FE properties (including the saturation plateau) should become stronger as $E_{\pi\pi^*}$ further reduces below 2.1 eV. Thus, with increasing parameters $w$ and $B_1$ we predict to observe stronger influence on a FN part of the $j$ vs. $F_\text{s}$ dependence by means of  increasing the local field enhancement and lowering the turn-on field. The parameter $E_{\pi\pi^*}$ is primarily responsible for current density saturation -- this is because $E_{\pi\pi^*}$ defines the depth of accumulation well near the surface/vacuum interface and therefore defines how fast it depletes.

As seen previously, in all cases field-independent mobility model formulation leads to overall larger saturation plateaus compared to those obtained by taking field-dependent mobility.

\section{\label{sec:comparison}Comparison to Experiment and Discussion}

\subsection{\label{sec:experiment}Experimental Results}

A detailed description of the methods and techniques used to evaluate the plan-view area of electron emission can be found in Ref.~\onlinecite{Chubenko_2017}. Here, we present a brief overview of the main ideas and provide crucial findings for the (N)UNCD/Ni/Mo/SS1 sample tested at the cathode-anode separation of 106 $\mu$m (the (N)UNCD film was grown as described elsewhere \cite{Quintero_2014, Baryshev_2014} on top of a 4.4 mm diameter stainless steel (SS) stub with a Ni/Mo buffer layer).

To visualize electron FE patterns, the YAG:CE/Mo (yttrium aluminum garnet crystal doped with cerium and coated with molybdenum) anode screen was used as an electron-collecting electrode. Green light is emitted when electrons hit the anode. This effect, which is known as cathodoluminescence, allows for mapping of FE patterns on the anode surface. The photocamera viewing the back side of the anode was used to collect the FE images, and current-voltage measurements were done concurrently. To estimate the area $S_\text{YAG}$ of FE patterns formed on the YAG screen, an image-processing algorithm was developed. A comprehensive description of the procedure and the implementation details are provided in the Supporting Information Section of our previous paper.\cite{Chubenko_2017} After the numerical processing of a full dataset of electron emission micrographs, it was found that the FE area strongly depends on the applied electric field through the continuously increasing number of FE sites and through the expansion of FE spots with an increasing electric field. 

\begin{figure}[t]
\centering
\includegraphics[width=3.4in]{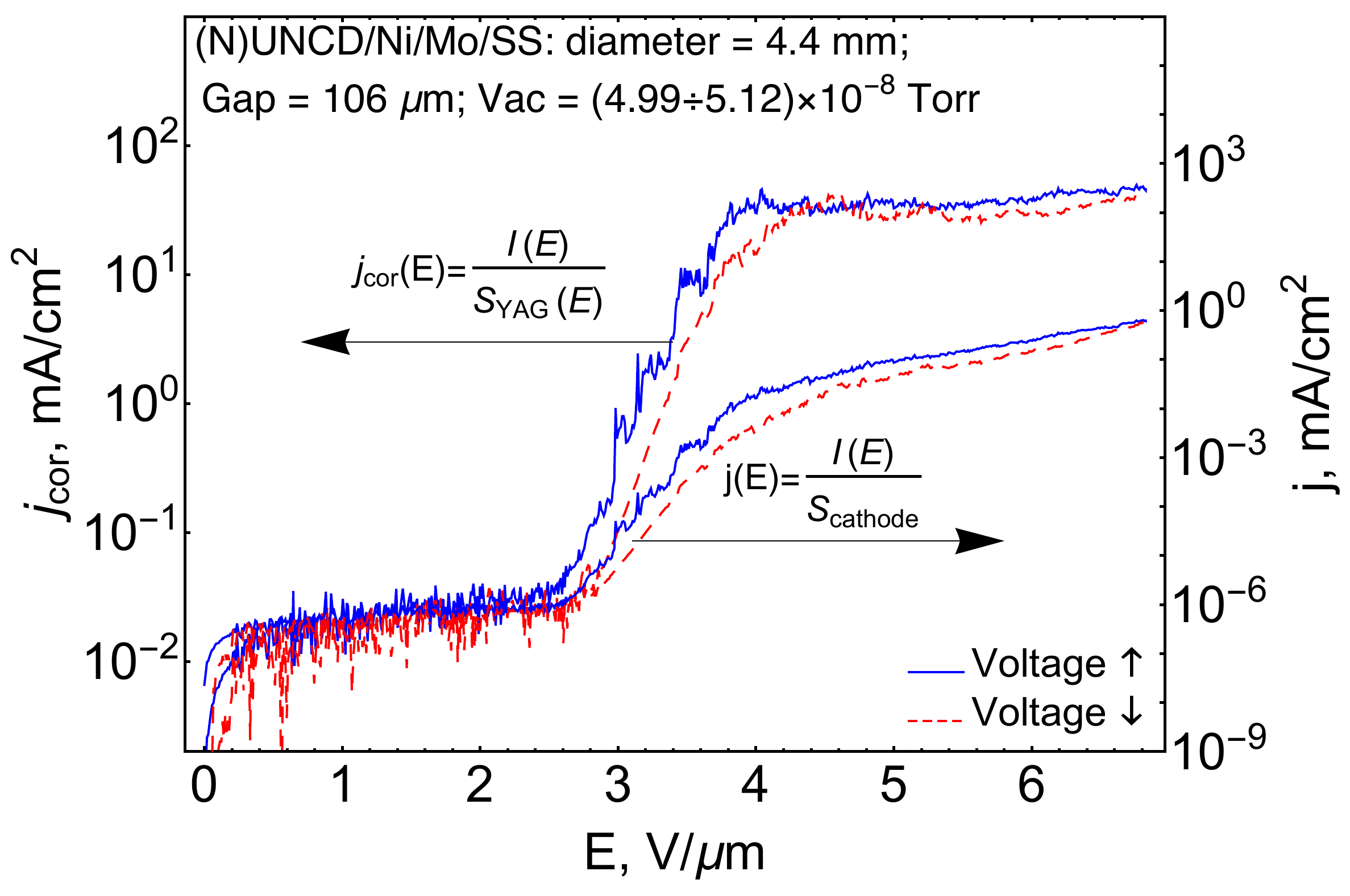}
\caption{Difference between the current density $j$, normalized by the total cathode area, and the current density $j_{\text{cor}}$, normalized by the field-dependent area of emission sites formed on an anode screen, as a function of the applied electric field.}
\label{jvsjcor}
\end{figure}

The maximal FE current measured from this particular sample was about $I\sim$ 100 $\mu$A at $E\sim$ 6.8 V~$\mu$m$^{-1}$. The corresponding maximal current density estimated by the conventional method, i.e. by normalizing the experimentally measured current over the entire cathode area $S_{\text{cathode}}$ = constant,
\begin{equation}
j(E)=\frac{I(E)}{S_{\text{cathode}}}
\end{equation}
was found to be $j\sim$ 1 mA~cm$^{-2}$ as shown in Fig.~\ref{jvsjcor}. This latter method appears to be unreliable to describe fundamental emissive properties of nanodiamond films since it does not account for the increase of the current due to the increase of the number of emission sites which, as described above, is a characteristic feature of (N)UNCD field emitters. Therefore, it leads to a wrong interpretation of experimental results when compared to theory.

The apparent underestimation of the field emission current density due to the overestimation of the actual field emission area can be reduced if the experimentally measured current is normalized by the dynamic FE area $S_{\text{YAG}}(E)$ defined as the total area of the FE sites formed on the YAG screen and estimated by the algorithm\cite{Chubenko_2017}
\begin{equation}
j_{\text{cor}}(E) = \frac{I(E)}{S_{\text{YAG}}(E)}.
\end{equation}
This new method leads to a quantitatively and qualitatively different behavior compared to that obtained by the conventional method. In a semi-$log$ plot (see Fig.~\ref{jvsjcor}), $j_{\text{cor}}$ rises linearly in a low-current-density region, demonstrating Fowler-Nordheim-like behavior, and saturates at about $j_{\text{cor}}\sim 100$ mA~cm$^{-2}$ starting from $E\sim 4$ V~$\mu$m$^{-1}$, further demonstrating a plateau. 

\begin{figure}[b]
\centering
\subfigure[][]{\includegraphics[width=1.6in]{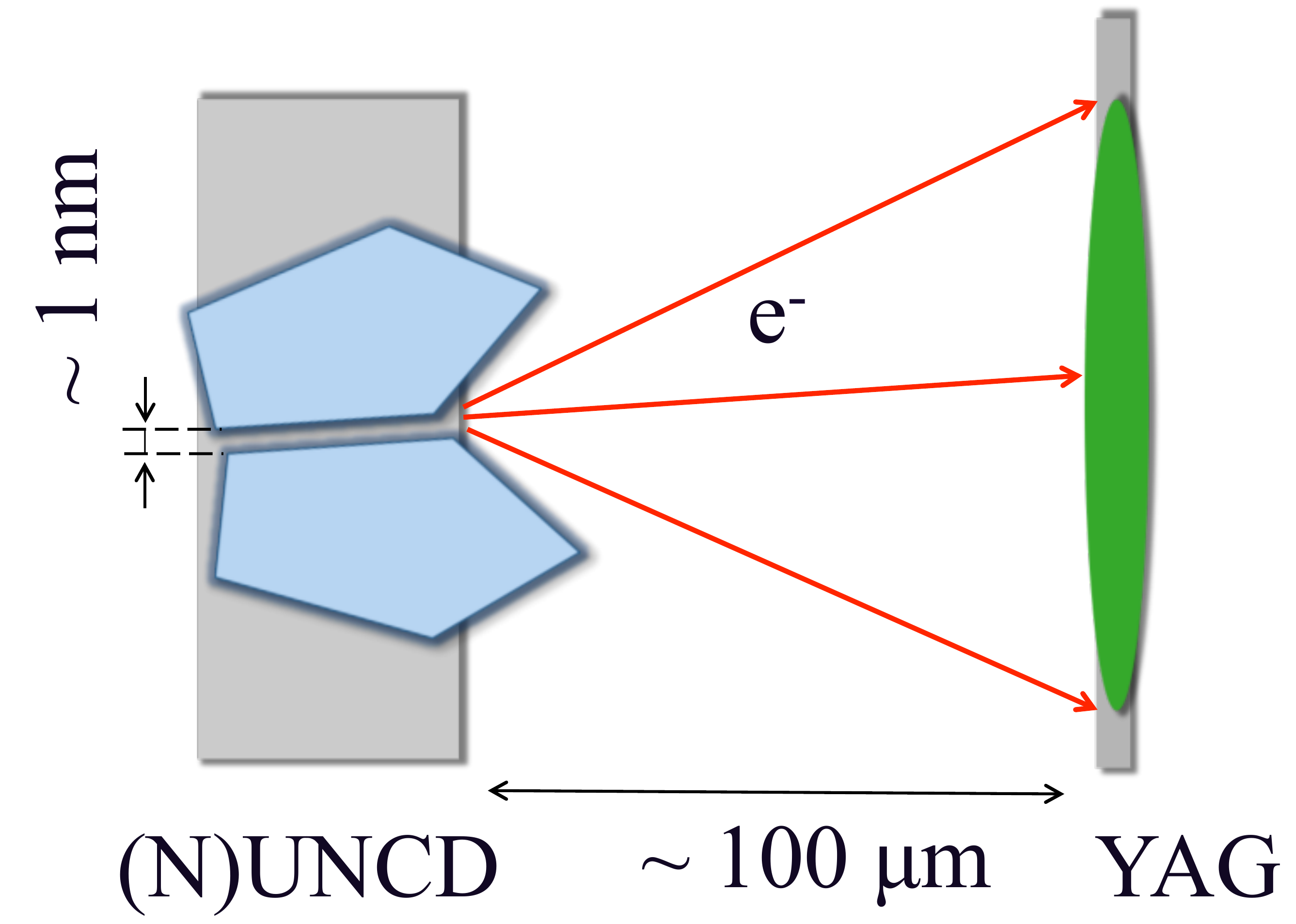}
\label{spot_cor_1}}
\subfigure[][]{\includegraphics[width=1.6in]{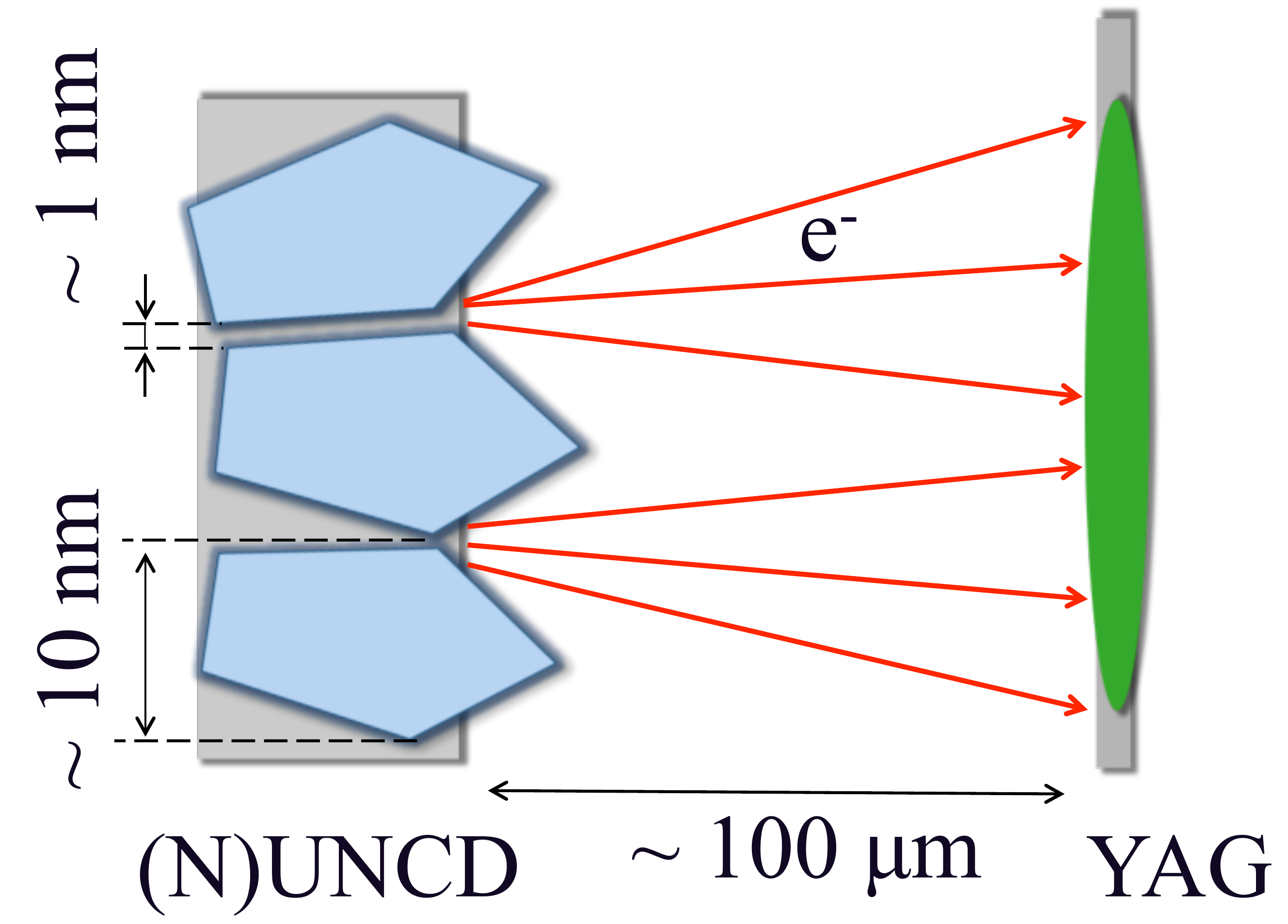}
\label{spot_cor_2}}
\caption{Formation of a field emission spot from the (N)UNCD electron emitter on the anode screen assuming that (a) every single spot on the anode plate is due to electron emission from a single grain boundary (this assumption gives the highest value of the current density) and (b) every single spot on the anode screen is due to several emitting grain boundaries.}
\label{spot_cor} 
\end{figure}

\begin{figure}[!h]
\centering
\includegraphics[width=3.2in]{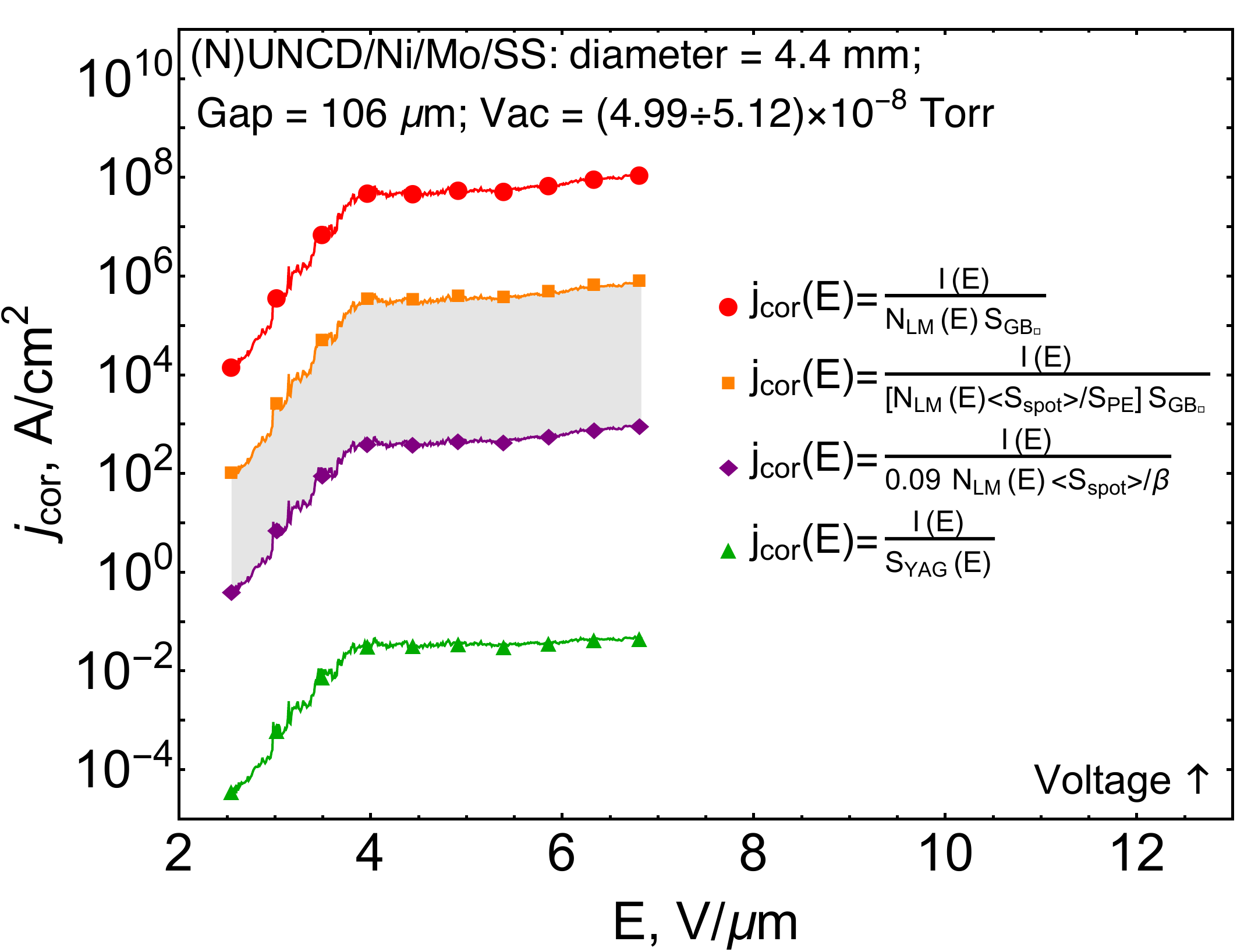}
\caption{Strategies for estimation of the field emission current density of (N)UNCD due to limited emission area. The purple curve was estimated for $\beta =1000$. The shaded area between the orange and the purple curves provides the experimental uncertainty window, whereas the red curve gives the maximal possible current density under the assumption that each FE spot on the anode plate corresponds to the emission from a single grain boundary. The curve at the bottom represents current density obtained in Ref.~\onlinecite{Chubenko_2017}.}
\label{jcor}
\end{figure}

From this we concluded\cite{Chubenko_2017} that the current-density saturation level of $\sim 100$ mA~cm$^{-2}$ represents a basic intrinsic property of (N)UNCD materials. However, the normalization factor $S_{\text{YAG}}(E)$ may still be unrealistic. In order to estimate the true FE current density, i.e. the current density obtained by normalizing the measured FE current by the cathode surface area from which the current was collected, additional corrections due to electron trajectories and/or fraction of emitting grain boundaries should be taken into account.\cite{thanks_Forbes} For this purpose, the following approximations can be made. If we assume that every single spot on the screen is formed by electron emission from a single grain boundary representing a square of approximate area 1 nm $\times$ 1 nm\footnote{Nitrogen incorporation results in a morphology change: the average grain size increases from about 4 to 16 nm and the grain-boundary width increases from about 0.5 to 2.2 nm as the nitrogen content in plasma is changed from 0$\%$ to 20$\%$.\cite{Birrell_2002} In the calculations, we use an average grain size 10 nm and the grain-boundary width 1 nm.} as shown in Fig.~\ref{spot_cor_1}, then the current density can be estimated as
\begin{equation}
j_{\text{cor}}(E) = \frac{I(E)}{N_{\text{LM}}(E)S_{\text{GB}_\Box}},
\end{equation}
where the number of local maxima, $N_\text{LM}$, is defined as a number of emission sites estimated by our algorithm and $S_{\text{GB}_\Box}= 10^{-14}$ cm$^2$. Under this assumption, the current density is obviously overestimated (see the upper curve in Fig.~\ref{jcor}), since it was experimentally shown \cite{Xu_1994} that a single FE spot on an anode is formed by several sub-emission sites originating from grain boundaries surrounding diamond grains (see the schematic in Fig.~\ref{spot_cor_2}). 
If we assume that the average radius of grains is about $R_\text{G}\sim5$ nm and the average width of grain boundaries is $W_{\text{GB}}\sim 1$ nm, the average area of a grain and the area of a surrounding grain boundary can be estimated as $S_\text{G}=\pi R_\text{G}^2$ and $S_{\text{GB}}=\pi((R_\text{G}+W_{\text{GB}}/2)^2-R_\text{G}^2)$, respectively. Then the surface fraction of grain boundaries can be simply found as $S_{\text{GB}}/(S_{\text{GB}}+S_\text{G})\approx 0.174$. Moreover, if we assume that in (N)UNCD films only 52$\%$ of grain boundaries are $sp^2$-bonded, as was estimated in Ref.~\onlinecite{Corrigan_2002}, we obtain a correction coefficient $\sim 0.09$. It should be noticed that, at high values of the applied electric field, some optical effects (e.g., blooming and/or other YAG screen post-glowing effects) may arise and lead to overestimation of the emission area. Nevertheless, our algorithm can identify the number of emission sites $N_{\text{LM}}$ with high accuracy. Therefore, we use the product of the average area of the emission sites detected on low-current micrographs ($<S_{\text{spot}}>\sim 1.3\times 10^{-5}$ cm$^2$), when the glowing effects are minimal, and the number of LM found on each micrograph (rather than the total found area $S_{\text{YAG}}(E)$). The current density can then be estimated as
\begin{equation}
j_{\text{cor}}(E) = \frac{I(E)}{0.09 N_{\text{LM}}(E)<S_{\text{spot}}>}.
\end{equation}
As mentioned earlier, the electric-field lines focus onto  the localized conductive (N)UNCD grain boundaries, providing large local electric fields, but also modifying electron trajectories. The latter causes the emitter-size change as compared to that seen on the YAG screen. In a simple approximation of straight-line trajectories, the correction due to this effect is given by the field enhancement factor $\beta$. The current density then becomes
\begin{equation}
j_{\text{cor}}(E) = \frac{I(E)}{0.09 N_{\text{LM}}(E)<S_{\text{spot}}>/\beta}.
\end{equation}
The resulting curve is shown in Fig.~\ref{jcor}.

\begin{figure}[b]
\centering
\includegraphics[width=2in]{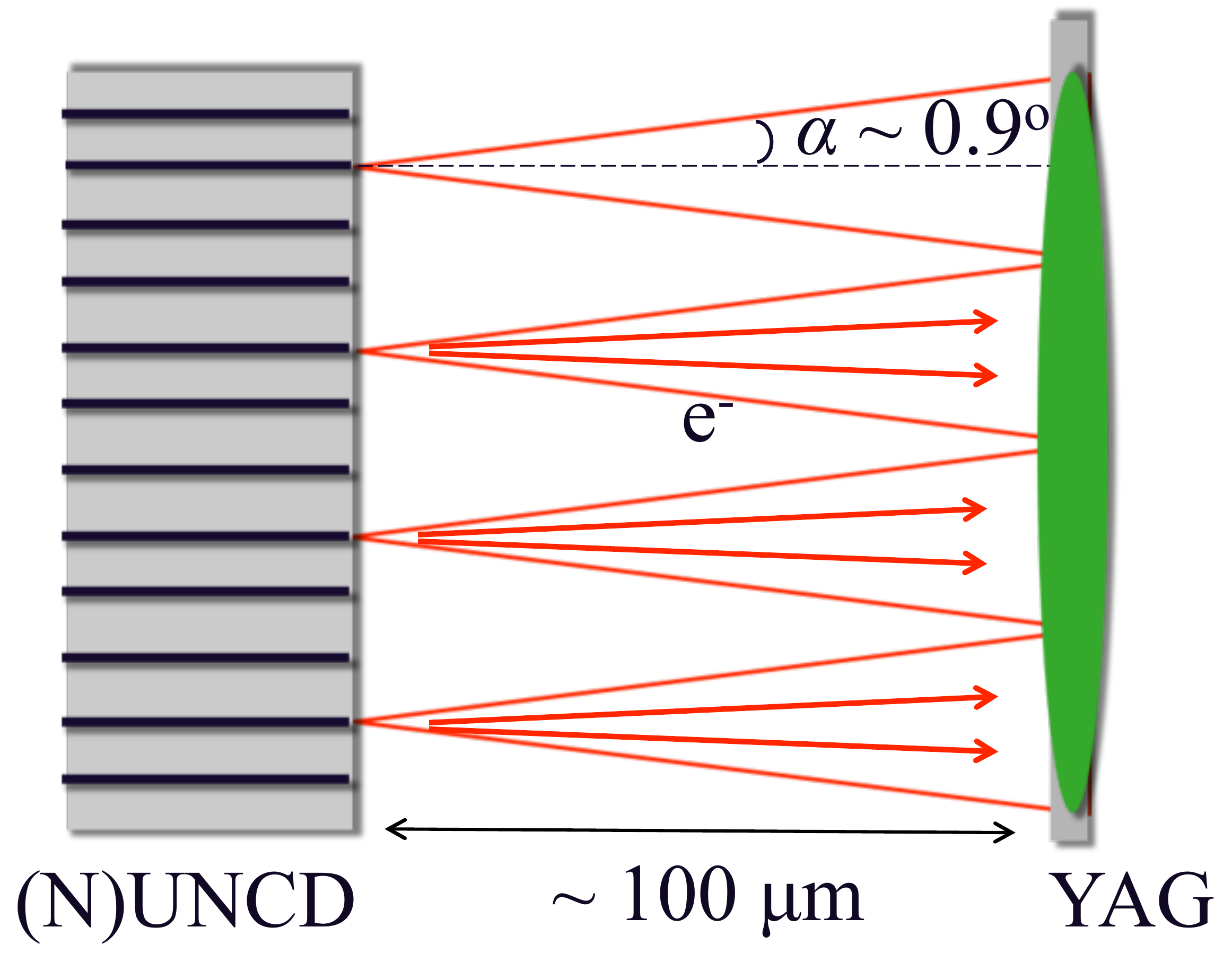}
\caption{In the point electron emitter approximation, minimum 150 emitters are required to form a spot $<S_{\text{spot}}>\sim 1.3\times 10^{-5}$ cm$^2$ on a YAG screen (or minimum $24\times 10^3$ emitters are required to form $S_{\text{YAG}}^{\text{\text{total}}}\sim 2\times 10^{-3}$ cm$^2$).}
\label{dot_emitters}
\end{figure}

Under these assumptions, the local FE area still could be overestimated since all surrounding grain boundaries are allowed to participate in electron emission. It was found from the mean transverse energy measurements \cite{Chen_2018} that $\sim$1 kV  electrons are emitted from the (N)UNCD films at the angle $\alpha\sim 0.9^\circ$ to the surface normal. If we assume that a grain boundary is a point electron emitter (see Fig.~\ref{dot_emitters}), then the FE area formed on an anode screen placed at a distance $d_{\text{gap}}=106$ $\mu$m from such emitter is $S_{\text{PE}} = \pi (d_{\text{gap}}\times \tan{\alpha})^2\sim 8.7\times 10^{-8}$ cm$^2$. The minimal number of emitters required to form an estimated FE area on a YAG screen can be found from the ratio $S_{\text{YAG}}/S_{\text{PE}}\approx N_{\text{LM}}(E)<S_{\text{spot}}>/S_{\text{PE}}$. Then the upper limit of the current density can be estimated by normalizing the current by this ratio and the approximate area of a grain boundary $S_{\text{GB}_\Box}$
\begin{equation}
j_{\text{cor}}(E) = \frac{I(E)}{[N_{\text{LM}}(E)<S_{\text{spot}}>/S_{\text{PE}}] S_{\text{GB}_\Box}}.
\end{equation}
The resulting curve is shown in Fig.~\ref{jcor}.

After all approximations are taken into account, we find the experimental current density lies within the shaded limits shown in Fig.~\ref{jcor}, i.e. somewhere between $10^3$ and $10^6$ A cm$^{-2}$. This result is in agreement with Ref.~\onlinecite{Zhu_1998}, where the experimental local current density of nanostructured diamond films was estimated to be more than $10^4$~A~cm$^{-2}$. A more accurate evaluation of the experimental current density requires a detailed simulation of the electron trajectories from UNCD surface in an electric field. However, the main difficulty in the estimation of the FE area comes from the inability to determine the fraction of emitting GBs which form an isolated emission spot on the YAG screen. 

\subsection{Correlation between Theory and Experiment}

\begin{table*}
\caption {Dependence of electronic and FE characteristics of (N)UNCD on the localization length $L_\text{loc}$.}
\begin{ruledtabular}
\begin{tabular}{ lccccc } 
  & $L_{\text{loc}}=5$ $\AA$ & $L_{\text{loc}}=10$ $\AA$ & $L_{\text{loc}}=20$ $\AA$ &$L_{\text{loc}}=40$ $\AA$&$L_{\text{loc}}=60$ $\AA$  \\
\hline
$g(E_\text{F})$, eV$^{-1}$cm$^{-3}$ &$1.218\times 10^{20}$ &$4.306\times 10^{19}$ &$1.522\times 10^{19}$ &$5.383\times 10^{18}$ &$2.930\times 10^{18}$ \\
$g(E_{\pi^*})$, eV$^{-1}$cm$^{-3}$ & $8.619\times 10^{20}$ &$3.047\times 10^{20}$ &$1.077\times 10^{20}$ &$3.809\times 10^{19}$ &$2.073\times 10^{19}$ \\
$n(y_\text{b})$, cm$^{-3}$ & $4.901\times 10^{18}$ &$1.733\times 10^{18}$ &$6.126\times 10^{17}$ &$2.166\times 10^{17}$ &$1.179\times 10^{17}$ \\
$n(0)$, cm$^{-3}$ &$1.089\times 10^{21}$ &$3.849\times 10^{20}$ &$1.361\times 10^{20}$ &$4.811\times 10^{19}$ &$2.619\times 10^{19}$\\
$\beta$ & $\sim 1850$ & $\sim 1170$ &$\sim 800$&$\sim 560$ & $\sim 440$ \\
$j_{\text{sat}}|_{\mu=\mu_0}$, A cm$^{-2}$ & $\sim 3\times 10^7$ &$\sim 7\times 10^6$ & $\sim 2\times 10^6$  & $\sim 5 \times 10^5$ & $\sim 2\times 10^5$  \\
$j_{\text{sat}}|_{\mu=\mu(\mathcal{F})}$, A cm$^{-2}$ 
& $\sim 6\times 10^5$ 
&$\sim 1.8\times 10^5$ 
& $\sim 5\times 10^4$  
& $\sim 1.6 \times 10^4$ & $\sim 8 \times 10^3$  \\
\end{tabular}
\end{ruledtabular}
\label{UNCD_localization}
\end{table*}

To calculate FE characteristics of highly conductive (N)UNCD films and compare against the obtained experimental results, we use the parameters listed in Table~\ref{UNCD_parameters}. In the literature,\cite{Nesladek_1996, Zammit_1998, Achatz_2006_APL} describing subgap absorption in polycrystalline carbonic materials, coefficients $B_1\propto K$ and $B_2\propto K$ are used as general fitting parameters. However, to estimate the carrier concentrations, $K$ needs to be included in Eqs.~\ref{n_pi_sigma} and \ref{p_pi_sigma} explicitly. $K$ contains information about the localization length, $L_{\text{loc}}$ (see Eq.~\ref{K}), which increases with increasing delocalization of the defect states. We use the localization length $L_{\text{loc}}$ as a free model parameter since its value is not known from experiment. The enhancement factor $\beta$ is used as an adjustable parameter when the theoretical curves, $j$ vs. $F_\text{s}$, are compared to the experimental results, $j$ vs. $E$, represented by the uncertainty gray-shaded range in Fig.~\ref{jcor}. Numerical results for some electronic and FE characteristics calculated as a function of the localization length $L_{\text{loc}}$ are summarized in Table~\ref{UNCD_localization}.

\begin{figure}[!]
\centering
\subfigure[][]{\includegraphics[width=3.2in]{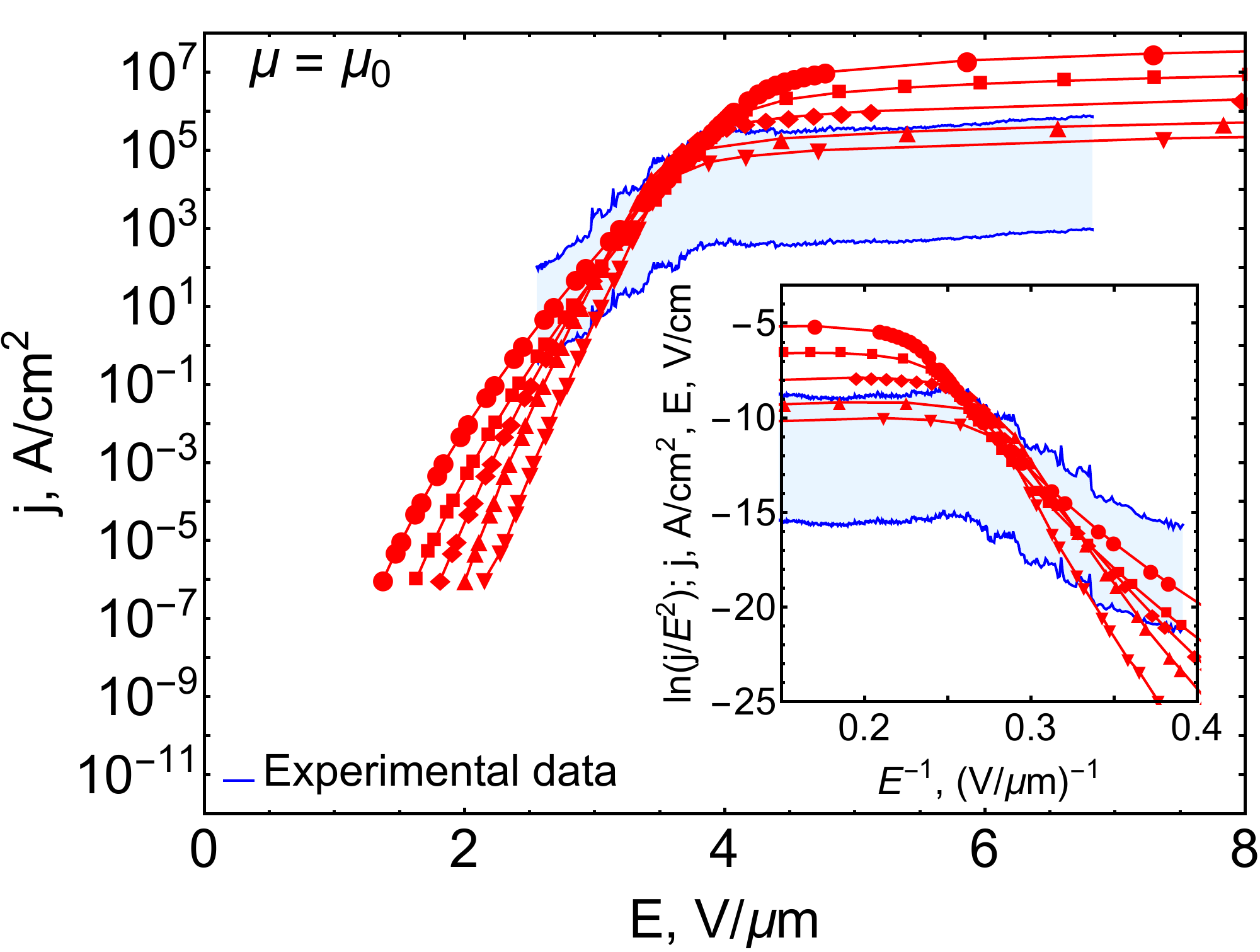}
\label{sat_mu0}}
\subfigure[][]{\includegraphics[width=3.2in]{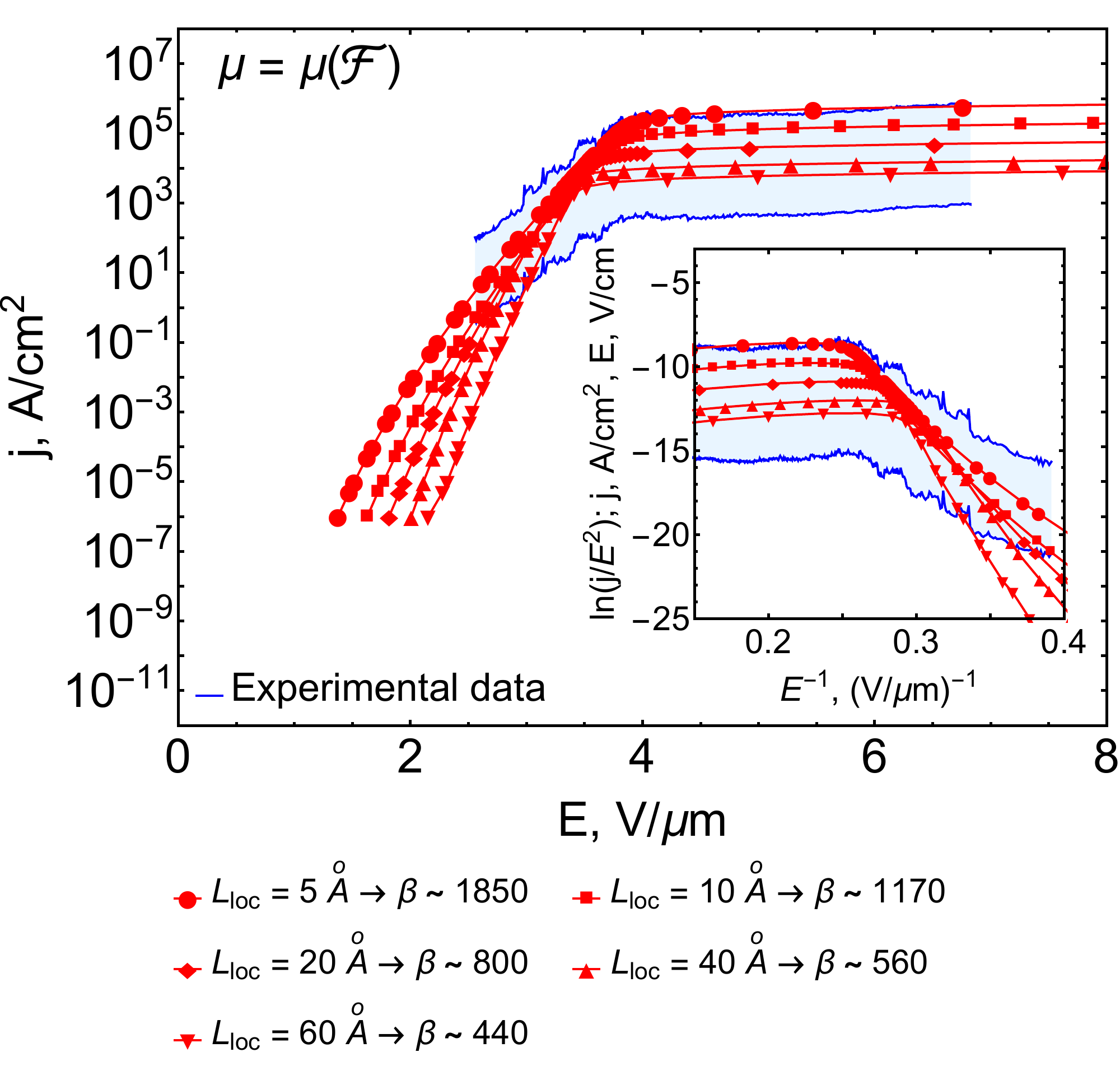}
\label{sat_mu}}
\caption{Field emission characteristics of (N)UNCD films calculated assuming (a) the field-independent mobility and (b) the field-dependent mobility and using the parameters from Table~\ref{UNCD_parameters} compared to experimental data analyzed in Section~\ref{sec:experiment}. The localization length $L_{\text{loc}}$ was used as a free model parameter.}
\label{UNCD_comp}
\end{figure}

Results for the field-independent mobility are shown in Fig.~\ref{sat_mu0}. For $L_{\text{loc}}= 5 - 10$ $\AA$,
the current-density saturation plateau reaches up to about $j_{\text{sat}}\sim 10^8 - 10^7$ A~cm$^{-2}$, a few orders of magnitude outside the experimental upper saturation limit. For the localization length $L_{\text{loc}}\sim 20 - 60$ $\AA$, indicating strong delocalization of the interband-gap states, the calculated saturation plateau overlaps with the experimental upper limit. Overall, it is found that our model overestimates the experimental current-density plateau in the regime of constant mobility.

When the dependence of the electron mobility on the applied electric field is taken into account, there are substantial differences that occur. The results are shown in Fig.~\ref{sat_mu}. Here, all calculated current-density curves lie within the experimentally determined boundaries, between $10^3$ and $10^6$ A cm$^{-2}$. This finding emphasizes the importance of knowing exact $\mu(\mathcal{F})$ relations since limited electron transport, as can be seen, is an important factor behind the saturation effect. Generally speaking, carrier mobility can be limited by ionized-impurity, acoustic and/or optical phonon, and electron-electron scattering. In polycrystalline materials like (N)UNCD that have conductivity through grain boundary networks, an additional type of scattering, scattering from a grain/grain boundary interface, could have significant effect. Detailed experimental studies of the electron-mobility response to strong electric fields would be necessary to further refine our calculations.

Fig.~\ref{UNCD_comp} also demonstrates that the saturation plateau is sensitive to the localization length. According to Eq.~\ref{n_pi_sigma}, $n_{\pi^*}\propto L_\text{loc}^{-3/2}$. Therefore, an increase in the localization length, with all other parameters unchanged, results in decreasing carrier concentration. This in turn restrains the current density at a lower value. Resulting from our model, the electron concentration in the bulk of the material is about $n(y_\text{b})\sim 10^{17}-10^{18}$ cm$^{-3}$, depending on the localization length $L_\text{loc}$ as shown in Table~\ref{UNCD_localization}. If the band bending, induced by the strong electric field, results in reduced separation between $E_\text{F}$ and $E_{\pi^*}$, such that position of the Fermi level coincides with a maximum of a $\pi^*$ band ($y=0$), then carrier concentration near the surface can increase up to $n(0) \sim 10^{20}-10^{21}$ cm$^{-3}$. This will give rise to the emission current in a metal-like manner. At some point, supply of electrons to the near-surface region becomes limited eventually resulting in the saturation behavior. 

Comparison to experiment suggests that the correct choice for $L_\text{loc}$ is between 10 and 40 $\AA$. This is in line with expectations because $L_\text{loc}=10-40$ $\AA$ matches the average grain-boundary width of highly N-incorporated UNCD,  and simultaneously places the calculated current saturation plateau in the middle of the $j$-fork and predicts the field enhancement factor close to the experimentally derived values. Therefore, the provided model fundamentally relates saturation effect and the turn-on field through the electron supply: fast deviation from the FN-like dependence and current-density saturation at moderate values is a signature of the limited electron supply that is a second important factor behind the saturation effect.

Based on the obtained insights, we conclude that FE characteristics of a large-area planar field emitter, such as (N)UNCD, suffer from intrinsic electronic properties rather than from screening the electric field due to high critical emitted current densities $j_\text{cr}\sim 10^7$ A cm$^{-2}$.\cite{Dyke_1953, Barbour_1953} Therefore, the Child-Langmuir law, a leading explanation for the saturation effect so far,\cite{Cahay_2014} cannot account for the FE current-density saturation in (N)UNCD. Further detailed  understanding of saturation in (N)UNCD will require more information on the band structure and transport via optical and high field transport experiments carried on samples with well controlled doping. Emission-area-resolved field emission experiments should be also further improved in terms of accuracy and precision to enable more reliable $j$ vs. $E$ dependences for validating theoretical work.

\section{\label{sec:conclusions}Conclusion}

The Stratton-Baskin-Lvov-Fursey formalism, commonly used to explain the FE properties of conventional semiconductors, has been adapted to study current-density saturation in (N)UNCD. We evaluate the role of the density-of-states structure and conclude that the behavior of the FE current density in this material can be qualitatively and quantitatively explained by the properties similar to those of conventional semiconductors, i.e. by ($i$) partial penetration of the electric field into the material, resulting in the formation of a space-charge layer underneath the surface, and ($ii$) electric-field-dependent carrier mobility. 

The quantitative results depend on the choice of density-of-states model parameters, degree of delocalization, and mobility. The shape of the current-density curve always demonstrates saturation behavior which we refer to the overall charge available in the bulk and at the accumulation layer near material/vacuum interface, and to the rate at which the charge can be moved across the material. The latter conclusions rest upon comparison of our model and experimental results. Specifically, we found that the space-charge effect alone causes the current-density saturation at the level of $j\sim 10^5-10^8$ A cm$^{-2}$, depending on the degree of delocalization of the interband-gap states. The calculated saturation plateau significantly reduces down to $j\sim 10^4-10^6$ A cm$^{-2}$ and matches well with experimental uncertainty range when the effect of field-dependent mobility is taken into considerations. Given the importance of the found mobility effect on saturation, the comprehensive study of electron transport in (N)UNCD in high electric fields is required.

\begin{acknowledgments}
OC was supported by The George Washington University in the form of a graduate student fellowship. SSB was supported by the U.S. National Science Foundation under Award No. PHY-1549132, the Center for Bright Beams and under Award No. PHY-1535639. SVB was supported by funding from the College of Engineering, Michigan State University, under Global Impact Initiative.

The authors thank Prof. R. G. Forbes and Prof. A. V. Arkhipov for valuable discussions. We are grateful to Dr. M. Lujan for his help with the manuscript.
\end{acknowledgments}

\bibliography{References}

\begin{thebibliography}{71}%
\makeatletter
\providecommand \@ifxundefined [1]{%
 \@ifx{#1\undefined}
}%
\providecommand \@ifnum [1]{%
 \ifnum #1\expandafter \@firstoftwo
 \else \expandafter \@secondoftwo
 \fi
}%
\providecommand \@ifx [1]{%
 \ifx #1\expandafter \@firstoftwo
 \else \expandafter \@secondoftwo
 \fi
}%
\providecommand \natexlab [1]{#1}%
\providecommand \enquote  [1]{``#1''}%
\providecommand \bibnamefont  [1]{#1}%
\providecommand \bibfnamefont [1]{#1}%
\providecommand \citenamefont [1]{#1}%
\providecommand \href@noop [0]{\@secondoftwo}%
\providecommand \href [0]{\begingroup \@sanitize@url \@href}%
\providecommand \@href[1]{\@@startlink{#1}\@@href}%
\providecommand \@@href[1]{\endgroup#1\@@endlink}%
\providecommand \@sanitize@url [0]{\catcode `\\12\catcode `\$12\catcode
  `\&12\catcode `\#12\catcode `\^12\catcode `\_12\catcode `\%12\relax}%
\providecommand \@@startlink[1]{}%
\providecommand \@@endlink[0]{}%
\providecommand \url  [0]{\begingroup\@sanitize@url \@url }%
\providecommand \@url [1]{\endgroup\@href {#1}{\urlprefix }}%
\providecommand \urlprefix  [0]{URL }%
\providecommand \Eprint [0]{\href }%
\providecommand \doibase [0]{http://dx.doi.org/}%
\providecommand \selectlanguage [0]{\@gobble}%
\providecommand \bibinfo  [0]{\@secondoftwo}%
\providecommand \bibfield  [0]{\@secondoftwo}%
\providecommand \translation [1]{[#1]}%
\providecommand \BibitemOpen [0]{}%
\providecommand \bibitemStop [0]{}%
\providecommand \bibitemNoStop [0]{.\EOS\space}%
\providecommand \EOS [0]{\spacefactor3000\relax}%
\providecommand \BibitemShut  [1]{\csname bibitem#1\endcsname}%
\let\auto@bib@innerbib\@empty
\bibitem [{\citenamefont {Wu}\ \emph {et~al.}(2000)\citenamefont {Wu},
  \citenamefont {Wang}, \citenamefont {Cao}, \citenamefont {Wang},\ and\
  \citenamefont {Jiang}}]{Wu_2000}%
  \BibitemOpen
  \bibfield  {author} {\bibinfo {author} {\bibfnamefont {K.}~\bibnamefont
  {Wu}}, \bibinfo {author} {\bibfnamefont {E.~G.}\ \bibnamefont {Wang}},
  \bibinfo {author} {\bibfnamefont {Z.~X.}\ \bibnamefont {Cao}}, \bibinfo
  {author} {\bibfnamefont {Z.~L.}\ \bibnamefont {Wang}}, \ and\ \bibinfo
  {author} {\bibfnamefont {X.}~\bibnamefont {Jiang}},\ }\href
  {https://doi.org/10.1063/1.1287602} {\bibfield  {journal} {\bibinfo
  {journal} {J. Appl. Phys.}\ }\textbf {\bibinfo {volume} {88}},\ \bibinfo
  {pages} {2967} (\bibinfo {year} {2000})}\BibitemShut {NoStop}%
\bibitem [{\citenamefont {Corrigan}\ \emph {et~al.}(2002)\citenamefont
  {Corrigan}, \citenamefont {Gruen}, \citenamefont {Krauss}, \citenamefont
  {Zapol},\ and\ \citenamefont {Chang}}]{Corrigan_2002}%
  \BibitemOpen
  \bibfield  {author} {\bibinfo {author} {\bibfnamefont {T.~D.}\ \bibnamefont
  {Corrigan}}, \bibinfo {author} {\bibfnamefont {D.~M.}\ \bibnamefont {Gruen}},
  \bibinfo {author} {\bibfnamefont {A.~R.}\ \bibnamefont {Krauss}}, \bibinfo
  {author} {\bibfnamefont {P.}~\bibnamefont {Zapol}}, \ and\ \bibinfo {author}
  {\bibfnamefont {R.~P.~H.}\ \bibnamefont {Chang}},\ }\href
  {https://doi.org/10.1016/S0925-9635(01)00517-9} {\bibfield  {journal}
  {\bibinfo  {journal} {Diamond and Rel. Mat.}\ }\textbf {\bibinfo {volume}
  {11}},\ \bibinfo {pages} {43} (\bibinfo {year} {2002})}\BibitemShut {NoStop}%
\bibitem [{\citenamefont {Ikeda}\ and\ \citenamefont
  {Teii}(2009)}]{Ikeda_2009}%
  \BibitemOpen
  \bibfield  {author} {\bibinfo {author} {\bibfnamefont {T.}~\bibnamefont
  {Ikeda}}\ and\ \bibinfo {author} {\bibfnamefont {K.}~\bibnamefont {Teii}},\
  }\href {https://doi.org/10.1063/1.3115767} {\bibfield  {journal} {\bibinfo
  {journal} {Appl. Phys. Lett.}\ }\textbf {\bibinfo {volume} {94}},\ \bibinfo
  {pages} {143102} (\bibinfo {year} {2009})}\BibitemShut {NoStop}%
\bibitem [{\citenamefont {Lin}\ \emph {et~al.}(2011)\citenamefont {Lin},
  \citenamefont {Sankaran}, \citenamefont {Chen}, \citenamefont {Lee},
  \citenamefont {Chen}, \citenamefont {Lin},\ and\ \citenamefont
  {Tai}}]{Lin_2011}%
  \BibitemOpen
  \bibfield  {author} {\bibinfo {author} {\bibfnamefont {Y.~C.}\ \bibnamefont
  {Lin}}, \bibinfo {author} {\bibfnamefont {K.~J.}\ \bibnamefont {Sankaran}},
  \bibinfo {author} {\bibfnamefont {Y.~C.}\ \bibnamefont {Chen}}, \bibinfo
  {author} {\bibfnamefont {C.~Y.}\ \bibnamefont {Lee}}, \bibinfo {author}
  {\bibfnamefont {H.~C.}\ \bibnamefont {Chen}}, \bibinfo {author}
  {\bibfnamefont {I.~N.}\ \bibnamefont {Lin}}, \ and\ \bibinfo {author}
  {\bibfnamefont {N.~H.}\ \bibnamefont {Tai}},\ }\href
  {https://doi.org/10.1016/j.diamond.2010.11.026} {\bibfield  {journal}
  {\bibinfo  {journal} {Diamond and Rel. Mat.}\ }\textbf {\bibinfo {volume}
  {20}},\ \bibinfo {pages} {191} (\bibinfo {year} {2011})}\BibitemShut
  {NoStop}%
\bibitem [{\citenamefont {Baryshev}\ \emph {et~al.}(2014)\citenamefont
  {Baryshev}, \citenamefont {Antipov}, \citenamefont {Shao}, \citenamefont
  {Jing}, \citenamefont {P{\'e}rez~Quintero}, \citenamefont {Qiu},
  \citenamefont {Liu}, \citenamefont {Gai}, \citenamefont {Kanareykin},\ and\
  \citenamefont {Sumant}}]{Baryshev_2014}%
  \BibitemOpen
  \bibfield  {author} {\bibinfo {author} {\bibfnamefont {S.~V.}\ \bibnamefont
  {Baryshev}}, \bibinfo {author} {\bibfnamefont {S.}~\bibnamefont {Antipov}},
  \bibinfo {author} {\bibfnamefont {J.}~\bibnamefont {Shao}}, \bibinfo {author}
  {\bibfnamefont {C.}~\bibnamefont {Jing}}, \bibinfo {author} {\bibfnamefont
  {K.~J.}\ \bibnamefont {P{\'e}rez~Quintero}}, \bibinfo {author} {\bibfnamefont
  {J.}~\bibnamefont {Qiu}}, \bibinfo {author} {\bibfnamefont {W.}~\bibnamefont
  {Liu}}, \bibinfo {author} {\bibfnamefont {W.}~\bibnamefont {Gai}}, \bibinfo
  {author} {\bibfnamefont {A.~D.}\ \bibnamefont {Kanareykin}}, \ and\ \bibinfo
  {author} {\bibfnamefont {A.~V.}\ \bibnamefont {Sumant}},\ }\href
  {https://doi.org/10.1063/1.4901723} {\bibfield  {journal} {\bibinfo
  {journal} {Appl. Phys. Lett.}\ }\textbf {\bibinfo {volume} {105}},\ \bibinfo
  {pages} {203505} (\bibinfo {year} {2014})}\BibitemShut {NoStop}%
\bibitem [{\citenamefont {Liao}\ \emph {et~al.}(1998)\citenamefont {Liao},
  \citenamefont {Zhang}, \citenamefont {Wang},\ and\ \citenamefont
  {Liao}}]{Liao_1998}%
  \BibitemOpen
  \bibfield  {author} {\bibinfo {author} {\bibfnamefont {M.}~\bibnamefont
  {Liao}}, \bibinfo {author} {\bibfnamefont {Z.}~\bibnamefont {Zhang}},
  \bibinfo {author} {\bibfnamefont {W.}~\bibnamefont {Wang}}, \ and\ \bibinfo
  {author} {\bibfnamefont {K.}~\bibnamefont {Liao}},\ }\href
  {https://doi.org/10.1063/1.368096} {\bibfield  {journal} {\bibinfo  {journal}
  {J. Appl. Phys.}\ }\textbf {\bibinfo {volume} {84}},\ \bibinfo {pages} {1081}
  (\bibinfo {year} {1998})}\BibitemShut {NoStop}%
\bibitem [{\citenamefont {Xu}, \citenamefont {Chen},\ and\ \citenamefont
  {Deng}(2000)}]{Xu_2000}%
  \BibitemOpen
  \bibfield  {author} {\bibinfo {author} {\bibfnamefont {N.~S.}\ \bibnamefont
  {Xu}}, \bibinfo {author} {\bibfnamefont {J.}~\bibnamefont {Chen}}, \ and\
  \bibinfo {author} {\bibfnamefont {S.~Z.}\ \bibnamefont {Deng}},\ }\href
  {https://doi.org/10.1063/1.126377} {\bibfield  {journal} {\bibinfo  {journal}
  {Appl. Phys. Lett.}\ }\textbf {\bibinfo {volume} {76}},\ \bibinfo {pages}
  {2463} (\bibinfo {year} {2000})}\BibitemShut {NoStop}%
\bibitem [{\citenamefont {Ducati}\ \emph {et~al.}(2002)\citenamefont {Ducati},
  \citenamefont {Barborini}, \citenamefont {Piseri}, \citenamefont {Milani},\
  and\ \citenamefont {Robertson}}]{Ducati_2002}%
  \BibitemOpen
  \bibfield  {author} {\bibinfo {author} {\bibfnamefont {C.}~\bibnamefont
  {Ducati}}, \bibinfo {author} {\bibfnamefont {E.}~\bibnamefont {Barborini}},
  \bibinfo {author} {\bibfnamefont {P.}~\bibnamefont {Piseri}}, \bibinfo
  {author} {\bibfnamefont {P.}~\bibnamefont {Milani}}, \ and\ \bibinfo {author}
  {\bibfnamefont {J.}~\bibnamefont {Robertson}},\ }\href
  {https://doi.org/10.1063/1.1512969} {\bibfield  {journal} {\bibinfo
  {journal} {J. Appl. Phys.}\ }\textbf {\bibinfo {volume} {92}},\ \bibinfo
  {pages} {5482} (\bibinfo {year} {2002})}\BibitemShut {NoStop}%
\bibitem [{\citenamefont {Varshney}\ \emph {et~al.}(2011)\citenamefont
  {Varshney}, \citenamefont {Venkateswara~Rao}, \citenamefont {Guinel},
  \citenamefont {Ishikawa}, \citenamefont {Weiner},\ and\ \citenamefont
  {Morell}}]{Varshney_2011}%
  \BibitemOpen
  \bibfield  {author} {\bibinfo {author} {\bibfnamefont {D.}~\bibnamefont
  {Varshney}}, \bibinfo {author} {\bibfnamefont {C.}~\bibnamefont
  {Venkateswara~Rao}}, \bibinfo {author} {\bibfnamefont {M.~J.~F.}\
  \bibnamefont {Guinel}}, \bibinfo {author} {\bibfnamefont {Y.}~\bibnamefont
  {Ishikawa}}, \bibinfo {author} {\bibfnamefont {B.~R.}\ \bibnamefont
  {Weiner}}, \ and\ \bibinfo {author} {\bibfnamefont {G.}~\bibnamefont
  {Morell}},\ }\href {https://doi.org/10.1063/1.3627370} {\bibfield  {journal}
  {\bibinfo  {journal} {J. Appl. Phys.}\ }\textbf {\bibinfo {volume} {110}},\
  \bibinfo {pages} {044324} (\bibinfo {year} {2011})}\BibitemShut {NoStop}%
\bibitem [{\citenamefont {Cahay}\ \emph {et~al.}(2014)\citenamefont {Cahay},
  \citenamefont {Murray}, \citenamefont {Back}, \citenamefont {Fairchild},
  \citenamefont {Boeckl}, \citenamefont {Bulmer}, \citenamefont {Koziol},
  \citenamefont {Gruen}, \citenamefont {Sparkes}, \citenamefont {Orozco},\ and\
  \citenamefont {O'Neill}}]{Cahay_2014}%
  \BibitemOpen
  \bibfield  {author} {\bibinfo {author} {\bibfnamefont {M.}~\bibnamefont
  {Cahay}}, \bibinfo {author} {\bibfnamefont {P.~T.}\ \bibnamefont {Murray}},
  \bibinfo {author} {\bibfnamefont {T.~C.}\ \bibnamefont {Back}}, \bibinfo
  {author} {\bibfnamefont {S.}~\bibnamefont {Fairchild}}, \bibinfo {author}
  {\bibfnamefont {J.}~\bibnamefont {Boeckl}}, \bibinfo {author} {\bibfnamefont
  {J.}~\bibnamefont {Bulmer}}, \bibinfo {author} {\bibfnamefont {K.~K.~K.}\
  \bibnamefont {Koziol}}, \bibinfo {author} {\bibfnamefont {G.}~\bibnamefont
  {Gruen}}, \bibinfo {author} {\bibfnamefont {M.}~\bibnamefont {Sparkes}},
  \bibinfo {author} {\bibfnamefont {F.}~\bibnamefont {Orozco}}, \ and\ \bibinfo
  {author} {\bibfnamefont {W.}~\bibnamefont {O'Neill}},\ }\href
  {https://doi.org/10.1063/1.4900787} {\bibfield  {journal} {\bibinfo
  {journal} {Appl. Phys. Lett.}\ }\textbf {\bibinfo {volume} {105}},\ \bibinfo
  {pages} {173107} (\bibinfo {year} {2014})}\BibitemShut {NoStop}%
\bibitem [{\citenamefont {Arthur}(1965)}]{Arthur_1965}%
  \BibitemOpen
  \bibfield  {author} {\bibinfo {author} {\bibfnamefont {J.~R.}\ \bibnamefont
  {Arthur}},\ }\href {https://doi.org/10.1063/1.1702953} {\bibfield  {journal}
  {\bibinfo  {journal} {J. Appl. Phys.}\ }\textbf {\bibinfo {volume} {36}},\
  \bibinfo {pages} {3221} (\bibinfo {year} {1965})}\BibitemShut {NoStop}%
\bibitem [{\citenamefont {Baskin}, \citenamefont {L'vov},\ and\ \citenamefont
  {Fursey}(1971)}]{Baskin_1971}%
  \BibitemOpen
  \bibfield  {author} {\bibinfo {author} {\bibfnamefont {L.~M.}\ \bibnamefont
  {Baskin}}, \bibinfo {author} {\bibfnamefont {O.~I.}\ \bibnamefont {L'vov}}, \
  and\ \bibinfo {author} {\bibfnamefont {G.~N.}\ \bibnamefont {Fursey}},\
  }\href {https://doi.org/10.1002/pssb.2220470105} {\bibfield  {journal}
  {\bibinfo  {journal} {Phys. Stat. Sol. (b)}\ }\textbf {\bibinfo {volume}
  {47}},\ \bibinfo {pages} {49} (\bibinfo {year} {1971})}\BibitemShut {NoStop}%
\bibitem [{\citenamefont {Serbun}\ \emph {et~al.}(2013)\citenamefont {Serbun},
  \citenamefont {Bornmann}, \citenamefont {Navitski}, \citenamefont {Muller},
  \citenamefont {Prommesberger}, \citenamefont {Langer}, \citenamefont {Dams},\
  and\ \citenamefont {Schreiner}}]{Serbun_2013}%
  \BibitemOpen
  \bibfield  {author} {\bibinfo {author} {\bibfnamefont {P.}~\bibnamefont
  {Serbun}}, \bibinfo {author} {\bibfnamefont {B.}~\bibnamefont {Bornmann}},
  \bibinfo {author} {\bibfnamefont {A.}~\bibnamefont {Navitski}}, \bibinfo
  {author} {\bibfnamefont {G.}~\bibnamefont {Muller}}, \bibinfo {author}
  {\bibfnamefont {C.}~\bibnamefont {Prommesberger}}, \bibinfo {author}
  {\bibfnamefont {C.}~\bibnamefont {Langer}}, \bibinfo {author} {\bibfnamefont
  {F.}~\bibnamefont {Dams}}, \ and\ \bibinfo {author} {\bibfnamefont
  {R.}~\bibnamefont {Schreiner}},\ }\href {https://doi.org/10.1116/1.4765088}
  {\bibfield  {journal} {\bibinfo  {journal} {J. Vac. Sci. Technol. B}\
  }\textbf {\bibinfo {volume} {31}},\ \bibinfo {pages} {02B101} (\bibinfo
  {year} {2013})}\BibitemShut {NoStop}%
\bibitem [{\citenamefont {Stratton}(1955)}]{Stratton_1955}%
  \BibitemOpen
  \bibfield  {author} {\bibinfo {author} {\bibfnamefont {R.}~\bibnamefont
  {Stratton}},\ }\href {https://doi.org/10.1088/0370-1301/68/10/307} {\bibfield
   {journal} {\bibinfo  {journal} {Proc. Phys. Soc. B}\ }\textbf {\bibinfo
  {volume} {68}},\ \bibinfo {pages} {746} (\bibinfo {year} {1955})}\BibitemShut
  {NoStop}%
\bibitem [{\citenamefont {Stratton}(1962)}]{Stratton_1962}%
  \BibitemOpen
  \bibfield  {author} {\bibinfo {author} {\bibfnamefont {R.}~\bibnamefont
  {Stratton}},\ }\href {https://doi.org/10.1103/PhysRev.125.67} {\bibfield
  {journal} {\bibinfo  {journal} {Phys. Rev.}\ }\textbf {\bibinfo {volume}
  {125}},\ \bibinfo {pages} {67} (\bibinfo {year} {1962})}\BibitemShut
  {NoStop}%
\bibitem [{\citenamefont {Fowler}\ and\ \citenamefont
  {Nordheim}(1928)}]{Fowler_1928}%
  \BibitemOpen
  \bibfield  {author} {\bibinfo {author} {\bibfnamefont {R.~H.}\ \bibnamefont
  {Fowler}}\ and\ \bibinfo {author} {\bibfnamefont {L.~W.}\ \bibnamefont
  {Nordheim}},\ }\href {https://doi.org/10.1098/rspa.1928.0091} {\bibfield
  {journal} {\bibinfo  {journal} {Proc. R. Soc. London, Ser. A}\ }\textbf
  {\bibinfo {volume} {119}},\ \bibinfo {pages} {173} (\bibinfo {year}
  {1928})}\BibitemShut {NoStop}%
\bibitem [{\citenamefont {Nordheim}(1928)}]{Nordheim_1928}%
  \BibitemOpen
  \bibfield  {author} {\bibinfo {author} {\bibfnamefont {L.~W.}\ \bibnamefont
  {Nordheim}},\ }\href {https://doi.org/10.1098/rspa.1928.0222} {\bibfield
  {journal} {\bibinfo  {journal} {Proc. R. Soc. London, Ser. A}\ }\textbf
  {\bibinfo {volume} {121}},\ \bibinfo {pages} {626} (\bibinfo {year}
  {1928})}\BibitemShut {NoStop}%
\bibitem [{\citenamefont {Murphy}\ and\ \citenamefont
  {Good~Jr}(1956)}]{Murphy_1956}%
  \BibitemOpen
  \bibfield  {author} {\bibinfo {author} {\bibfnamefont {E.~L.}\ \bibnamefont
  {Murphy}}\ and\ \bibinfo {author} {\bibfnamefont {R.~H.}\ \bibnamefont
  {Good~Jr}},\ }\href {https://doi.org/10.1103/PhysRev.102.1464} {\bibfield
  {journal} {\bibinfo  {journal} {Phys. Rev.}\ }\textbf {\bibinfo {volume}
  {102}},\ \bibinfo {pages} {1464} (\bibinfo {year} {1956})}\BibitemShut
  {NoStop}%
\bibitem [{\citenamefont {Huang}, \citenamefont {Qin},\ and\ \citenamefont
  {Zhang}(1997)}]{Huang_1997}%
  \BibitemOpen
  \bibfield  {author} {\bibinfo {author} {\bibfnamefont {Q.-A.}\ \bibnamefont
  {Huang}}, \bibinfo {author} {\bibfnamefont {M.}~\bibnamefont {Qin}}, \ and\
  \bibinfo {author} {\bibfnamefont {B.}~\bibnamefont {Zhang}},\ }\href
  {https://doi.org/10.1063/1.365304} {\bibfield  {journal} {\bibinfo  {journal}
  {J. Appl. Phys.}\ }\textbf {\bibinfo {volume} {81}} (\bibinfo {year}
  {1997})}\BibitemShut {NoStop}%
\bibitem [{\citenamefont {Liu}, \citenamefont {Chiang},\ and\ \citenamefont
  {Heritage}(2006)}]{Liu_2006}%
  \BibitemOpen
  \bibfield  {author} {\bibinfo {author} {\bibfnamefont {K.~X.}\ \bibnamefont
  {Liu}}, \bibinfo {author} {\bibfnamefont {C.-J.}\ \bibnamefont {Chiang}}, \
  and\ \bibinfo {author} {\bibfnamefont {J.~P.}\ \bibnamefont {Heritage}},\
  }\href {https://doi.org/10.1063/1.2168031} {\bibfield  {journal} {\bibinfo
  {journal} {J. Appl. Phys.}\ }\textbf {\bibinfo {volume} {99}},\ \bibinfo
  {pages} {034502} (\bibinfo {year} {2006})}\BibitemShut {NoStop}%
\bibitem [{\citenamefont {Kingston}\ and\ \citenamefont
  {Neustadter}(1955)}]{Kingston_1955}%
  \BibitemOpen
  \bibfield  {author} {\bibinfo {author} {\bibfnamefont {R.~H.}\ \bibnamefont
  {Kingston}}\ and\ \bibinfo {author} {\bibfnamefont {S.~F.}\ \bibnamefont
  {Neustadter}},\ }\href {https://doi.org/10.1063/1.1722077} {\bibfield
  {journal} {\bibinfo  {journal} {J. Appl. Phys.}\ }\textbf {\bibinfo {volume}
  {26}},\ \bibinfo {pages} {718} (\bibinfo {year} {1955})}\BibitemShut
  {NoStop}%
\bibitem [{\citenamefont {Seiwatz}\ and\ \citenamefont
  {Green}(1958)}]{Seiwatz_1958}%
  \BibitemOpen
  \bibfield  {author} {\bibinfo {author} {\bibfnamefont {R.}~\bibnamefont
  {Seiwatz}}\ and\ \bibinfo {author} {\bibfnamefont {M.}~\bibnamefont
  {Green}},\ }\href {https://doi.org/10.1063/1.1723358} {\bibfield  {journal}
  {\bibinfo  {journal} {J. Appl. Phys.}\ }\textbf {\bibinfo {volume} {29}},\
  \bibinfo {pages} {1034} (\bibinfo {year} {1958})}\BibitemShut {NoStop}%
\bibitem [{\citenamefont {Lin}\ and\ \citenamefont {Liou}(1998)}]{Lin_1998}%
  \BibitemOpen
  \bibfield  {author} {\bibinfo {author} {\bibfnamefont {L.-T.~S.}\
  \bibnamefont {Lin}}\ and\ \bibinfo {author} {\bibfnamefont {Y.}~\bibnamefont
  {Liou}},\ }\href {https://doi.org/10.1063/1.367190} {\bibfield  {journal}
  {\bibinfo  {journal} {J. Appl. Phys.}\ }\textbf {\bibinfo {volume} {83}},\
  \bibinfo {pages} {4303} (\bibinfo {year} {1998})}\BibitemShut {NoStop}%
\bibitem [{\citenamefont {Robertson}\ and\ \citenamefont
  {O'Reilly}(1987)}]{Robertson_1987}%
  \BibitemOpen
  \bibfield  {author} {\bibinfo {author} {\bibfnamefont {J.}~\bibnamefont
  {Robertson}}\ and\ \bibinfo {author} {\bibfnamefont {E.~P.}\ \bibnamefont
  {O'Reilly}},\ }\href {https://doi.org/10.1103/PhysRevB.35.2946} {\bibfield
  {journal} {\bibinfo  {journal} {Phys. Rev. B}\ }\textbf {\bibinfo {volume}
  {35}},\ \bibinfo {pages} {2946} (\bibinfo {year} {1987})}\BibitemShut
  {NoStop}%
\bibitem [{\citenamefont {Zapol}\ \emph {et~al.}(2001)\citenamefont {Zapol},
  \citenamefont {Sternberg}, \citenamefont {Curtiss}, \citenamefont
  {Frauenheim},\ and\ \citenamefont {Gruen}}]{Zapol_2001}%
  \BibitemOpen
  \bibfield  {author} {\bibinfo {author} {\bibfnamefont {P.}~\bibnamefont
  {Zapol}}, \bibinfo {author} {\bibfnamefont {M.}~\bibnamefont {Sternberg}},
  \bibinfo {author} {\bibfnamefont {L.~A.}\ \bibnamefont {Curtiss}}, \bibinfo
  {author} {\bibfnamefont {T.}~\bibnamefont {Frauenheim}}, \ and\ \bibinfo
  {author} {\bibfnamefont {D.~M.}\ \bibnamefont {Gruen}},\ }\href
  {https://doi.org/10.1103/PhysRevB.65.045403} {\bibfield  {journal} {\bibinfo
  {journal} {Phys. Rev. B}\ }\textbf {\bibinfo {volume} {65}},\ \bibinfo
  {pages} {045403} (\bibinfo {year} {2001})}\BibitemShut {NoStop}%
\bibitem [{\citenamefont {Forbes}(2009)}]{Forbes_2009}%
  \BibitemOpen
  \bibfield  {author} {\bibinfo {author} {\bibfnamefont {R.~G.}\ \bibnamefont
  {Forbes}},\ }\href {https://doi.org/10.1116/1.3137964} {\bibfield  {journal}
  {\bibinfo  {journal} {J. Vac. Sci. Technol. B}\ }\textbf {\bibinfo {volume}
  {27}},\ \bibinfo {pages} {1200} (\bibinfo {year} {2009})}\BibitemShut
  {NoStop}%
\bibitem [{\citenamefont {Xu}, \citenamefont {Tzeng},\ and\ \citenamefont
  {Latham}(1993)}]{Xu_1993}%
  \BibitemOpen
  \bibfield  {author} {\bibinfo {author} {\bibfnamefont {N.~S.}\ \bibnamefont
  {Xu}}, \bibinfo {author} {\bibfnamefont {Y.}~\bibnamefont {Tzeng}}, \ and\
  \bibinfo {author} {\bibfnamefont {R.~V.}\ \bibnamefont {Latham}},\ }\href
  {https://doi.org/10.1088/0022-3727/26/10/035} {\bibfield  {journal} {\bibinfo
   {journal} {J. Phys. D: Appl. Phys.}\ }\textbf {\bibinfo {volume} {26}},\
  \bibinfo {pages} {1776} (\bibinfo {year} {1993})}\BibitemShut {NoStop}%
\bibitem [{\citenamefont {Xu}, \citenamefont {Tzeng},\ and\ \citenamefont
  {Latham}(1994)}]{Xu_1994}%
  \BibitemOpen
  \bibfield  {author} {\bibinfo {author} {\bibfnamefont {N.~S.}\ \bibnamefont
  {Xu}}, \bibinfo {author} {\bibfnamefont {Y.}~\bibnamefont {Tzeng}}, \ and\
  \bibinfo {author} {\bibfnamefont {R.~V.}\ \bibnamefont {Latham}},\ }\href
  {https://doi.org/10.1088/0022-3727/27/9/027} {\bibfield  {journal} {\bibinfo
  {journal} {J. Phys. D: Appl. Phys.}\ }\textbf {\bibinfo {volume} {27}},\
  \bibinfo {pages} {1988} (\bibinfo {year} {1994})}\BibitemShut {NoStop}%
\bibitem [{\citenamefont {Zhu}, \citenamefont {Kochanski},\ and\ \citenamefont
  {Jin}(1998)}]{Zhu_1998}%
  \BibitemOpen
  \bibfield  {author} {\bibinfo {author} {\bibfnamefont {W.}~\bibnamefont
  {Zhu}}, \bibinfo {author} {\bibfnamefont {G.~P.}\ \bibnamefont {Kochanski}},
  \ and\ \bibinfo {author} {\bibfnamefont {S.}~\bibnamefont {Jin}},\ }\href
  {https://doi.org/10.1126/science.282.5393.1471} {\bibfield  {journal}
  {\bibinfo  {journal} {Science}\ }\textbf {\bibinfo {volume} {282}},\ \bibinfo
  {pages} {1471} (\bibinfo {year} {1998})}\BibitemShut {NoStop}%
\bibitem [{\citenamefont {Chubenko}\ \emph
  {et~al.}(2017{\natexlab{a}})\citenamefont {Chubenko}, \citenamefont
  {Baturin}, \citenamefont {Kovi}, \citenamefont {Sumant},\ and\ \citenamefont
  {Baryshev}}]{Chubenko_2017}%
  \BibitemOpen
  \bibfield  {author} {\bibinfo {author} {\bibfnamefont {O.}~\bibnamefont
  {Chubenko}}, \bibinfo {author} {\bibfnamefont {S.~S.}\ \bibnamefont
  {Baturin}}, \bibinfo {author} {\bibfnamefont {K.~K.}\ \bibnamefont {Kovi}},
  \bibinfo {author} {\bibfnamefont {A.~V.}\ \bibnamefont {Sumant}}, \ and\
  \bibinfo {author} {\bibfnamefont {S.~V.}\ \bibnamefont {Baryshev}},\ }\href
  {https://doi.org/10.1021/acsami.7b07062} {\bibfield  {journal} {\bibinfo
  {journal} {ACS Appl. Mater. Interfaces}\ }\textbf {\bibinfo {volume} {9}},\
  \bibinfo {pages} {33229} (\bibinfo {year} {2017}{\natexlab{a}})}\BibitemShut
  {NoStop}%
\bibitem [{\citenamefont {Chubenko}\ \emph
  {et~al.}(2017{\natexlab{b}})\citenamefont {Chubenko}, \citenamefont
  {Baturin}, \citenamefont {Sumant}, \citenamefont {Zinovev}, \citenamefont
  {Kovi},\ and\ \citenamefont {Baryshev}}]{Chubenko_2017_IVNC}%
  \BibitemOpen
  \bibfield  {author} {\bibinfo {author} {\bibfnamefont {O.}~\bibnamefont
  {Chubenko}}, \bibinfo {author} {\bibfnamefont {S.~S.}\ \bibnamefont
  {Baturin}}, \bibinfo {author} {\bibfnamefont {A.~V.}\ \bibnamefont {Sumant}},
  \bibinfo {author} {\bibfnamefont {A.~V.}\ \bibnamefont {Zinovev}}, \bibinfo
  {author} {\bibfnamefont {K.~K.}\ \bibnamefont {Kovi}}, \ and\ \bibinfo
  {author} {\bibfnamefont {S.~V.}\ \bibnamefont {Baryshev}},\ }in\ \href
  {https://doi.org/10.1109/IVNC.2017.8051543} {\emph {\bibinfo {booktitle}
  {30th International Vacuum Nanoelectronics Conference (IVNC)}}}\ (\bibinfo
  {year} {2017})\ pp.\ \bibinfo {pages} {46--47}\BibitemShut {NoStop}%
\bibitem [{\citenamefont {Chubenko}\ \emph
  {et~al.}(2017{\natexlab{c}})\citenamefont {Chubenko}, \citenamefont
  {Afanasev}, \citenamefont {Baturin},\ and\ \citenamefont
  {Baryshev}}]{Chubenko_2017_IVNC_2}%
  \BibitemOpen
  \bibfield  {author} {\bibinfo {author} {\bibfnamefont {O.}~\bibnamefont
  {Chubenko}}, \bibinfo {author} {\bibfnamefont {A.}~\bibnamefont {Afanasev}},
  \bibinfo {author} {\bibfnamefont {S.~S.}\ \bibnamefont {Baturin}}, \ and\
  \bibinfo {author} {\bibfnamefont {S.~V.}\ \bibnamefont {Baryshev}},\ }in\
  \href {https://doi.org/10.1109/IVNC.2017.8051647} {\emph {\bibinfo
  {booktitle} {30th International Vacuum Nanoelectronics Conference (IVNC)}}}\
  (\bibinfo {year} {2017})\ pp.\ \bibinfo {pages} {284--285}\BibitemShut
  {NoStop}%
\bibitem [{\citenamefont {Li}(2006)}]{Li_Semiconductor}%
  \BibitemOpen
  \bibfield  {author} {\bibinfo {author} {\bibfnamefont {S.~S.}\ \bibnamefont
  {Li}},\ }\href@noop {} {\emph {\bibinfo {title} {Semiconductor Physical
  Electronics}}},\ \bibinfo {edition} {2nd}\ ed.\ (\bibinfo  {publisher}
  {Springer Science \& Business Media},\ \bibinfo {year} {2006})\BibitemShut
  {NoStop}%
\bibitem [{\citenamefont {Masetti}, \citenamefont {Severi},\ and\ \citenamefont
  {Solmi}(1983)}]{Masetti_1983}%
  \BibitemOpen
  \bibfield  {author} {\bibinfo {author} {\bibfnamefont {G.}~\bibnamefont
  {Masetti}}, \bibinfo {author} {\bibfnamefont {M.}~\bibnamefont {Severi}}, \
  and\ \bibinfo {author} {\bibfnamefont {S.}~\bibnamefont {Solmi}},\ }\href
  {https://doi.org/10.1109/T-ED.1983.21207} {\bibfield  {journal} {\bibinfo
  {journal} {IEEE Trans. Electron Devices}\ }\textbf {\bibinfo {volume}
  {ED-30}},\ \bibinfo {pages} {764} (\bibinfo {year} {1983})}\BibitemShut
  {NoStop}%
\bibitem [{\citenamefont {Reggiani}\ \emph {et~al.}(2002)\citenamefont
  {Reggiani}, \citenamefont {Valdinoci}, \citenamefont {Colalongo},
  \citenamefont {Rudan}, \citenamefont {Baccarani}, \citenamefont {Stricker},
  \citenamefont {Illien}, \citenamefont {Felber}, \citenamefont {Fichtner},\
  and\ \citenamefont {Zullino}}]{Reggiani_2002}%
  \BibitemOpen
  \bibfield  {author} {\bibinfo {author} {\bibfnamefont {S.}~\bibnamefont
  {Reggiani}}, \bibinfo {author} {\bibfnamefont {M.}~\bibnamefont {Valdinoci}},
  \bibinfo {author} {\bibfnamefont {L.}~\bibnamefont {Colalongo}}, \bibinfo
  {author} {\bibfnamefont {M.}~\bibnamefont {Rudan}}, \bibinfo {author}
  {\bibfnamefont {G.}~\bibnamefont {Baccarani}}, \bibinfo {author}
  {\bibfnamefont {A.~D.}\ \bibnamefont {Stricker}}, \bibinfo {author}
  {\bibfnamefont {F.}~\bibnamefont {Illien}}, \bibinfo {author} {\bibfnamefont
  {N.}~\bibnamefont {Felber}}, \bibinfo {author} {\bibfnamefont
  {W.}~\bibnamefont {Fichtner}}, \ and\ \bibinfo {author} {\bibfnamefont
  {L.}~\bibnamefont {Zullino}},\ }\href {https://doi.org/10.1109/16.987121}
  {\bibfield  {journal} {\bibinfo  {journal} {IEEE Trans. Electron Devices}\
  }\textbf {\bibinfo {volume} {49}},\ \bibinfo {pages} {490} (\bibinfo {year}
  {2002})}\BibitemShut {NoStop}%
\bibitem [{\citenamefont {Caughey}\ and\ \citenamefont
  {Thomas}(1967)}]{Caughey_1967}%
  \BibitemOpen
  \bibfield  {author} {\bibinfo {author} {\bibfnamefont {D.~M.}\ \bibnamefont
  {Caughey}}\ and\ \bibinfo {author} {\bibfnamefont {R.~E.}\ \bibnamefont
  {Thomas}},\ }\href {https://doi.org/10.1109/PROC.1967.6123} {\bibfield
  {journal} {\bibinfo  {journal} {Proc. IEEE}\ }\textbf {\bibinfo {volume}
  {55}},\ \bibinfo {pages} {2192} (\bibinfo {year} {1967})}\BibitemShut
  {NoStop}%
\bibitem [{\citenamefont {Swirhun}, \citenamefont {Del~Alamo},\ and\
  \citenamefont {Swanson}(1986)}]{Swirhun_1986_EDL}%
  \BibitemOpen
  \bibfield  {author} {\bibinfo {author} {\bibfnamefont {S.~E.}\ \bibnamefont
  {Swirhun}}, \bibinfo {author} {\bibfnamefont {J.~A.}\ \bibnamefont
  {Del~Alamo}}, \ and\ \bibinfo {author} {\bibfnamefont {R.~M.}\ \bibnamefont
  {Swanson}},\ }\href {https://doi.org/10.1109/EDL.1986.26333} {\bibfield
  {journal} {\bibinfo  {journal} {Electron Device Lett.}\ }\textbf {\bibinfo
  {volume} {7}},\ \bibinfo {pages} {168} (\bibinfo {year} {1986})}\BibitemShut
  {NoStop}%
\bibitem [{\citenamefont {Swirhun}, \citenamefont {Kwark},\ and\ \citenamefont
  {Swanson}(1986)}]{Swirhun_1986}%
  \BibitemOpen
  \bibfield  {author} {\bibinfo {author} {\bibfnamefont {S.~E.}\ \bibnamefont
  {Swirhun}}, \bibinfo {author} {\bibfnamefont {Y.-H.}\ \bibnamefont {Kwark}},
  \ and\ \bibinfo {author} {\bibfnamefont {R.~M.}\ \bibnamefont {Swanson}},\
  }in\ \href {https://doi.org/10.1109/IEDM.1986.191101} {\emph {\bibinfo
  {booktitle} {International Electron Devices Meeting}}}\ (\bibinfo {year}
  {1986})\ pp.\ \bibinfo {pages} {24--27}\BibitemShut {NoStop}%
\bibitem [{\citenamefont {Ryder}(1953)}]{Ryder_1953}%
  \BibitemOpen
  \bibfield  {author} {\bibinfo {author} {\bibfnamefont {E.~J.}\ \bibnamefont
  {Ryder}},\ }\href {https://doi.org/10.1103/PhysRev.90.766} {\bibfield
  {journal} {\bibinfo  {journal} {Phys. Rev.}\ }\textbf {\bibinfo {volume}
  {90}},\ \bibinfo {pages} {766} (\bibinfo {year} {1953})}\BibitemShut
  {NoStop}%
\bibitem [{\citenamefont {Gunn}(1956)}]{Gunn_1956}%
  \BibitemOpen
  \bibfield  {author} {\bibinfo {author} {\bibfnamefont {J.~B.}\ \bibnamefont
  {Gunn}},\ }\href {https://doi.org/10.1080/00207215608937008} {\bibfield
  {journal} {\bibinfo  {journal} {J. Electronics}\ }\textbf {\bibinfo {volume}
  {2}},\ \bibinfo {pages} {87} (\bibinfo {year} {1956})}\BibitemShut {NoStop}%
\bibitem [{\citenamefont {Prior}(1959)}]{Prior_1959}%
  \BibitemOpen
  \bibfield  {author} {\bibinfo {author} {\bibfnamefont {A.~C.}\ \bibnamefont
  {Prior}},\ }\href {https://doi.org/10.1016/0022-3697(60)90034-2} {\bibfield
  {journal} {\bibinfo  {journal} {J. Phys. Chem. Solids}\ }\textbf {\bibinfo
  {volume} {12}},\ \bibinfo {pages} {175} (\bibinfo {year} {1959})}\BibitemShut
  {NoStop}%
\bibitem [{\citenamefont {Schweitzer}\ and\ \citenamefont
  {Seeger}(1965)}]{Schweitzer_1965}%
  \BibitemOpen
  \bibfield  {author} {\bibinfo {author} {\bibfnamefont {D.}~\bibnamefont
  {Schweitzer}}\ and\ \bibinfo {author} {\bibfnamefont {K.}~\bibnamefont
  {Seeger}},\ }\href {https://doi.org/10.1007/BF01380796} {\bibfield  {journal}
  {\bibinfo  {journal} {Zeitschrift f{\"u}r Physik A Hadrons and Nuclei}\
  }\textbf {\bibinfo {volume} {183}},\ \bibinfo {pages} {207} (\bibinfo {year}
  {1965})}\BibitemShut {NoStop}%
\bibitem [{\citenamefont {Shockley}(1951)}]{Shockley_1951}%
  \BibitemOpen
  \bibfield  {author} {\bibinfo {author} {\bibfnamefont {W.}~\bibnamefont
  {Shockley}},\ }\href {https://doi.org/10.1002/j.1538-7305.1951.tb03693.x}
  {\bibfield  {journal} {\bibinfo  {journal} {Bell System Tech. J.}\ }\textbf
  {\bibinfo {volume} {30}},\ \bibinfo {pages} {990} (\bibinfo {year}
  {1951})}\BibitemShut {NoStop}%
\bibitem [{\citenamefont {Forbes}(2006)}]{Forbes_2006}%
  \BibitemOpen
  \bibfield  {author} {\bibinfo {author} {\bibfnamefont {R.~G.}\ \bibnamefont
  {Forbes}},\ }\href {https://doi.org/10.1063/1.2354582} {\bibfield  {journal}
  {\bibinfo  {journal} {Appl. Phys. Lett.}\ }\textbf {\bibinfo {volume} {89}},\
  \bibinfo {pages} {113122} (\bibinfo {year} {2006})}\BibitemShut {NoStop}%
\bibitem [{\citenamefont {Karabutov}\ \emph {et~al.}(1999)\citenamefont
  {Karabutov}, \citenamefont {Frolov}, \citenamefont {Pimenov},\ and\
  \citenamefont {Konov}}]{Karabutov_1999}%
  \BibitemOpen
  \bibfield  {author} {\bibinfo {author} {\bibfnamefont {A.~V.}\ \bibnamefont
  {Karabutov}}, \bibinfo {author} {\bibfnamefont {V.~D.}\ \bibnamefont
  {Frolov}}, \bibinfo {author} {\bibfnamefont {S.~M.}\ \bibnamefont {Pimenov}},
  \ and\ \bibinfo {author} {\bibfnamefont {V.~I.}\ \bibnamefont {Konov}},\
  }\href {https://doi.org/10.1016/S0925-9635(98)00308-2} {\bibfield  {journal}
  {\bibinfo  {journal} {Diamond and Rel. Mat.}\ }\textbf {\bibinfo {volume}
  {8}},\ \bibinfo {pages} {763} (\bibinfo {year} {1999})}\BibitemShut {NoStop}%
\bibitem [{\citenamefont {Harniman}\ \emph {et~al.}(2015)\citenamefont
  {Harniman}, \citenamefont {Fox}, \citenamefont {Janssen}, \citenamefont
  {Drijkoningen}, \citenamefont {Haenen},\ and\ \citenamefont
  {May}}]{Harniman_2015}%
  \BibitemOpen
  \bibfield  {author} {\bibinfo {author} {\bibfnamefont {R.~L.}\ \bibnamefont
  {Harniman}}, \bibinfo {author} {\bibfnamefont {O.~J.~L.}\ \bibnamefont
  {Fox}}, \bibinfo {author} {\bibfnamefont {W.}~\bibnamefont {Janssen}},
  \bibinfo {author} {\bibfnamefont {S.}~\bibnamefont {Drijkoningen}}, \bibinfo
  {author} {\bibfnamefont {K.}~\bibnamefont {Haenen}}, \ and\ \bibinfo {author}
  {\bibfnamefont {P.~W.}\ \bibnamefont {May}},\ }\href
  {https://doi.org/10.1016/j.carbon.2015.06.082} {\bibfield  {journal}
  {\bibinfo  {journal} {Carbon}\ }\textbf {\bibinfo {volume} {94}},\ \bibinfo
  {pages} {386} (\bibinfo {year} {2015})}\BibitemShut {NoStop}%
\bibitem [{\citenamefont {Chen}\ \emph {et~al.}(2001)\citenamefont {Chen},
  \citenamefont {Gruen}, \citenamefont {Krauss}, \citenamefont {Corrigan},
  \citenamefont {Witek},\ and\ \citenamefont {Swain}}]{Chen_2001}%
  \BibitemOpen
  \bibfield  {author} {\bibinfo {author} {\bibfnamefont {Q.}~\bibnamefont
  {Chen}}, \bibinfo {author} {\bibfnamefont {D.~M.}\ \bibnamefont {Gruen}},
  \bibinfo {author} {\bibfnamefont {A.~R.}\ \bibnamefont {Krauss}}, \bibinfo
  {author} {\bibfnamefont {T.~D.}\ \bibnamefont {Corrigan}}, \bibinfo {author}
  {\bibfnamefont {M.}~\bibnamefont {Witek}}, \ and\ \bibinfo {author}
  {\bibfnamefont {G.~M.}\ \bibnamefont {Swain}},\ }\href
  {https://doi.org/10.1149/1.1344550} {\bibfield  {journal} {\bibinfo
  {journal} {J. Electrochem. Soc.}\ }\textbf {\bibinfo {volume} {148}},\
  \bibinfo {pages} {E44} (\bibinfo {year} {2001})}\BibitemShut {NoStop}%
\bibitem [{\citenamefont {Beloborodov}\ \emph {et~al.}(2006)\citenamefont
  {Beloborodov}, \citenamefont {Zapol}, \citenamefont {Gruen},\ and\
  \citenamefont {Curtiss}}]{Beloborodov_2006}%
  \BibitemOpen
  \bibfield  {author} {\bibinfo {author} {\bibfnamefont {I.~S.}\ \bibnamefont
  {Beloborodov}}, \bibinfo {author} {\bibfnamefont {P.}~\bibnamefont {Zapol}},
  \bibinfo {author} {\bibfnamefont {D.~M.}\ \bibnamefont {Gruen}}, \ and\
  \bibinfo {author} {\bibfnamefont {L.~A.}\ \bibnamefont {Curtiss}},\ }\href
  {https://doi.org/10.1103/PhysRevB.74.235434} {\bibfield  {journal} {\bibinfo
  {journal} {Phys. Rev. B}\ }\textbf {\bibinfo {volume} {74}},\ \bibinfo
  {pages} {235434} (\bibinfo {year} {2006})}\BibitemShut {NoStop}%
\bibitem [{\citenamefont {Achatz}\ \emph
  {et~al.}(2006{\natexlab{a}})\citenamefont {Achatz}, \citenamefont {Williams},
  \citenamefont {Bruno}, \citenamefont {Gruen}, \citenamefont {Garrido},\ and\
  \citenamefont {Stutzmann}}]{Achatz_2006}%
  \BibitemOpen
  \bibfield  {author} {\bibinfo {author} {\bibfnamefont {P.}~\bibnamefont
  {Achatz}}, \bibinfo {author} {\bibfnamefont {O.~A.}\ \bibnamefont
  {Williams}}, \bibinfo {author} {\bibfnamefont {P.}~\bibnamefont {Bruno}},
  \bibinfo {author} {\bibfnamefont {D.~M.}\ \bibnamefont {Gruen}}, \bibinfo
  {author} {\bibfnamefont {J.~A.}\ \bibnamefont {Garrido}}, \ and\ \bibinfo
  {author} {\bibfnamefont {M.}~\bibnamefont {Stutzmann}},\ }\href
  {https://doi.org/10.1103/PhysRevB.74.155429} {\bibfield  {journal} {\bibinfo
  {journal} {Phys. Rev. B}\ }\textbf {\bibinfo {volume} {74}},\ \bibinfo
  {pages} {155429} (\bibinfo {year} {2006}{\natexlab{a}})}\BibitemShut
  {NoStop}%
\bibitem [{\citenamefont {Bhattacharyya}\ \emph {et~al.}(2001)\citenamefont
  {Bhattacharyya}, \citenamefont {Auciello}, \citenamefont {Birrell},
  \citenamefont {Carlisle}, \citenamefont {Curtiss}, \citenamefont {Goyette},\
  and\ \citenamefont {Zapol}}]{Bhattacharyya_2001}%
  \BibitemOpen
  \bibfield  {author} {\bibinfo {author} {\bibfnamefont {S.}~\bibnamefont
  {Bhattacharyya}}, \bibinfo {author} {\bibfnamefont {O.}~\bibnamefont
  {Auciello}}, \bibinfo {author} {\bibfnamefont {J.}~\bibnamefont {Birrell}},
  \bibinfo {author} {\bibfnamefont {J.~A.}\ \bibnamefont {Carlisle}}, \bibinfo
  {author} {\bibfnamefont {L.~A.}\ \bibnamefont {Curtiss}}, \bibinfo {author}
  {\bibfnamefont {A.~R. K. J. S. A.~S.}\ \bibnamefont {Goyette}, \bibfnamefont
  {A.~N.) D. M.~Gruen}}, \ and\ \bibinfo {author} {\bibfnamefont
  {P.}~\bibnamefont {Zapol}},\ }\href {https://doi.org/10.1063/1.1400761}
  {\bibfield  {journal} {\bibinfo  {journal} {Appl. Phys. Lett.}\ }\textbf
  {\bibinfo {volume} {79}},\ \bibinfo {pages} {1441} (\bibinfo {year}
  {2001})}\BibitemShut {NoStop}%
\bibitem [{\citenamefont {Birrell}\ \emph {et~al.}(2002)\citenamefont
  {Birrell}, \citenamefont {Carlisle}, \citenamefont {Auciello}, \citenamefont
  {Gruen},\ and\ \citenamefont {Gibson}}]{Birrell_2002}%
  \BibitemOpen
  \bibfield  {author} {\bibinfo {author} {\bibfnamefont {J.}~\bibnamefont
  {Birrell}}, \bibinfo {author} {\bibfnamefont {J.~A.}\ \bibnamefont
  {Carlisle}}, \bibinfo {author} {\bibfnamefont {O.}~\bibnamefont {Auciello}},
  \bibinfo {author} {\bibfnamefont {D.~M.}\ \bibnamefont {Gruen}}, \ and\
  \bibinfo {author} {\bibfnamefont {J.~M.}\ \bibnamefont {Gibson}},\ }\href
  {https://doi.org/10.1063/1.1503153} {\bibfield  {journal} {\bibinfo
  {journal} {Appl. Phys. Lett.}\ }\textbf {\bibinfo {volume} {81}},\ \bibinfo
  {pages} {2235} (\bibinfo {year} {2002})}\BibitemShut {NoStop}%
\bibitem [{\citenamefont {Williams}\ \emph {et~al.}(2004)\citenamefont
  {Williams}, \citenamefont {Curat}, \citenamefont {Gerbi}, \citenamefont
  {Gruen},\ and\ \citenamefont {Jackman}}]{Williams_2004}%
  \BibitemOpen
  \bibfield  {author} {\bibinfo {author} {\bibfnamefont {O.~A.}\ \bibnamefont
  {Williams}}, \bibinfo {author} {\bibfnamefont {S.}~\bibnamefont {Curat}},
  \bibinfo {author} {\bibfnamefont {J.~E.}\ \bibnamefont {Gerbi}}, \bibinfo
  {author} {\bibfnamefont {D.~M.}\ \bibnamefont {Gruen}}, \ and\ \bibinfo
  {author} {\bibfnamefont {R.~B.}\ \bibnamefont {Jackman}},\ }\href
  {https://doi.org/10.1063/1.1785288} {\bibfield  {journal} {\bibinfo
  {journal} {Appl. Phys. Lett.}\ }\textbf {\bibinfo {volume} {85}},\ \bibinfo
  {pages} {1680} (\bibinfo {year} {2004})}\BibitemShut {NoStop}%
\bibitem [{\citenamefont {Ikeda}\ \emph {et~al.}(2008)\citenamefont {Ikeda},
  \citenamefont {Teii}, \citenamefont {Casiraghi}, \citenamefont {Robertson},\
  and\ \citenamefont {Ferrari}}]{Ikeda_2008}%
  \BibitemOpen
  \bibfield  {author} {\bibinfo {author} {\bibfnamefont {T.}~\bibnamefont
  {Ikeda}}, \bibinfo {author} {\bibfnamefont {K.}~\bibnamefont {Teii}},
  \bibinfo {author} {\bibfnamefont {C.}~\bibnamefont {Casiraghi}}, \bibinfo
  {author} {\bibfnamefont {J.}~\bibnamefont {Robertson}}, \ and\ \bibinfo
  {author} {\bibfnamefont {A.~C.}\ \bibnamefont {Ferrari}},\ }\href
  {https://doi.org/10.1063/1.2990061} {\bibfield  {journal} {\bibinfo
  {journal} {J. Appl. Phys.}\ }\textbf {\bibinfo {volume} {104}},\ \bibinfo
  {pages} {073720} (\bibinfo {year} {2008})}\BibitemShut {NoStop}%
\bibitem [{\citenamefont {Ilie}\ \emph {et~al.}(2000)\citenamefont {Ilie},
  \citenamefont {Ferrari}, \citenamefont {Yagi},\ and\ \citenamefont
  {Robertson}}]{Ilie_2000}%
  \BibitemOpen
  \bibfield  {author} {\bibinfo {author} {\bibfnamefont {A.}~\bibnamefont
  {Ilie}}, \bibinfo {author} {\bibfnamefont {A.~C.}\ \bibnamefont {Ferrari}},
  \bibinfo {author} {\bibfnamefont {T.}~\bibnamefont {Yagi}}, \ and\ \bibinfo
  {author} {\bibfnamefont {J.}~\bibnamefont {Robertson}},\ }\href
  {https://doi.org/10.1063/1.126430} {\bibfield  {journal} {\bibinfo  {journal}
  {Appl. Phys. Lett.}\ }\textbf {\bibinfo {volume} {76}},\ \bibinfo {pages}
  {2627} (\bibinfo {year} {2000})}\BibitemShut {NoStop}%
\bibitem [{\citenamefont {Carey}, \citenamefont {Forrest},\ and\ \citenamefont
  {Silva}(2001)}]{Carey_2001}%
  \BibitemOpen
  \bibfield  {author} {\bibinfo {author} {\bibfnamefont {J.~D.}\ \bibnamefont
  {Carey}}, \bibinfo {author} {\bibfnamefont {R.~D.}\ \bibnamefont {Forrest}},
  \ and\ \bibinfo {author} {\bibfnamefont {S.~R.~P.}\ \bibnamefont {Silva}},\
  }\href {https://doi.org/10.1063/1.1366369} {\bibfield  {journal} {\bibinfo
  {journal} {Appl. Phys. Lett.}\ }\textbf {\bibinfo {volume} {78}},\ \bibinfo
  {pages} {2339} (\bibinfo {year} {2001})}\BibitemShut {NoStop}%
\bibitem [{\citenamefont {Dasgupta}\ \emph {et~al.}(1991)\citenamefont
  {Dasgupta}, \citenamefont {Demichelis}, \citenamefont {Pirri},\ and\
  \citenamefont {Tagliaferro}}]{Dasgupta_1991}%
  \BibitemOpen
  \bibfield  {author} {\bibinfo {author} {\bibfnamefont {D.}~\bibnamefont
  {Dasgupta}}, \bibinfo {author} {\bibfnamefont {F.}~\bibnamefont
  {Demichelis}}, \bibinfo {author} {\bibfnamefont {C.~F.}\ \bibnamefont
  {Pirri}}, \ and\ \bibinfo {author} {\bibfnamefont {A.}~\bibnamefont
  {Tagliaferro}},\ }\href {https://doi.org/10.1103/PhysRevB.43.2131} {\bibfield
   {journal} {\bibinfo  {journal} {Phys. Rev. B}\ }\textbf {\bibinfo {volume}
  {43}},\ \bibinfo {pages} {2131} (\bibinfo {year} {1991})}\BibitemShut
  {NoStop}%
\bibitem [{\citenamefont {Nesl{\'a}dek}\ \emph {et~al.}(1996)\citenamefont
  {Nesl{\'a}dek}, \citenamefont {Meykens}, \citenamefont {Stals}, \citenamefont
  {Van{\v{e}}{\v{c}}ek},\ and\ \citenamefont {Rosa}}]{Nesladek_1996}%
  \BibitemOpen
  \bibfield  {author} {\bibinfo {author} {\bibfnamefont {M.}~\bibnamefont
  {Nesl{\'a}dek}}, \bibinfo {author} {\bibfnamefont {K.}~\bibnamefont
  {Meykens}}, \bibinfo {author} {\bibfnamefont {L.~M.}\ \bibnamefont {Stals}},
  \bibinfo {author} {\bibfnamefont {M.}~\bibnamefont {Van{\v{e}}{\v{c}}ek}}, \
  and\ \bibinfo {author} {\bibfnamefont {J.}~\bibnamefont {Rosa}},\ }\href
  {https://doi.org/10.1103/PhysRevB.54.5552} {\bibfield  {journal} {\bibinfo
  {journal} {Phys. Rev. B}\ }\textbf {\bibinfo {volume} {54}},\ \bibinfo
  {pages} {5552} (\bibinfo {year} {1996})}\BibitemShut {NoStop}%
\bibitem [{\citenamefont {Zammit}\ \emph {et~al.}(1998)\citenamefont {Zammit},
  \citenamefont {Madhusoodanan}, \citenamefont {Marinelli}, \citenamefont
  {Mercuri},\ and\ \citenamefont {Foglietta}}]{Zammit_1998}%
  \BibitemOpen
  \bibfield  {author} {\bibinfo {author} {\bibfnamefont {U.}~\bibnamefont
  {Zammit}}, \bibinfo {author} {\bibfnamefont {K.~N.}\ \bibnamefont
  {Madhusoodanan}}, \bibinfo {author} {\bibfnamefont {M.}~\bibnamefont
  {Marinelli}}, \bibinfo {author} {\bibfnamefont {F.}~\bibnamefont {Mercuri}},
  \ and\ \bibinfo {author} {\bibfnamefont {S.}~\bibnamefont {Foglietta}},\
  }\href {https://doi.org/10.1103/PhysRevB.57.4518} {\bibfield  {journal}
  {\bibinfo  {journal} {Phys. Rev. B}\ }\textbf {\bibinfo {volume} {57}},\
  \bibinfo {pages} {4518} (\bibinfo {year} {1998})}\BibitemShut {NoStop}%
\bibitem [{\citenamefont {Tauc}(1974)}]{Tauc}%
  \BibitemOpen
  \bibfield  {author} {\bibinfo {author} {\bibfnamefont {J.}~\bibnamefont
  {Tauc}},\ }in\ \href@noop {} {\emph {\bibinfo {booktitle} {Amorphous and
  Liquid Semiconductors}}}\ (\bibinfo  {publisher} {Springer},\ \bibinfo {year}
  {1974})\ pp.\ \bibinfo {pages} {159--220}\BibitemShut {NoStop}%
\bibitem [{\citenamefont {Yu}\ and\ \citenamefont {Cardona}(2010)}]{Yu}%
  \BibitemOpen
  \bibfield  {author} {\bibinfo {author} {\bibfnamefont {P.~Y.}\ \bibnamefont
  {Yu}}\ and\ \bibinfo {author} {\bibfnamefont {M.}~\bibnamefont {Cardona}},\
  }\href@noop {} {\emph {\bibinfo {title} {Fundamentals of Semiconductors:
  Physics and Materials Properties}}},\ \bibinfo {edition} {4th}\ ed.\
  (\bibinfo  {publisher} {Springer},\ \bibinfo {year} {2010})\BibitemShut
  {NoStop}%
\bibitem [{\citenamefont {AlZahrani}\ and\ \citenamefont
  {Srivastava}(2009)}]{Alzahrani_2009}%
  \BibitemOpen
  \bibfield  {author} {\bibinfo {author} {\bibfnamefont {A.~Z.}\ \bibnamefont
  {AlZahrani}}\ and\ \bibinfo {author} {\bibfnamefont {G.~P.}\ \bibnamefont
  {Srivastava}},\ }\href {http://dx.doi.org/10.1590/S0103-97332009000600013}
  {\bibfield  {journal} {\bibinfo  {journal} {Brazilian J. Phys.}\ }\textbf
  {\bibinfo {volume} {39}},\ \bibinfo {pages} {694} (\bibinfo {year}
  {2009})}\BibitemShut {NoStop}%
\bibitem [{\citenamefont {Pelton}\ \emph {et~al.}(1998)\citenamefont {Pelton},
  \citenamefont {O’Leary}, \citenamefont {Gaspari},\ and\ \citenamefont
  {Zukotynski}}]{Pelton_1998}%
  \BibitemOpen
  \bibfield  {author} {\bibinfo {author} {\bibfnamefont {M.}~\bibnamefont
  {Pelton}}, \bibinfo {author} {\bibfnamefont {S.~K.}\ \bibnamefont
  {O’Leary}}, \bibinfo {author} {\bibfnamefont {F.}~\bibnamefont {Gaspari}},
  \ and\ \bibinfo {author} {\bibfnamefont {S.}~\bibnamefont {Zukotynski}},\
  }\href {https://doi.org/10.1063/1.366793} {\bibfield  {journal} {\bibinfo
  {journal} {J. Appl. Phys.}\ }\textbf {\bibinfo {volume} {83}},\ \bibinfo
  {pages} {1029} (\bibinfo {year} {1998})}\BibitemShut {NoStop}%
\bibitem [{\citenamefont {Achatz}\ \emph
  {et~al.}(2006{\natexlab{b}})\citenamefont {Achatz}, \citenamefont {Garrido},
  \citenamefont {Stutzmann}, \citenamefont {Williams}, \citenamefont {Gruen},
  \citenamefont {Kromka},\ and\ \citenamefont
  {Steinm{\"u}ller}}]{Achatz_2006_APL}%
  \BibitemOpen
  \bibfield  {author} {\bibinfo {author} {\bibfnamefont {P.}~\bibnamefont
  {Achatz}}, \bibinfo {author} {\bibfnamefont {J.~A.}\ \bibnamefont {Garrido}},
  \bibinfo {author} {\bibfnamefont {M.}~\bibnamefont {Stutzmann}}, \bibinfo
  {author} {\bibfnamefont {O.~A.}\ \bibnamefont {Williams}}, \bibinfo {author}
  {\bibfnamefont {D.~M.}\ \bibnamefont {Gruen}}, \bibinfo {author}
  {\bibfnamefont {A.}~\bibnamefont {Kromka}}, \ and\ \bibinfo {author}
  {\bibfnamefont {D.}~\bibnamefont {Steinm{\"u}ller}},\ }\href
  {https://doi.org/10.1063/1.2183366} {\bibfield  {journal} {\bibinfo
  {journal} {Appl. Phys. Lett.}\ }\textbf {\bibinfo {volume} {88}},\ \bibinfo
  {pages} {101908} (\bibinfo {year} {2006}{\natexlab{b}})}\BibitemShut
  {NoStop}%
\bibitem [{\citenamefont {Csencsits}\ \emph {et~al.}(1996)\citenamefont
  {Csencsits}, \citenamefont {Zuiker}, \citenamefont {Gruen},\ and\
  \citenamefont {Krauss}}]{Csencsits_1996}%
  \BibitemOpen
  \bibfield  {author} {\bibinfo {author} {\bibfnamefont {R.}~\bibnamefont
  {Csencsits}}, \bibinfo {author} {\bibfnamefont {C.~D.}\ \bibnamefont
  {Zuiker}}, \bibinfo {author} {\bibfnamefont {D.~M.}\ \bibnamefont {Gruen}}, \
  and\ \bibinfo {author} {\bibfnamefont {A.~R.}\ \bibnamefont {Krauss}},\
  }\href {https://doi.org/10.4028/www.scientific.net/SSP.51-52.261} {\bibfield
  {journal} {\bibinfo  {journal} {Solid State Phenomena}\ }\textbf {\bibinfo
  {volume} {51}},\ \bibinfo {pages} {261} (\bibinfo {year} {1996})}\BibitemShut
  {NoStop}%
\bibitem [{\citenamefont {Bhattacharyya}\ and\ \citenamefont
  {Silva}(2005)}]{Bhattacharyya_2005}%
  \BibitemOpen
  \bibfield  {author} {\bibinfo {author} {\bibfnamefont {S.}~\bibnamefont
  {Bhattacharyya}}\ and\ \bibinfo {author} {\bibfnamefont {S.~R.~P.}\
  \bibnamefont {Silva}},\ }\href {https://doi.org/10.1016/j.tsf.2004.11.125}
  {\bibfield  {journal} {\bibinfo  {journal} {Thin Solid Films}\ }\textbf
  {\bibinfo {volume} {482}},\ \bibinfo {pages} {94} (\bibinfo {year}
  {2005})}\BibitemShut {NoStop}%
\bibitem [{\citenamefont {P{\'e}rez~Quintero}\ \emph
  {et~al.}(2014)\citenamefont {P{\'e}rez~Quintero}, \citenamefont {Antipov},
  \citenamefont {Sumant}, \citenamefont {Jing},\ and\ \citenamefont
  {Baryshev}}]{Quintero_2014}%
  \BibitemOpen
  \bibfield  {author} {\bibinfo {author} {\bibfnamefont {K.~J.}\ \bibnamefont
  {P{\'e}rez~Quintero}}, \bibinfo {author} {\bibfnamefont {S.}~\bibnamefont
  {Antipov}}, \bibinfo {author} {\bibfnamefont {A.~V.}\ \bibnamefont {Sumant}},
  \bibinfo {author} {\bibfnamefont {C.}~\bibnamefont {Jing}}, \ and\ \bibinfo
  {author} {\bibfnamefont {S.~V.}\ \bibnamefont {Baryshev}},\ }\href
  {https://doi.org/10.1063/1.4896418} {\bibfield  {journal} {\bibinfo
  {journal} {Appl. Phys. Lett.}\ }\textbf {\bibinfo {volume} {105}},\ \bibinfo
  {pages} {123103} (\bibinfo {year} {2014})}\BibitemShut {NoStop}%
\bibitem [{\citenamefont {Conwell}(1967)}]{Conwell_High}%
  \BibitemOpen
  \bibfield  {author} {\bibinfo {author} {\bibfnamefont {E.~M.}\ \bibnamefont
  {Conwell}},\ }\href@noop {} {\emph {\bibinfo {title} {High Field Transport in
  Semiconductors}}},\ Supplement 9\ (\bibinfo  {publisher} {Academic Press},\
  \bibinfo {year} {1967})\BibitemShut {NoStop}%
\bibitem [{\citenamefont {Dyke}\ and\ \citenamefont
  {Trolan}(1953)}]{Dyke_1953}%
  \BibitemOpen
  \bibfield  {author} {\bibinfo {author} {\bibfnamefont {W.~P.}\ \bibnamefont
  {Dyke}}\ and\ \bibinfo {author} {\bibfnamefont {J.~K.}\ \bibnamefont
  {Trolan}},\ }\href {https://doi.org/10.1103/PhysRev.89.799} {\bibfield
  {journal} {\bibinfo  {journal} {Phys. Rev.}\ }\textbf {\bibinfo {volume}
  {89}},\ \bibinfo {pages} {799} (\bibinfo {year} {1953})}\BibitemShut
  {NoStop}%
\bibitem [{\citenamefont {Barbour}\ \emph {et~al.}(1953)\citenamefont
  {Barbour}, \citenamefont {Dolan}, \citenamefont {Trolan}, \citenamefont
  {Martin},\ and\ \citenamefont {Dyke}}]{Barbour_1953}%
  \BibitemOpen
  \bibfield  {author} {\bibinfo {author} {\bibfnamefont {J.~P.}\ \bibnamefont
  {Barbour}}, \bibinfo {author} {\bibfnamefont {W.~W.}\ \bibnamefont {Dolan}},
  \bibinfo {author} {\bibfnamefont {J.~K.}\ \bibnamefont {Trolan}}, \bibinfo
  {author} {\bibfnamefont {E.~E.}\ \bibnamefont {Martin}}, \ and\ \bibinfo
  {author} {\bibfnamefont {W.~P.}\ \bibnamefont {Dyke}},\ }\href
  {https://doi.org/10.1103/PhysRev.92.45} {\bibfield  {journal} {\bibinfo
  {journal} {Phys. Rev.}\ }\textbf {\bibinfo {volume} {92}},\ \bibinfo {pages}
  {45} (\bibinfo {year} {1953})}\BibitemShut {NoStop}%
\bibitem [{tha()}]{thanks_Forbes}%
  \BibitemOpen
  \href@noop {} {}\bibinfo {note} {R. G. Forbes and A. V. Arkhipov, private
  communication (2017).}\BibitemShut {Stop}%
\bibitem [{\citenamefont {Chen}\ \emph {et~al.}()\citenamefont {Chen},
  \citenamefont {Adhikari}, \citenamefont {Spentzious}, \citenamefont {Kovi},
  \citenamefont {Antipov}, \citenamefont {Jing}, \citenamefont {Schroeder},\
  and\ \citenamefont {Baryshev}}]{Chen_2018}%
  \BibitemOpen
  \bibfield  {author} {\bibinfo {author} {\bibfnamefont {G.}~\bibnamefont
  {Chen}}, \bibinfo {author} {\bibfnamefont {G.}~\bibnamefont {Adhikari}},
  \bibinfo {author} {\bibfnamefont {L.}~\bibnamefont {Spentzious}}, \bibinfo
  {author} {\bibfnamefont {K.~K.}\ \bibnamefont {Kovi}}, \bibinfo {author}
  {\bibfnamefont {S.}~\bibnamefont {Antipov}}, \bibinfo {author} {\bibfnamefont
  {C.}~\bibnamefont {Jing}}, \bibinfo {author} {\bibfnamefont {W.~A.}\
  \bibnamefont {Schroeder}}, \ and\ \bibinfo {author} {\bibfnamefont {S.~V.}\
  \bibnamefont {Baryshev}},\ }\href {https://arxiv.org/abs/1812.00323}
  {\bibinfo  {journal} {\tt arXiv:1812.00323 [cond-mat.mtrl-sci]}\
  }\BibitemShut {NoStop}%
\end{thebibliography}%

\end{document}